\documentclass[useAMS,usenatbib]{mn2e}

\usepackage{graphicx}
\usepackage{graphics}
\usepackage{amsmath}
\usepackage{amssymb}
\usepackage{hyperref}
\usepackage{float}
\usepackage{caption}
\usepackage{color}
\usepackage{epstopdf}

\DeclareGraphicsExtensions{.ps,.pdf,.eps}

\title[ Selecting background galaxies in cluster weak-lensing analyses] {Selecting background galaxies in weak-lensing analysis of galaxy clusters
\thanks{Based  on data collected at the following facilities: Subaru Telescope  (University of Tokyo) and 
obtained from the SMOKA \citep{baba}, which is operated by the Astronomy Data Center, National Astronomical Observatory of Japan;
the Large Binocular Telescope, an international collaboration among institutions in the United States, Italy and Germany.
}
}

\author[I. Formicola et al.]
  {I.~Formicola,$^1$ M.~Radovich,$^2$ M.~Meneghetti,$^1$ P.~Mazzotta,$^3$ A.~Grado,$^4$ C.~Giocoli$^5$\\
  $^1$INAF-Osservatorio Astronomico di Bologna, via Ranzani 1, 40127 Bologna, Italy\\
  $^2$INAF-Osservatorio Astronomico di Padova, vicolo dellÕOsservatorio 5, 35122 Padova, Italy\\
  $^3$Dipartimento di Fisica, Universit\`a ``Tor Vergata", via della Ricerca Scientifica 1, 00133 Roma, Italy\\
  $^4$INAF-Osservatorio Astronomico di Capodimonte, via Moiariello 16, 80131 Napoli, Italy\\
  $^5$Aix Marseille Universit\'e, CNRS, LAM (Laboratoire d'Astrophysique de Marseille), UMR 7326,  13388, Marseille,  France}

\begin{document}
\date{Accepted 2016 February 26}

\pagerange{\pageref{firstpage}--\pageref{lastpage}} \pubyear{2002}

\maketitle

\label{firstpage}

\begin{abstract}
In this paper, we present a new method to select the faint, background galaxies used to derive the mass of galaxy clusters by  weak lensing.
 The method is based on the simultaneous
analysis of the shear signal, that should be consistent with zero for the foreground, unlensed galaxies, and of the colors of the galaxies: photometric data from the COSMic evOlution Survey  are used to train the color selection.
In order to validate this methodology, we test it against a set of state-of-the-art image simulations of mock  galaxy clusters in different redshift [$0.23-0.45$] and mass \textbf{[$0.5-1.55\times10^{15}M_\odot$]}  ranges, mimicking medium-deep multicolor imaging observations (e.g. SUBARU, LBT).\\
 The performance of our method in terms of contamination by unlensed sources is comparable to a selection based on photometric redshifts, which however requires a good spectral coverage and is thus much more observationally  demanding. The application of our method to simulations gives an  average ratio between estimated and true masses of $\sim 0.98 \pm 0.09$.
\\ As a further test, we finally  apply our method to real data, and compare our results with other weak lensing mass estimates in the literature: for  this purpose we choose the cluster Abell 2219 ($z=0.228$), for which multi-band (\textit{BVRi}) data are publicly available.
\end{abstract}


\section{Introduction}
Being the largest gravitationally bound structures in the Universe, galaxy clusters are the most powerful gravitational lenses on the sky. Their lensing signal is often detectable up to a few Mpc from the cluster center. While strong lensing events such as multiple images of distant galaxies and gravitational arcs occur in the cluster cores, at larger radii lensing by clusters appears in the so-called weak-lensing regime. 
In this case, the lensing induced distortions of the galaxy shapes are tiny and the lensing signal is detectable only by averaging over ensembles of 
a sufficient number of lensed galaxies. The measurement of their coherent distortions can be used to constrain the gravitational potential of the cluster and therefore to map its  mass distribution  \citep{bartelmann_01}.
\\Gravitational lensing offers several advantages compared to other methods to measure cluster masses. Indeed observations of the X-ray emission by the cluster intra-cluster-medium or the kinematics of the cluster galaxies can be used to measure masses only under the assumption of some sort of equilibrium between the baryons and the cluster gravitational potential.
 On the other end, there are several possible sources of biases in the weak lensing mass estimates  due to  
 residual instrumental distortions, dilution, wrong assumptions on the source redshift distribution,  large-scale-structure and halo triaxiality 
  which should be properly taken into account in the analysis \citep{henk_2003,corless,ania,henk_2011,carlo_2014,henk_2015, sereno_2015}.
\\  Image simulations probed to be very helpful for understanding some of the limitations of the lensing analysis and for quantifying and possibly reduce these biases. The weak lensing community has been involved in a series of large simulation-based challenges to identify the best performing algorithms for shear measurements \citep{erben_2001, heymans, massey, bridle, kitch}. Despite the fact that none of these experiments addressed the issue of shear measurability in regimes of relatively large shear signal (such as around clusters), these simulations resulted to be extremely useful to quantify  measurement biases and to calibrate the methods.
\cite{massimo_2010} used the image simulation software SkyLens \citep{massimo_2008} to mimic SUBARU and HST observations of a set of mock galaxy clusters. 
 These simulated observations were then processed with a Kaiser-Squires-Broadhurst (KSB) weak lensing pipeline to measure the ellipticities of background galaxies  \citep{kaiser_1995, luppino_1997, henk_1998}. Comparing different techniques to convert the shear measurements into mass estimates, it was possible to quantify the scatter induced by triaxiality and substructure. In a subsequent work, \cite{elena_2012} used analogous image simulations to investigate the dependence of accurate mass measurements on the cluster environment. 
 \\ In this paper, we specifically address the issue  
of the correct identification of background galaxies for the weak lensing analysis. Unlensed sources  in the  sample used for the shape measurement cause 
the  dilution of the lensing signal  which is stronger at small radii  due to the higher number density of cluster galaxies, leading to  underestimate the cluster mass.
\\Several approaches have been proposed to achieve this goal, which generally employ multi-band photometry.  
When photometric data are  available in only two bands, the  selection of background sources is generally done by identifying the galaxies redder than the cluster red-sequence \citep{broad_2005}. A suitable color cut has to be applied in order to avoid the contamination by faint red cluster galaxies which can result from  dust reddening, intrinsic scatter in galaxy colors and measurement errors \citep{okabe_2013}.
However, 
the number density of background galaxies obtained by such selection could not allow us to reach the accuracy in the  
 statistical analysis required by the weak lensing analysis  of galaxy clusters.
 Galaxies bluer than the cluster red-sequence, chosen by  
 properly defining a color offset, can also be included as background galaxies but are highly contaminated by foreground  cluster members \citep{elinor,joy,okabe_2013}. 
 If more than two bands are available,
  a more careful selection can be attempted by identifying the regions of the color-color (CC hereafter) diagrams populated by background galaxies \citep{elinor,umetsu_2010,umetsu_2012,elinor_2013}.
 \\Here, we propose a novel approach which optimises the selection of lensed background sources by employing the photometry of galaxies in the COSMic evOlution Survey (COSMOS).
 The colors of  the COSMOS  galaxies are used as training set for the 
separation of  background/foreground galaxies in the images of the galaxy clusters  under study.
 The method is based on the assumption that the galaxies in the COSMOS data-set \citep{ilbert} are  representative of the colors and of the redshift distribution of the global galaxy population in the universe 
 and that they allow us
 to characterize the distribution in the CC space of galaxies in different redshift ranges.   Simultaneously,  we tune the color cuts on the basis of the
amplitude of the shear signal of the  background selected galaxies.
We use simulations made with the SkyLens software to validate this method assuming different color information and different lens masses and redshifts. We also made comparison with others approach, such as the selections based on the identification of the cluster red sequence and on photometric redshifts.
Finally, for each selection method we derive an estimate of the cluster mass, which we compare to the true mass of the simulated cluster. While testing the performance of the selection method, we also process the simulated data as done for real observations, including the process of measuring the shear from the galaxy images. This serves as a further validation test for the weak-lensing analysis pipeline that we  recently used in \cite{mario_2015}.
\\To test our approach on real observations, we apply our method to the galaxy cluster Abell 2219, for which medium-deep SUBARU and LBT archival observations in the  \textit{BVRi} bands are available. Abell 2219 is  a very well known cluster for which several independent weak lensing analyses have been published  \citep{bardeau_07,hoekstra_07,okabe}.
\\The paper is structured as follows:
in Sect. \ref{sec_sim}, we describe the image simulations used to validate the method.
 The code used to derive the galaxy ellipticities is described in   Sect. \ref{sect_ksb}. 
 The selection method developed  is outlined  in  Sect.  \ref{sel_met} and compared with the method proposed by \citet{elinor}.
 In  Sect. \ref{Analysis of simulated data},
  we describe  the application of the selection method  to  simulated clusters using different combinations of photometric bands, cluster redshift and mass.
  In Sect. \ref{cont_bia}, we quantify the  contamination of the selections, and in Sect. \ref{mas_fit_sim_new}, we discuss how well the weak lensing masses reproduce the true masses of the simulated galaxy clusters. 
  In the second part of the paper, we show the analysis of  Abell 2219 (Sect. \ref{sec_a2219_prop}). More precisely, we give an overview of  the properties of Abell 2219 and of the previous lensing analyses of this cluster. Sect.~\ref{sect_image_reduction} describes our data reduction.
  Sect.~\ref{sect_ksb_abell} outlines the selection of the background galaxies and the derivation of the cluster mass, which is subsequently compared to previous results in the literature. Finally, we summarize our conclusions in Sect. \ref{conclus}. 
  Throughout this paper,  we assume a $\Lambda$CDM cosmology with  $\Omega_\Lambda = 0.7$, $\Omega_M=0.3$, $H_0=70$ km s$^{-1} $Mpc$^{-1}$.

 \section{Lensing Simulations}
\label{sec_sim}
To test and validate the selection method, we use realistic image simulations produced with SkyLens \citep{massimo_2008, massimo_2010, elena_2012}.
The code uses real galaxies taken from the Hubble Ultra-Deep-Field to create mock observations of patches of the sky with virtually any telescope.
 Being coupled with a ray-tracing code, it allows us to include the lensing effects from any mass distribution along the line of sight. As shown in \cite{massimo_2010} and in \cite{elena_2012}, it is particularly suitable to simulate weak and strong lensing effects by galaxy clusters.
\\The mock galaxy clusters used here are generated with the code MOKA \citep{carlo_2011}. The cluster mass distributions have multiple components, namely a triaxial halo, modelled with a Navarro-Frenk-White (NFW) profile  \citep{navarro},  substructures described by truncated Singular Isothermal Spheres  \citep{ben}, and  a  brightest central galaxy (BCG) modelled with an Hernquist density profile \citep{hern}. More details about MOKA can be found in \cite{carlo_2011}.
\\To evaluate the performance of the selection algorithm we need to populate the field-of-view around a mock galaxy cluster with  cluster members, foreground and background galaxies. Most importantly, galaxy colors have to be as much realistic as possible. MOKA populates the dark matter sub-haloes using the Halo Occupation Distribution (HOD) technique. Stellar masses and $B$-band luminosities are assigned to each galaxy accordingly to the mass of the dark matter (sub-)halo within which it formed, following \cite{wang_06}. The morphological type and the spectral energy distribution (SED) are defined on the basis of the stellar mass and such to reproduce observed morphology-density and morphology-radius relations \citep{van_der}. We verified that colors reproduce the expected red-sequences for clusters over a large range of redshifts.
\\The photometry of foreground and background galaxies is computed starting from the best-fitting SEDs determined  by \cite{coe_2006} for the HUDF,
	and convolving them with filter transmission.
\\The redshift distribution of the input sources also matches that of the HUDF. Given the exposure times and the observational set-up used in our simulations, the median redshift  is $z_m\sim1$.
All simulations in this work have a FOV of $30'\times30'$.  For simplicity, we adopt a spatially constant gaussian PSF with FWHM $0.6''$. 
\\We produced several sets of images with different cluster mass and redshift, and combination of filters which we analyze in Sect \ref{Analysis of simulated data}.
\\One set of simulations mimics the archival Subaru and LBT observations available for Abell 2219 which is analyzed in the second part of the paper. 
More precisely, we simulate images in the \textit{BVRi} bands with exposure times as listed in Table~\ref{tab1_a2219}. We reproduce the telescope throughputs, according to informations retrieved from the Subaru and LBT telescope websites.
In this case, the cluster simulated with MOKA has the same redshift of Abell 2219 and its virial mass  is chosen to be  $1 \times10^{15}M_\odot$, consistent with  previous mass measurements found in the  literature for this cluster. We assume a concentration of 3.8, consistent with values expected for clusters of this mass on the basis of recent concentration-mass relations \citep{massimo_2014}.
Using this set of simulations, we explore the effect of several selection techniques on the mass reconstruction,
based on different filter combinations.
 Hereafter, the analysis of this dataset with our selection method is coded with case 1 and case 2 if the 
  photometry in \textit{BVR} and \textit{BVRi} bands is used respectively, while  analyses of this dataset with alternative  selection methods are coded with case 1-2. 
\\Then we investigate  how the method works  under different  cluster redshifts. In this case, we mimic Subaru Suprime-Cam observations in the  \textit{gri} bands. 
Here we consider the case $M_{\rm vir}$ =  $1.55 \times10^{15}M_\odot$, $c_{\rm vir}$ = 3.8, and three cluster redshifts: $z_{\rm l}$ = $0.23, 0.35, 0.45$ which are coded as cases 3,4 and 5, respectively.
\\Finally, we investigate the mass dependence on the selection method. In this case we simulate Subaru Suprime-Cam observations in the  \textit{gri} bands
setting  $z_{\rm l}$ = $0.23$, $c_{\rm vir}$ = 3.8 while we vary the cluster mass:  $M_{vir}=0.5 \times10^{15}M_\odot$ and $M_{vir} =0.75 \times10^{15}M_\odot$; the cases 6 and 7 respectively. All simulations with photometry in \textit{gri} bands have an 
exposure time in the  \textit{g} and  \textit{i} bands  of $1800$s while we double it in the  \textit{r} band, which is used for the shear measurement. 
The setups of the simulations  are listed in Table \ref{tab_cont_set}.
\\ We  derive the multiband photometry running  SExtractor in dual-mode, with the   \textit{R} or  \textit{r}  band images  used as  detection images. 
 In the next section, we describe the shear measurement on these images that closely resembles that applied to the data analysis of A2219.

\begin{table}
\resizebox{\hsize}{!}{ 
\begin{tabular}{cccc}
\hline
\hline
Simulation  & Exp. Time & $z_{l}$ & Mass \\
&  $s$&&    $[10^{15} M_\odot]$ \\
\hline
\hline
case 1  & \textit{B}(720); \textit{V}(1080); \textit{R}(3330) & $0.23$ & $1$ \\ 
case 2  &\textit{B}(720); \textit{V}(1080); \textit{R}(3330); \textit{i}(3005.8) & $0.23$ & $1$ \\  
case 3  &\textit{g}(1800); \textit{r}(3600); \textit{i}(1800) & $0.23$ & $1.55$  \\
case 4 &\textit{g}(1800); \textit{r}(3600); \textit{i}(1800) & $0.35$ & $1.55$ \\
case 5   &\textit{g}(1800); \textit{r}(3600); \textit{i}(1800) & $0.45$ & $1.55$ \\ 
case 6 &\textit{g}(1800); \textit{r}(3600); \textit{i}(1800)  & $0.23$ & $0.5$ \\  
case 7  &\textit{g}(1800); \textit{r}(3600); \textit{i}(1800)  & $0.23$ & $0.75$ \\ 
\hline
 \end{tabular}
 }
\caption{Summary of the setups of the simulations produced.}
\label{tab_cont_set}
\end{table}

   %%%%%%%%%%%%%%%%%%%%%%%%%%%%%%%%%%%%%%%%%%%%%%%%%%%%%%%%%%%%%%%%%%%%%%%%%%%%%%%%%%%%%%%%%%%
   \section{Shear Measurements}
\label{sect_ksb}
The observed shear signal produced by the gravitational field of the  clusters
is measured using the KSB approach \citep{kaiser_1995, luppino_1997, henk_1998}.
This method uses stars to 
remove from galaxy ellipticities the PSF contribute due to the atmosphere and the telescope optics,
giving a shear estimator uncontaminated by systematics.
These distortions  can be taken into account considering the PSF as made of two components:
one isotropic,  which
 smears images mimicking the effect introduced by seeing, and one anisotropic (due to guiding errors, coaddition, optics, bad focusing etc.) that introduces anisotropic distortions which mimic a lensing signal.
 The effect of the PSF anisotropy  depends on the galaxy size and is stronger for the smallest sources.
In the formalism described below, source ellipticity is a complex (two components) quantity.
 The correction of the observed ellipticity $e_{obs}$ for the anisotropic component, as prescribed by the KSB method, is given by:
\begin{equation}
 \label{eq_1.0.0}
 e_{aniso}=e_{obs}-P^{sm} p,
\end{equation}
\begin{equation}
 \label{eq_1.1.0}
 p =  e_{obs}^* P^{{sm*}^{-1}}, 
\end{equation}
where $p$ measures the PSF anisotropy and  $P^{sm}$ is the smear polarizability tensor given in  \cite{ henk_1998}. 
The tensor $p$ can be estimated from images of stellar objects 
(the quantities with $^*$ symbol are computed on stars).
The relation between the intrinsic ellipticity $e$ of a galaxy and the reduced shear $g = \gamma/(1-\kappa)$ will be: 
\begin{equation}
 \label{eq_1.2.0}
e_{aniso} = e + P^{\gamma} g, 
\end{equation}
where the tensor $P^{\gamma}$, called the pre-seeing shear polarizability tensor \citep{luppino_1997}, describes the effect of seeing.  $P^{\gamma}$ has the following expression:
\begin{equation}
 \label{eq_1.4.0}
P^{\gamma}= P^{sh}-P^{sm}P^{sh*}{P^{sm*}}^{-1} = P^{sh}-P^{sm} q, 
\end{equation}
with $q=P^{sh*}{P^{sm*}}^{-1}$ and $P^{sh}$ being the shear polarizability tensor  \citep{henk_1998}.
\\We use the KSB implementation described in \cite{mario_2015}. 
The main feature of our pipeline is that the PSF is modelled  as a polynomial function of the position, evaluated with
 the PSFex software \citep{bertin_2011}, which extracts models of the PSF from the image.
The $p$ and $q$ terms are derived at each galaxy's position using this PSF model.
For each galaxy, the output quantity $e_{iso} = e_{aniso} {P^{\gamma}}^{-1}$ is computed and used to derive the average reduced shear,
being $\langle g \rangle = \langle e_{iso} \rangle$ under the assumption that the contribution of the intrinsic ellipticities vanishes when  averaging  over a large number of galaxies.
\\The main quantities described by the KSB approach are weighted through  a gaussian window function of scale length  $\theta$ to beat down the  noise.
The size of the window function has a crucial role since, if it is not properly chosen, noise will bias the ellipticity measurements.
We find the size of the window function that maximizes
 the ellipticity signal-to-noise ratio (Eq. \ref {eq_1}), defined by the Eq. 16 in  \cite{erben_2001}:
\begin{equation}
 \label{eq_1}
 SNe(\theta)= \frac{\int I(\theta)W(|\theta|)d^2 \theta}{\sigma_{sky}\sqrt {\int W^2(|\theta|)d^2 \theta}};
\end{equation}
as described by \cite{joy}.\\
\cite{henk_1998} 
first pointed out 
 that quantities  from stellar objects should be calculated with the same 
scale as the object to be corrected.
This is done in our implementation of the KSB analysis.
The selection of stars used to correct the distortions introduced by the PSF components is made 
in  the magnitude MAG AUTO  \footnote{SExtractor parameter indicating Kron-like elliptical aperture magnitude.} vs. $\delta$ = MU MAX -MAG AUTO space, where MU MAX is the SExtractor parameter indicating the peak surface brightness above the background.
The  quantity  $\delta$ is used as the estimator of the  object size to separate stars   from galaxies. When we analyze simulations, we use stars with  \textit{r} or  \textit{R} magnitudes in the range $[15, 19]$ to correct for the effects introduced by PSF anisotropy and seeing.
 \\The tangential and the cross components of the reduced shear, $g_{T}$ and $g_{\times}$,  are 
 obtained from the quantities $e_{1iso}$ and $e_{2iso}$ used, respectively, as  $g_{1}$ and  $g_{2}$ 
  in Eq. \ref{eq_1.0} and Eq. \ref{eqq_1.0}:
\begin{equation}
 \label{eq_1.0}
 g_{T(i)}=- g_{1(i)}\cos 2\varphi -  g_{2(i)}\sin 2\varphi;
 \end{equation}
 \begin{equation}
 \label{eqq_1.0}
 g_{\times(i)}=- g_{1(i)}\sin 2\varphi +  g_{2(i)}\cos 2\varphi.
\end{equation}
Here $(i)$ indicates the $i$-th galaxy and $\varphi$ the position angle determined assuming the position of the BCG as the  cluster center.
The radial profiles are obtained by averaging both  shear components in annuli  centered on the BCG.
\\We  exclude  from the catalog  galaxies with  SNe $< 5$  (our choice of SNe $<5$ corresponds to roughly SNR $<10$, if we use the definition  SNR = FLUX AUTO/FLUXERR AUTO, where   FLUX AUTO and FLUXERR AUTO are  SExtractor quantities)
and  galaxies with corrected ellipticities  $e_{1iso} > 1$ or $e_{2iso}$  $>1$, for which the ellipticity measurement is not meaningful.
The performances of our KSB code are extensively discussed in \cite{mario_2015}.
	In addition,	
	 in the appendix  we give an estimate of the bias in the shear measurement
using the set of simulations tailored on observations of Abell 2219 as described in the Sect. \ref{sec_sim}.
We find that  our KSB pipeline on average underestimates the measured shear by about $5\%$ with respect to the input values.
Therefore we apply a calibration factor of 1.05 to the measured ellipticities and use the  corrected ellipticities to derive cluster masses.

%

%
%%%%%%%%%%%%%%%%%%%%%%%%%%%%%%%%%%%%%%%%%%%%%%%%%%%%%%%%%%%%%%%%%%%%%%%%%%%%%%%%%%%%%%%%%

\section{Selection of the background sample}
\label{sel_met}

In this section,   we describe the technique that we developed for the selection in CC space assuming that observations in at least three  bands can be used.
\\ \cite{elinor} did an extensive work on the usage of colors to identify the background galaxies for the weak lensing analysis.
  Using  SUBARU Suprime-Cam observations of  the three clusters A1703 ($z=0.258$), A370 ($z=0.375$), RXJ 1347-11 ($z=0.451$),
  they derive the    CC diagrams of the  galaxies in these fields. More precisely, they used  \textit{r}-\textit{i} vs  \textit{g}-\textit{r} (A1703);   \textit{R}-\textit{z} vs  \textit{B}-\textit{R} (A370) and   \textit{R}-\textit{z} vs  \textit{V}-\textit{R} (RXJ 1347-11).
  \\In \cite{elinor} (Fig. 1) the cluster member population is identified  in CC space adding the information on the
   mean distance of all objects from the cluster center in a given cell of the CC diagram. 
The region in the CC space, where the mean distance  is lower, corresponds to an overdensity of galaxies   comprising the cluster red sequence and blue later type cluster members.
The identification of foreground and background populations is  done  looking at the galaxy overdensities in CC space and at the corresponding weak lensing signal. In particular, CC diagrams (Fig. 2 in \cite{elinor}) appear to be characterized by  a clump 
of  red objects with a rising weak lensing signal at small radii corresponding to background sources 
 in the redshift range  $0.5 < z< 2$. This clump is located in the  lower right side of the CC diagrams.
 Two others clumps  lying blueward of the cluster in CC space are populated by blue objects. The redder one  is located near the center  of the CC diagrams and is populated by foreground unlensed sources since they show  a little weak lensing signal  and no clustering near the cluster center
 in their 
 spatial distribution. Galaxies belonging to the bluer clump are also unclustered near to the cluster center and have a weak lensing signal similar to that of the red background galaxies indicating that these blue objects are background galaxies in the redshift range  $1 < z< 2.5$. This clump is located in the left side of the CC diagrams.
\\Here we propose a variation of the method adopted by \cite{elinor}.
More precisely, we introduce two novelties in the selection process, namely:
\begin{enumerate}
\item we use the galaxies in the COSMOS, including their colors and photometric redshifts, to identify the regions of the CC diagrams populated by the galaxies at redshift higher than the cluster.
COSMOS 
 colors    are  used   as a training set for the separation of background/foreground galaxies in the images of the galaxy clusters under study.
 We use the COSMOS photometric redshift catalog of  \cite{ilbert}. 
The redshift measurements are based on the photometry in 30 broad, intermediate and narrow bands spanning a range of wavelengths from the UV to the mid-IR. Based on a subsample of spectroscopically confirmed sources, the accuracy  of the photo-zs at  magnitudes \textit{i}$^{+}_{AB} < 24$ and redshifts $z < 1.25$  is $\sigma_{\Delta z/(1+z_s )} = 0.012$, where  $\Delta z= z_s-z_p$ is the difference between the spectroscopic and
 the photometric redshifts. Photo-zs up to $z \sim 2$ are less accurate: $\sigma_{\Delta z/(1+z_s )} = 0.06$ at \textit{i}$^{+}_{AB} \sim 24$. 
The source redshift distribution is characterized by median redshifts  $z_m = 0.66$ and $z_m= 1.06$ for sources in the magnitude ranges 22 $< i^{+}_{AB}< 22.5$ and 24.5 $< i^{+}_{AB} < 25$, respectively. 
\item 
 A  fine-tuning in the selection of background galaxies is done by maximizing the amplitude of the shear profile, in particular in the  inner radial bins, where the dilution of the shear  by cluster members is higher.
\end{enumerate}
The implementation of our method is based on the following steps:

\begin{enumerate}
    \item We determine on the COSMOS  ($col1, col2$) CC plane the
    line ($a \cdot col1+ col2+c=0$) separating background and foreground galaxies at the cluster's redshift.
    This line is defined as the tangent to the ellipse fitting the
20\% density contour in the foreground galaxies' CC
distribution, parallel to the ellipse major axis.
    \item For each $i$-galaxy observed in the cluster field with colors $col1_i$, $col2_i$, we compute its distance from this line:
    \begin{equation}
    \label{retta1}
    d_i=\frac{a \cdot col1_i+ col2_i +c}{\sqrt{a^2 + 1}} \;;
    \end{equation}
    and make an initial separation between foreground ($d_i \ge 0$) and background($d_i<0$) galaxies.
    \item In a similar way, we can define in the COSMOS CC plane additional color cuts that allow us to improve the separation of foreground and background galaxies at the cluster's redshift  as detailed later in Sect.  \ref{Analysis of simulated data}.
    \item We define a threshold in magnitude below which uncertainties in the measured fluxes are too high to derive reliable colors, but which most likely come from background galaxies: galaxies fainter than this threshold are included in the background sample.
    \item We measure the shear signal (Eq.  \ref{eq_1.0}) from the background galaxies and fine-tune the $a, c$ coefficients to maximize its amplitude. To this end, we randomly vary $a$ around  $\pm 50\%$ of its initial value, while we randomly vary $c$ of $\pm 0.2$ around its initial value. 
Then  we make a new selection of background galaxies $d_i(a_n, c_n) < 0$,   with $a_n$ and $c_n$ being the coefficient values in the $n$-th realization:  as a result of this iterative procedure, we select the values of $a, c$ that  maximize the amplitude of the shear signal  in the three inner radial bins.   Simultaneously, we also check  that the tangential shear profile of the foreground and cluster member galaxies is consistent with zero at all radii.
    \end{enumerate}
   Despite the limited size of the COSMOS field, several tests show that using the COSMOS colors as starting point for the identification of distant sources is safe. For example, if we apply the color cuts based on the COSMOS photometry to select high redshift galaxies in the  Hubble Ultra-Deep-Field, we find a low contamination by low redshift sources.
The method can be also applied  to data observed with a different filter set.
In the examples discussed later in this paper, we use the software ZEBRA \citep{feldman} to derive the SEDs of the
COSMOS galaxies and we use them to calculate
 magnitudes in other filters  by convolving with the filter transmission curves. \\ In  Sect. \ref{Analysis of simulated data}, we show the application of this approach to simulated images of clusters in different redshift and mass ranges exploring how it works with some combinations of filters.
The fraction of background/foreground misidentified galaxies is derived and discussed in the Sect. \ref{cont_bia}.

 %%%%%%%%%%%%%%%%%%%%%%%%%%%%%%%%%%%%%%%%%%%%%%%%%%%%%%%%%%%%%%%%%

 \section{Analysis of simulated data}
 \label{Analysis of simulated data}
In this section, we show the results of the application of our selection method  to simulations. First, we consider different combinations of cluster redshifts ($z_l$) and filters: $z_l$ = $0.23$ ($BVR$, $BVRi$, and $gri$; cases 1, 2, 3), $z_l$ = $0.35$ ($gri$; case 4) and $z_l$ = $0.45$ ($gri$; case 5). In  these cases, the mass of the lens is $M_{\rm vir}=10^{15}M_\odot$ (cases 1, 2) and $M_{\rm vir} =1.55\times  10^{15}M_\odot$ (case 3, 4, 5). Cases 1, 2 are tested also against alternative methods to identify the background galaxies. Finally, we complement these simulations with additional two test cases, where we keep the lens redshift fixed at $z_l=0.23$ and assume the lens mass to be $M_{vir}=0.5\times10^{15}\,M_\odot$ and $M_{vir}=0.75\times10^{15}\,M_\odot$ (cases 6 and 7). The filter combination used in these last two cases is the same used in the test case 3.  
 
 \subsection{Selection process}
 \label{sel_met_sim_uf}
\begin{figure}
  \begin{center}
          \includegraphics[width=80mm]{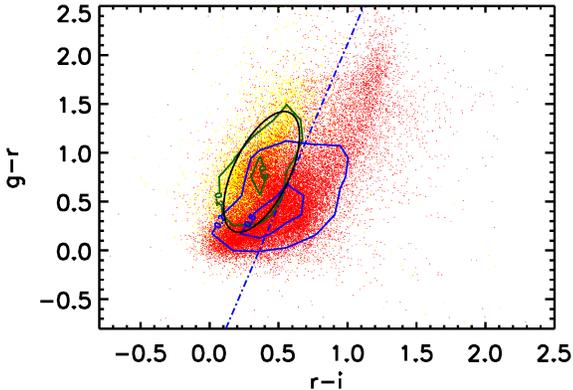}
\protect \caption{First step of the selection process to be applied to simulated data: for the test case 5, outlined in Sect.~\ref{sel_met_sim_uf}, we show  the CC diagram of COSMOS galaxies and the identification of background galaxies based on the photometric redshift estimates of \protect \cite{ilbert}.  The  galaxies at redshift higher and lower than the cluster redshift ($z_l$ = $0.45$) are indicated by red and yellow dots, respectively (the corresponding number density contours are given by the  blue and green solid lines).}
 \label{fig_cos_abel}
 \end{center}
\end{figure}

\begin{figure*}
  \begin{center}
\includegraphics[scale=.3]{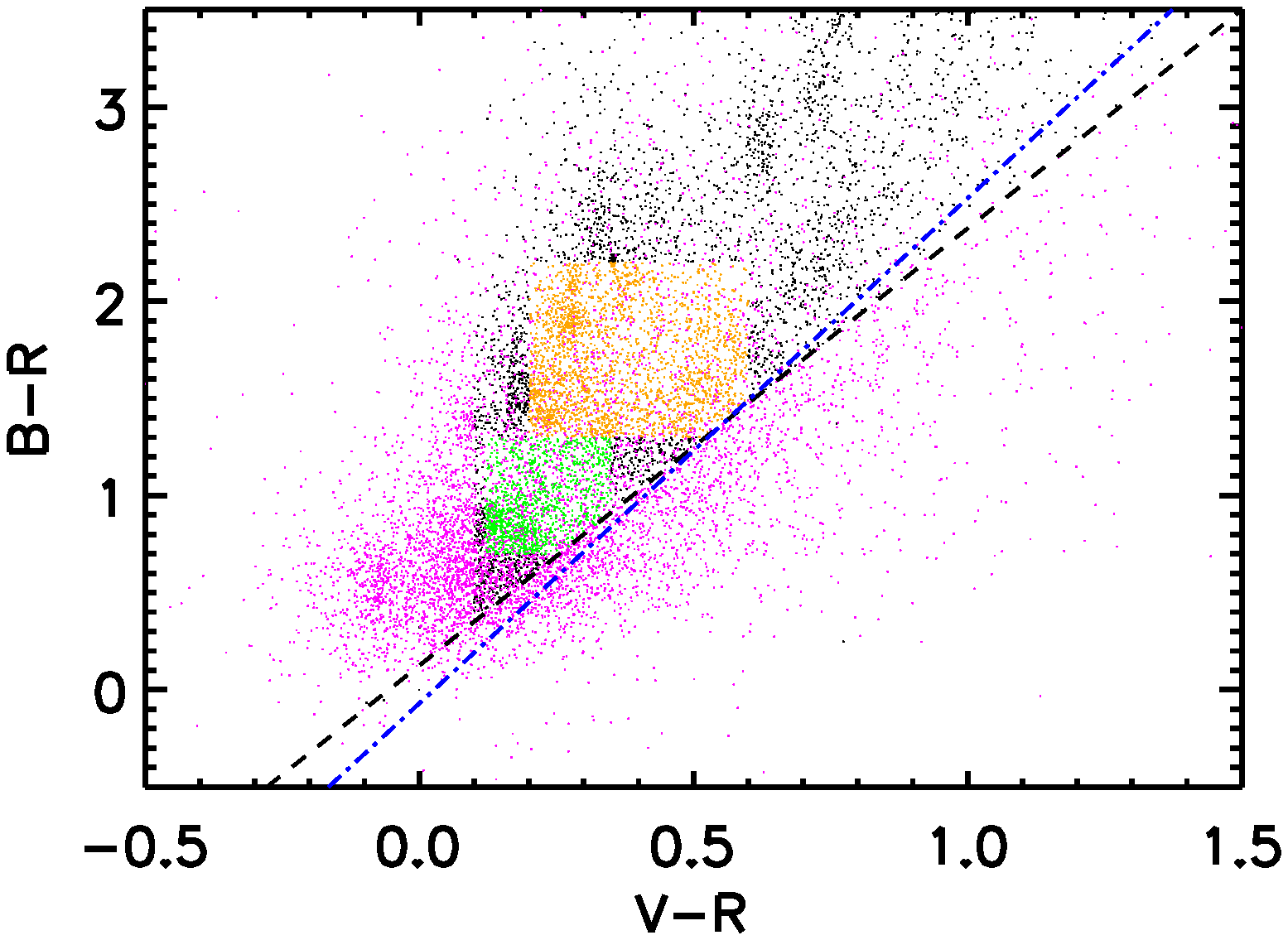}
\includegraphics[scale=.3]{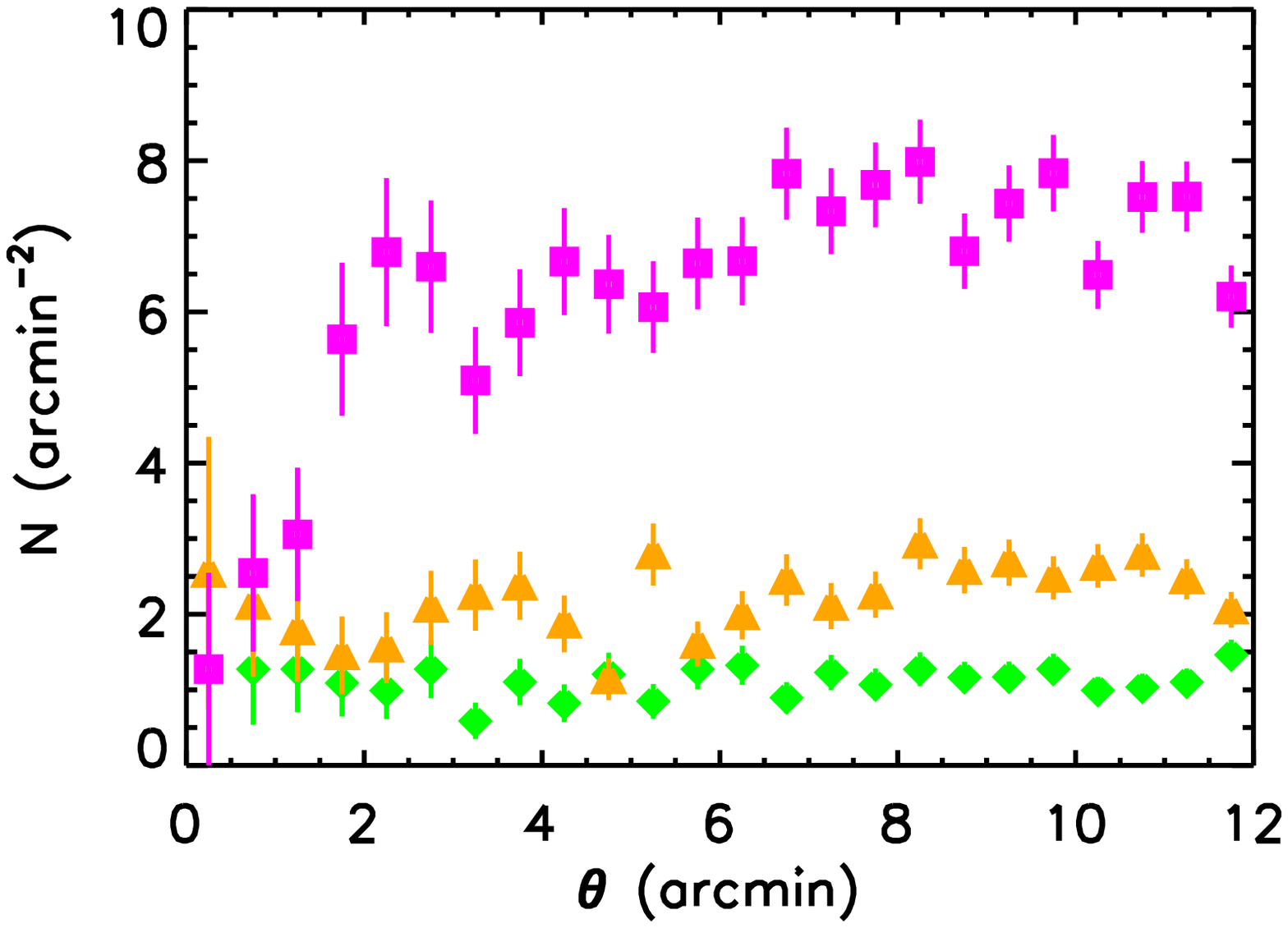}
 \includegraphics[scale=.3]{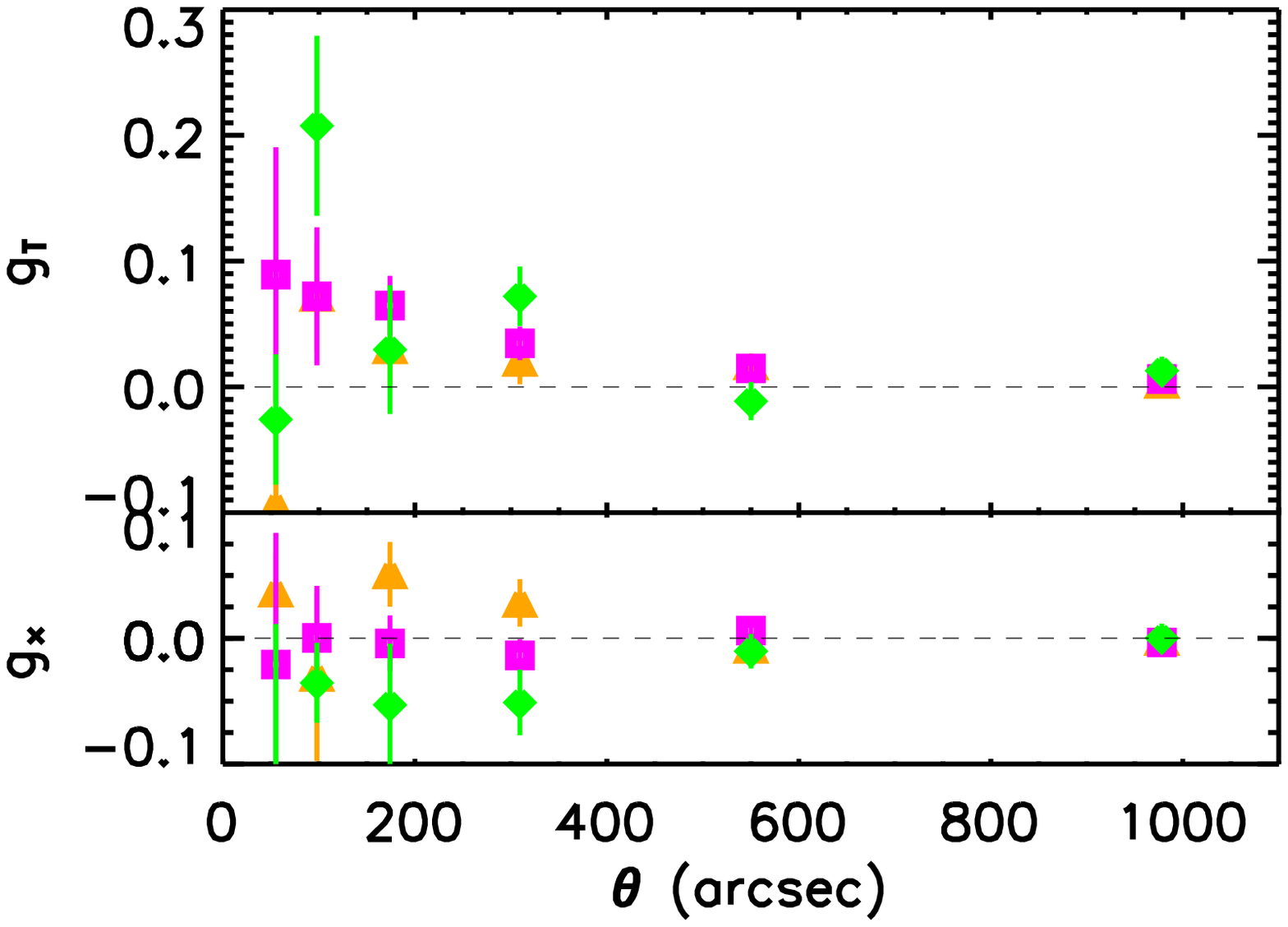}
 
   \includegraphics[scale=.3]{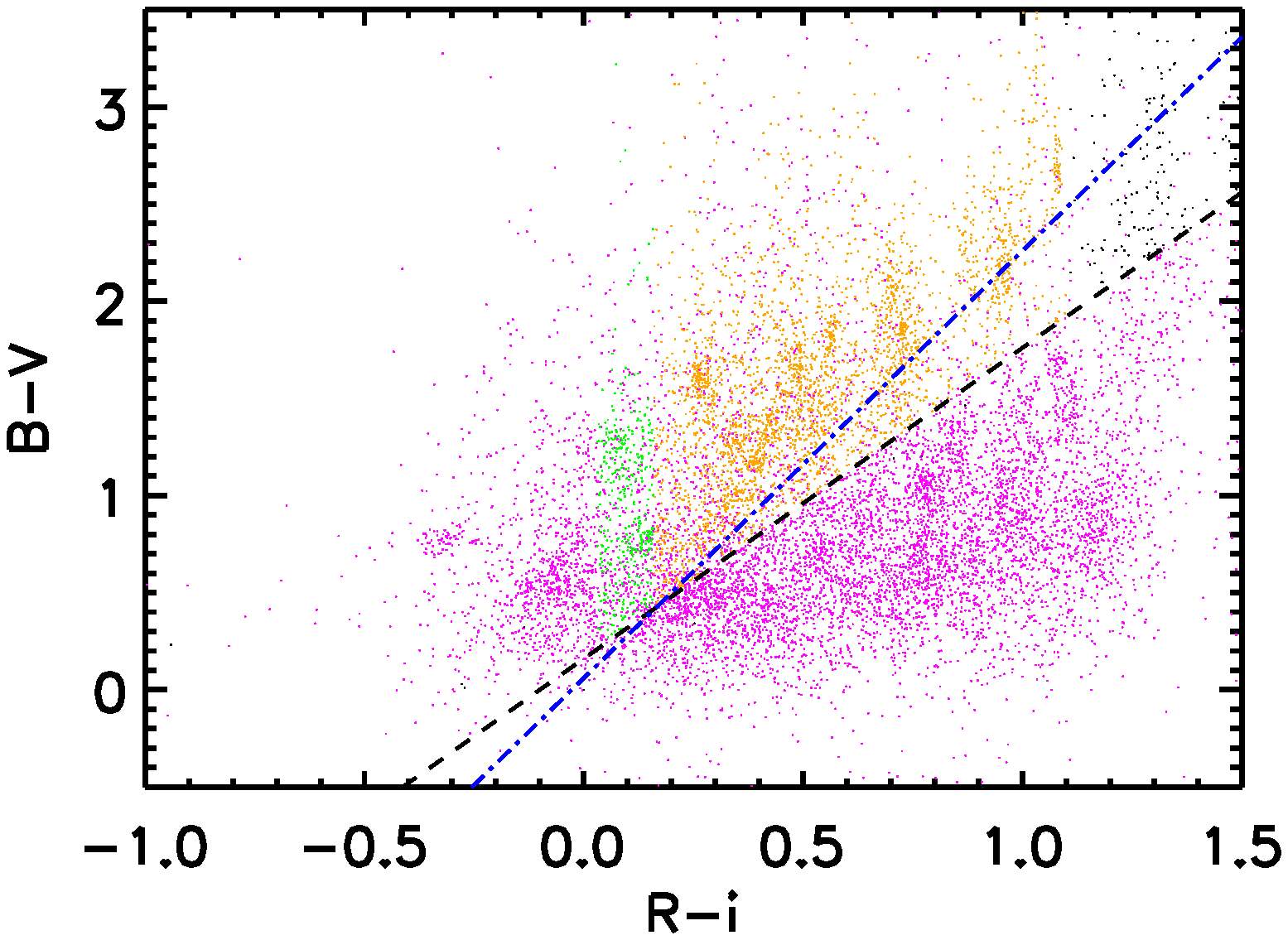}
\includegraphics[scale=.3]{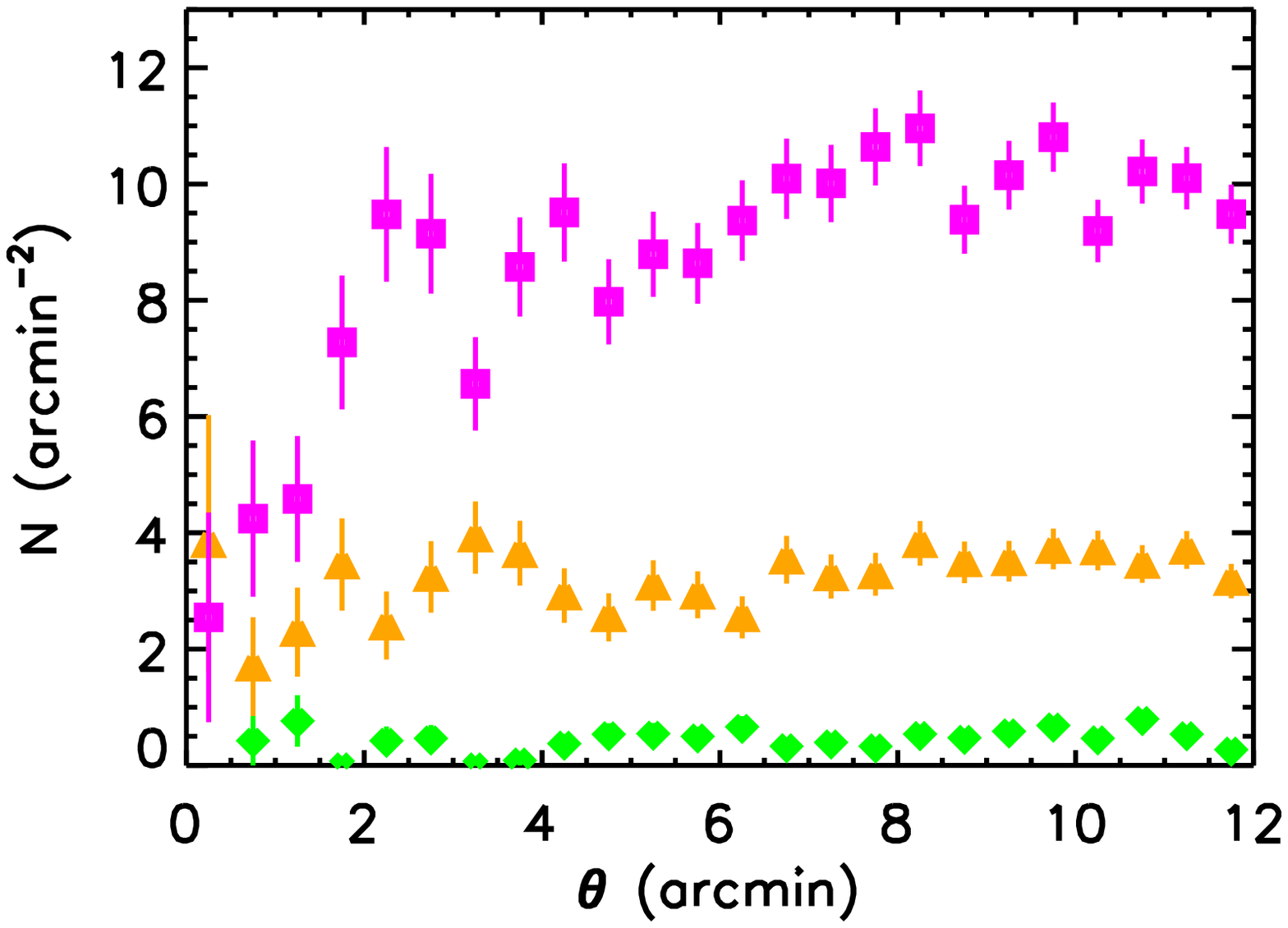}
    \includegraphics[scale=.3]{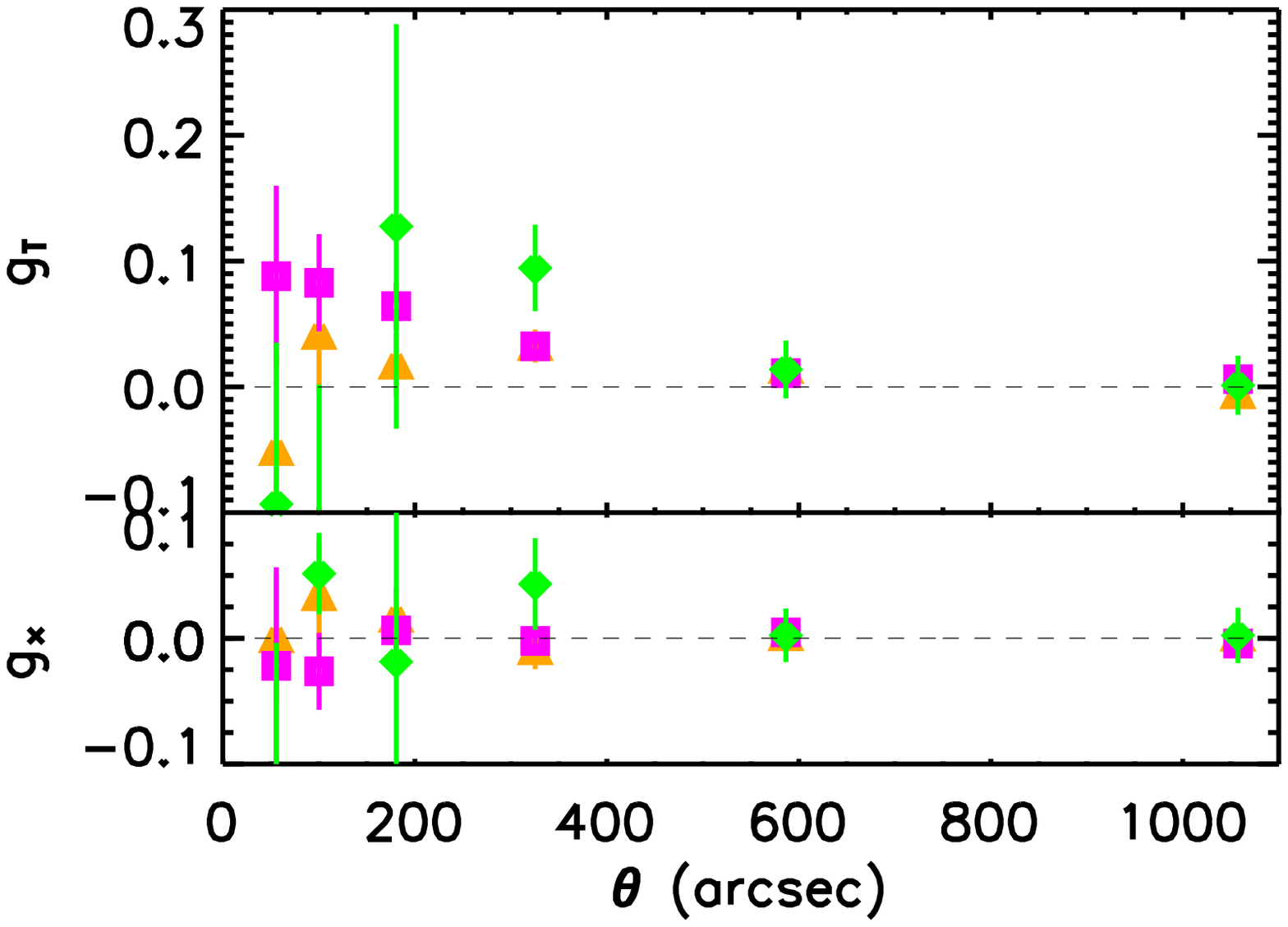}
    
     \includegraphics[scale=.3]{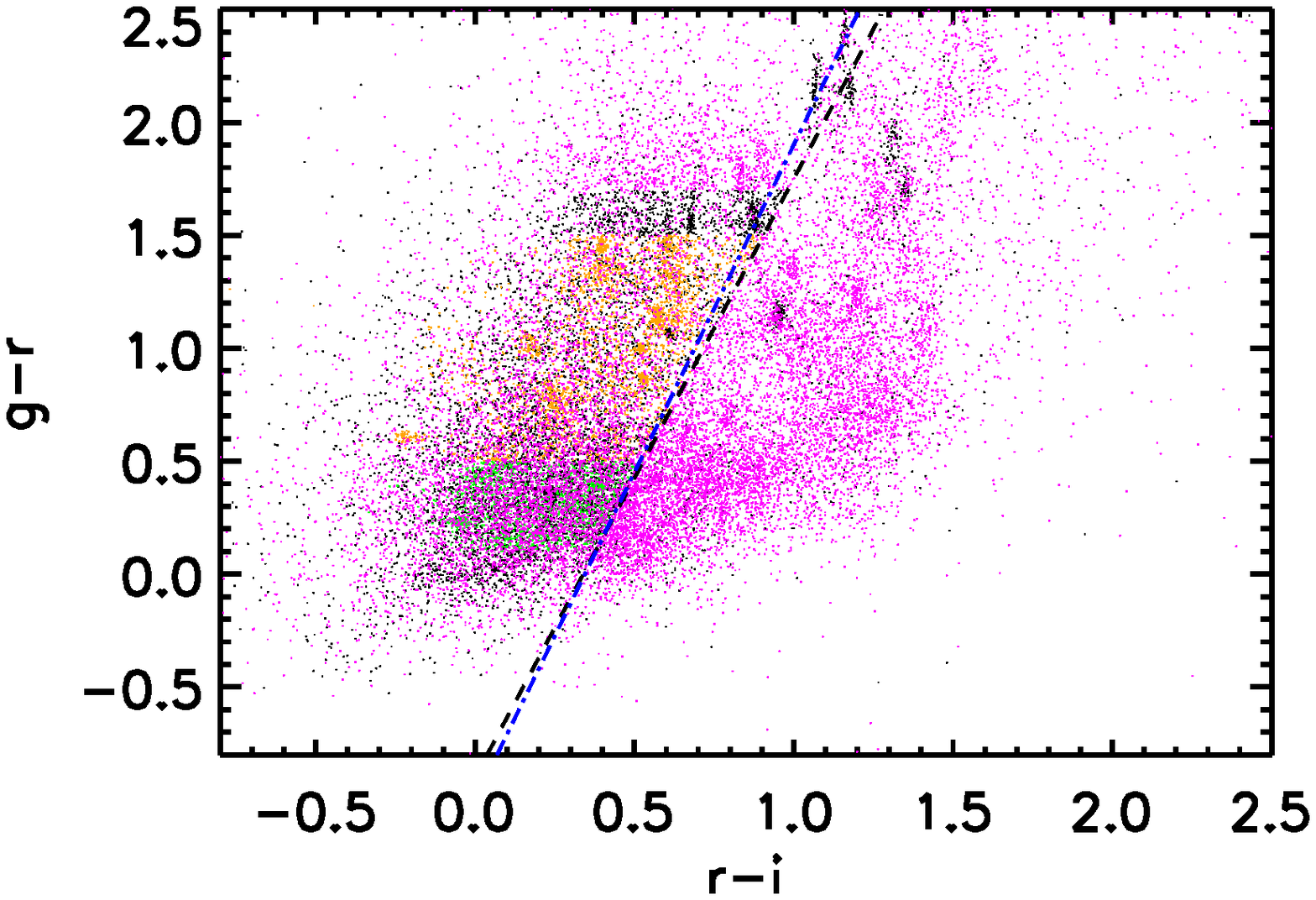} 
     \includegraphics[scale=.3]{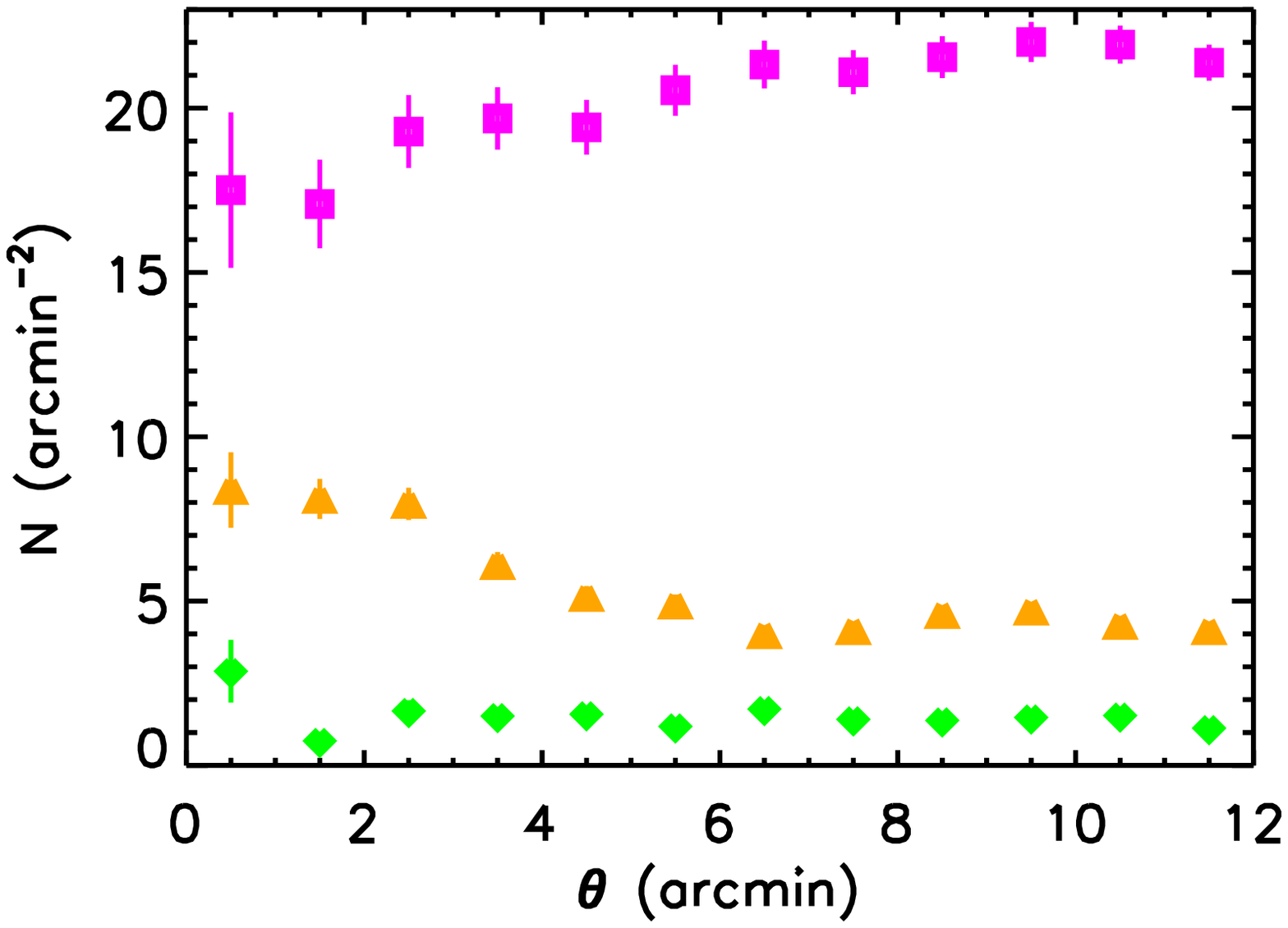}
  \includegraphics[scale=.3]{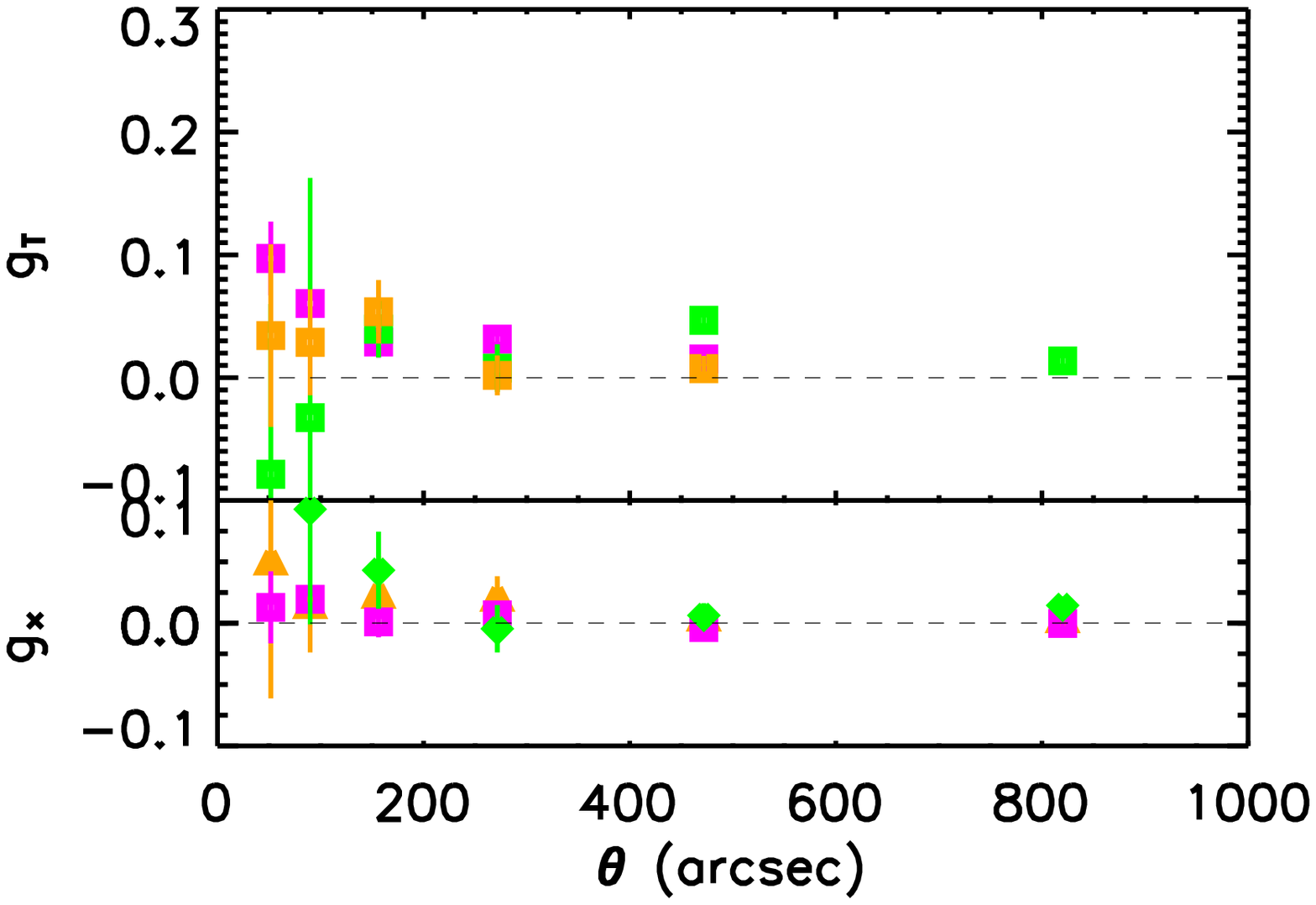} 
  
  \includegraphics[scale=.3]{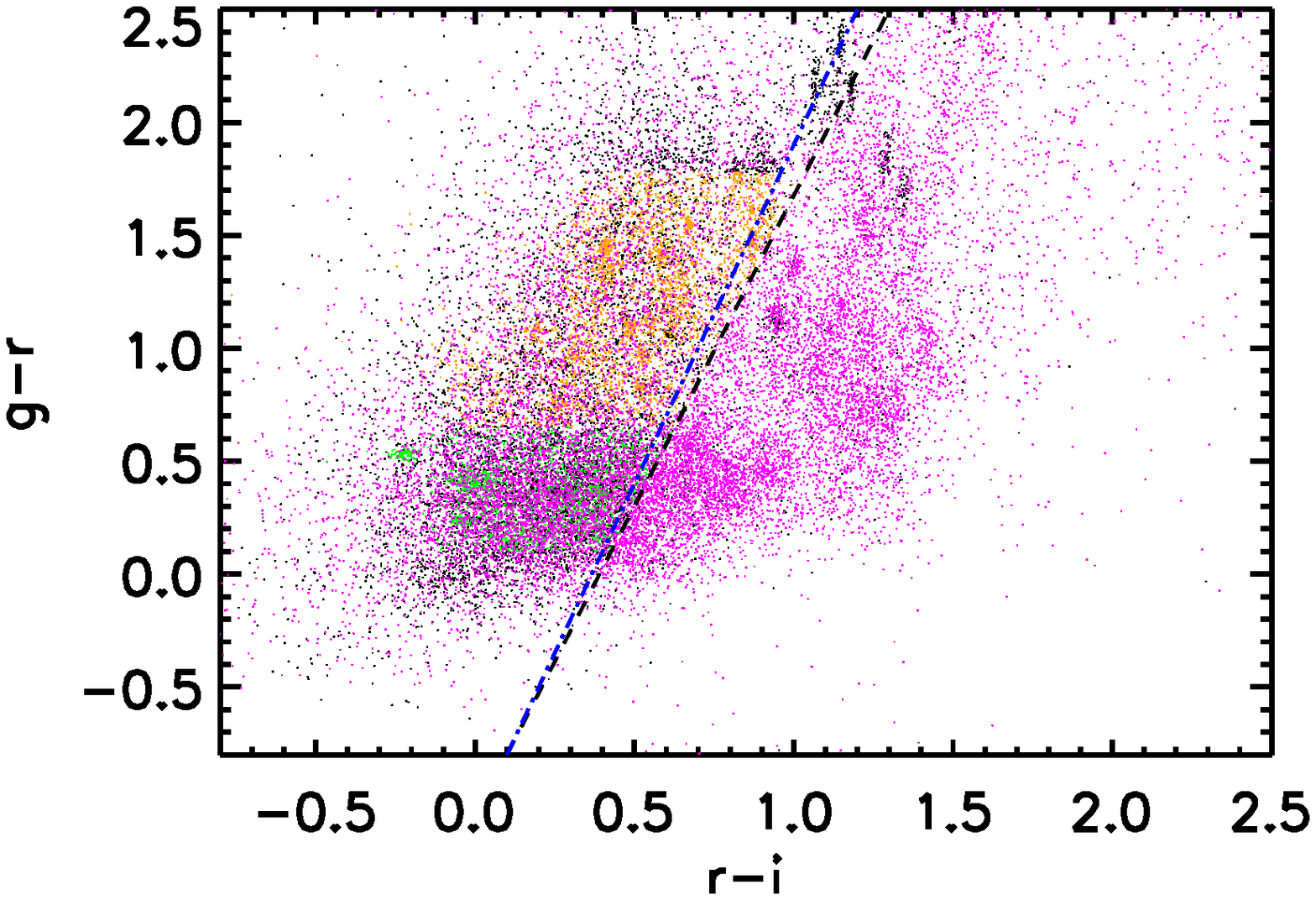}
  \includegraphics[scale=.3]{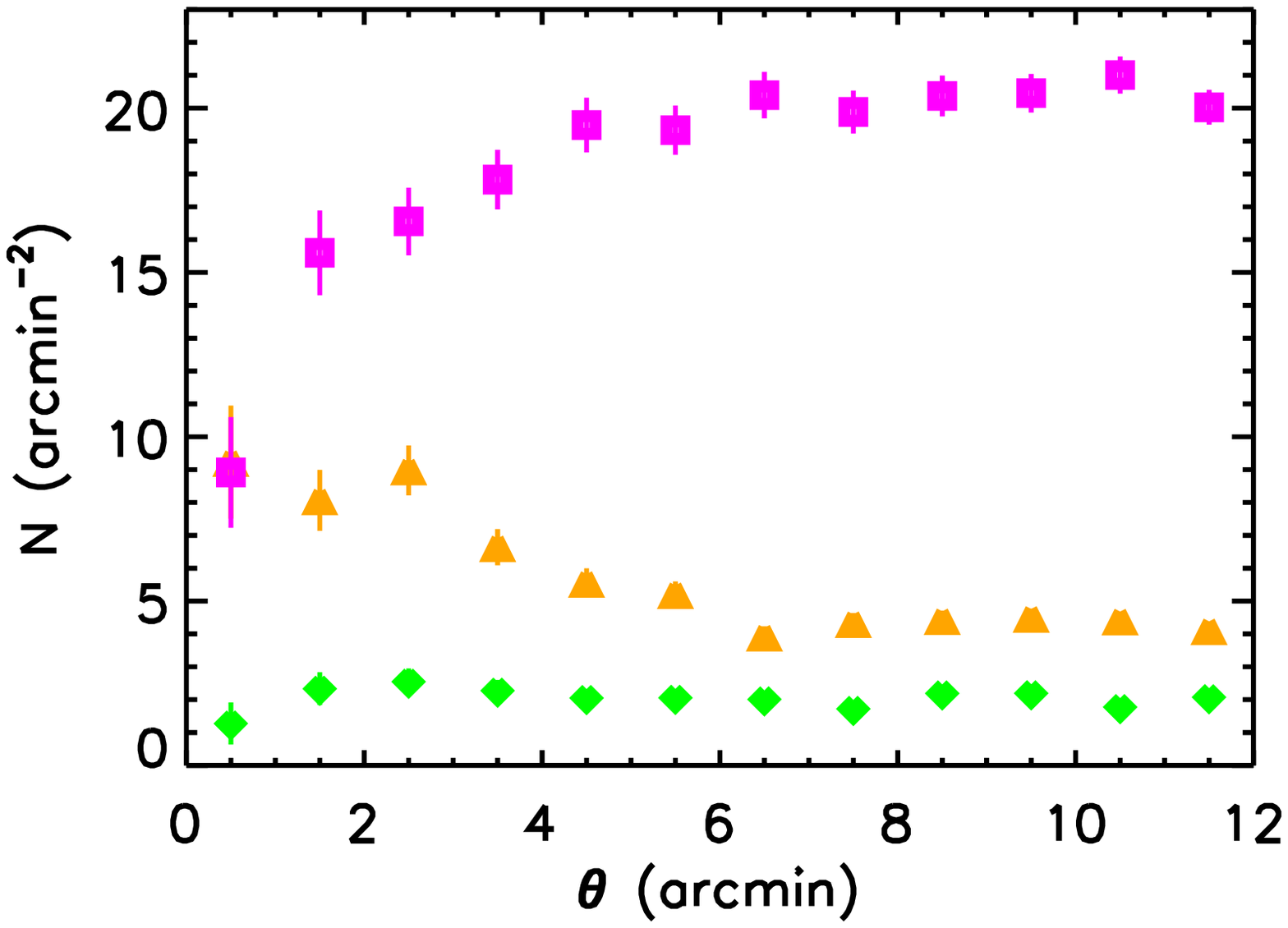}
  \includegraphics[scale=.3]{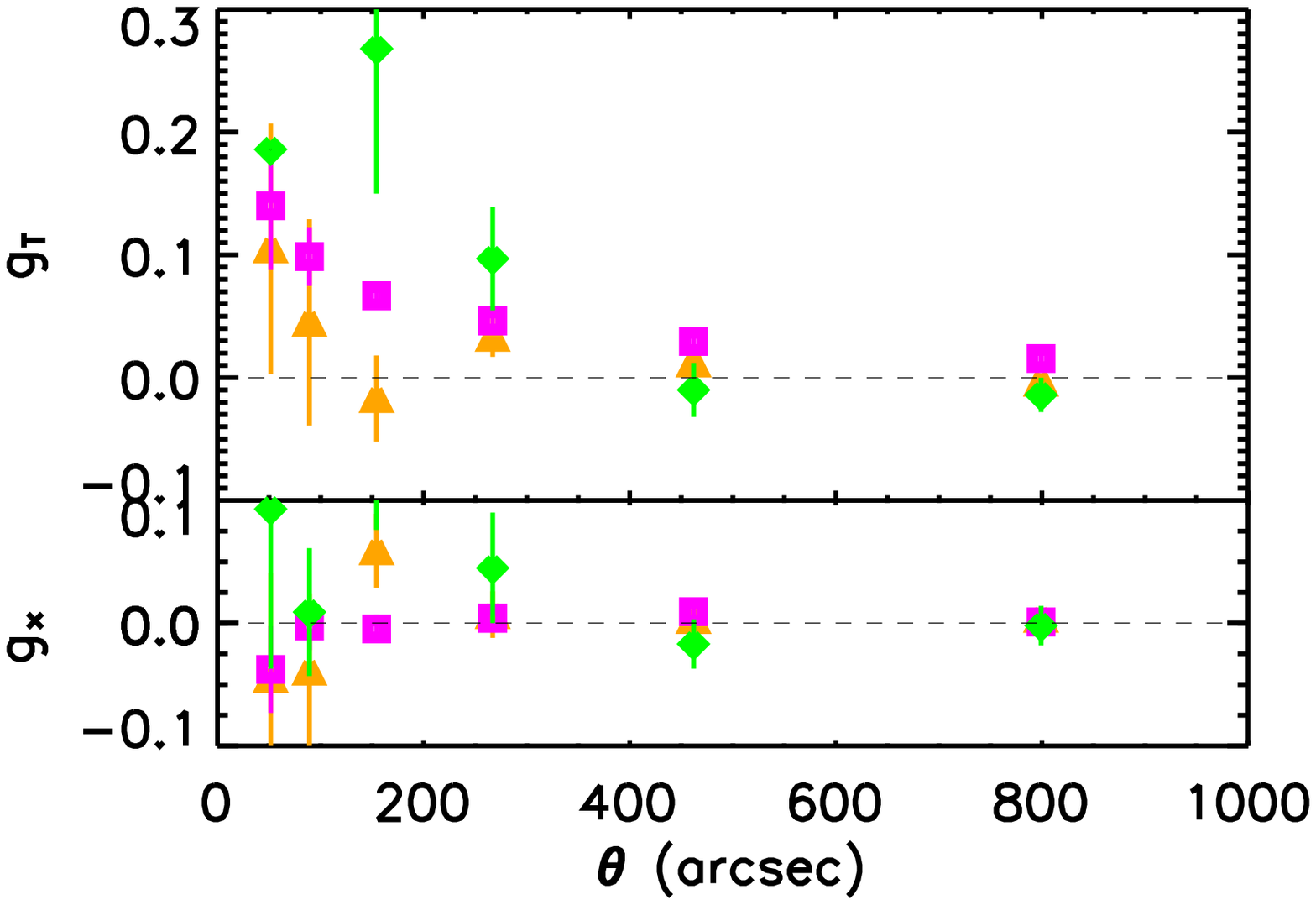}
 
     \includegraphics[scale=.3]{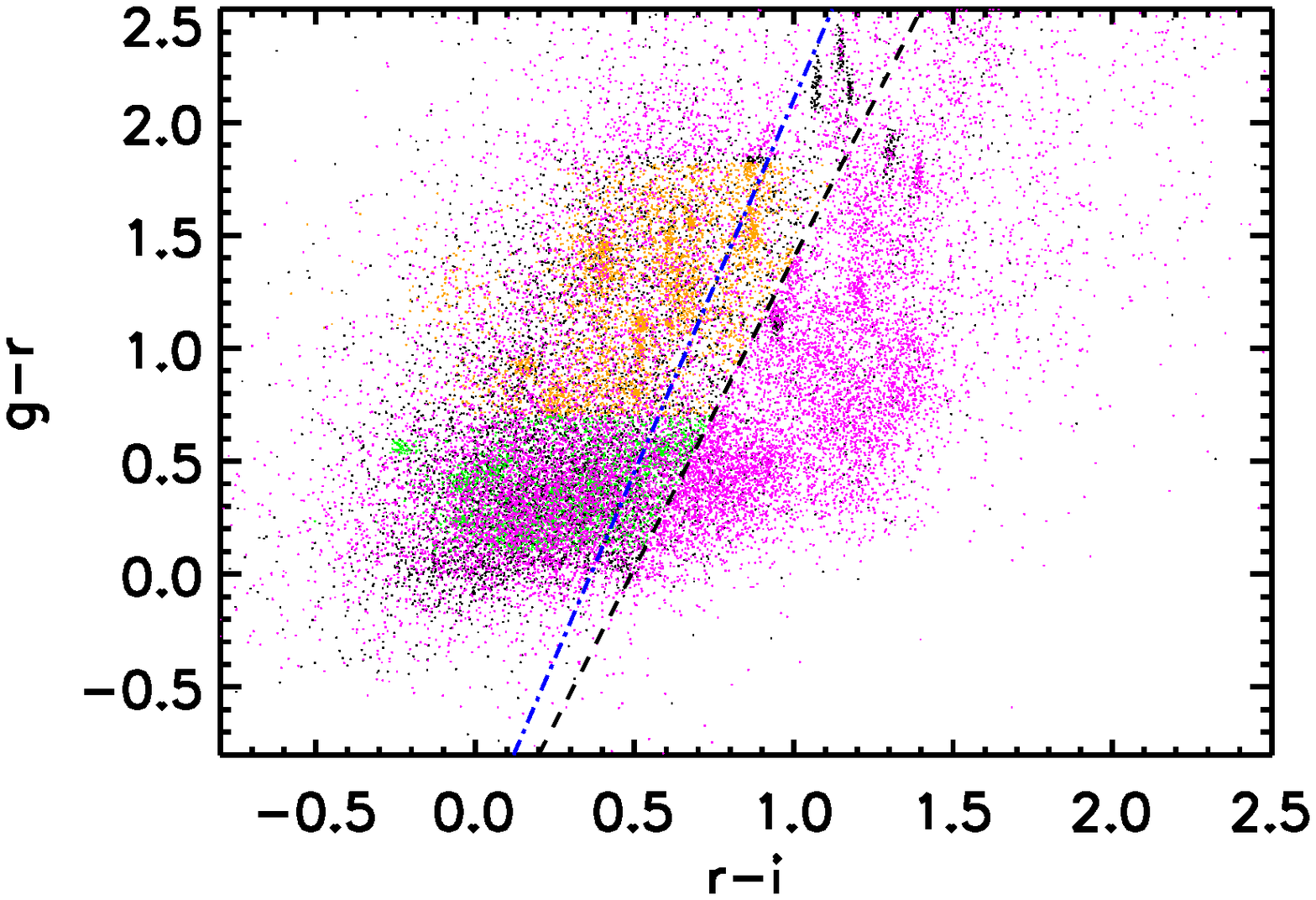}
    \includegraphics[scale=.3]{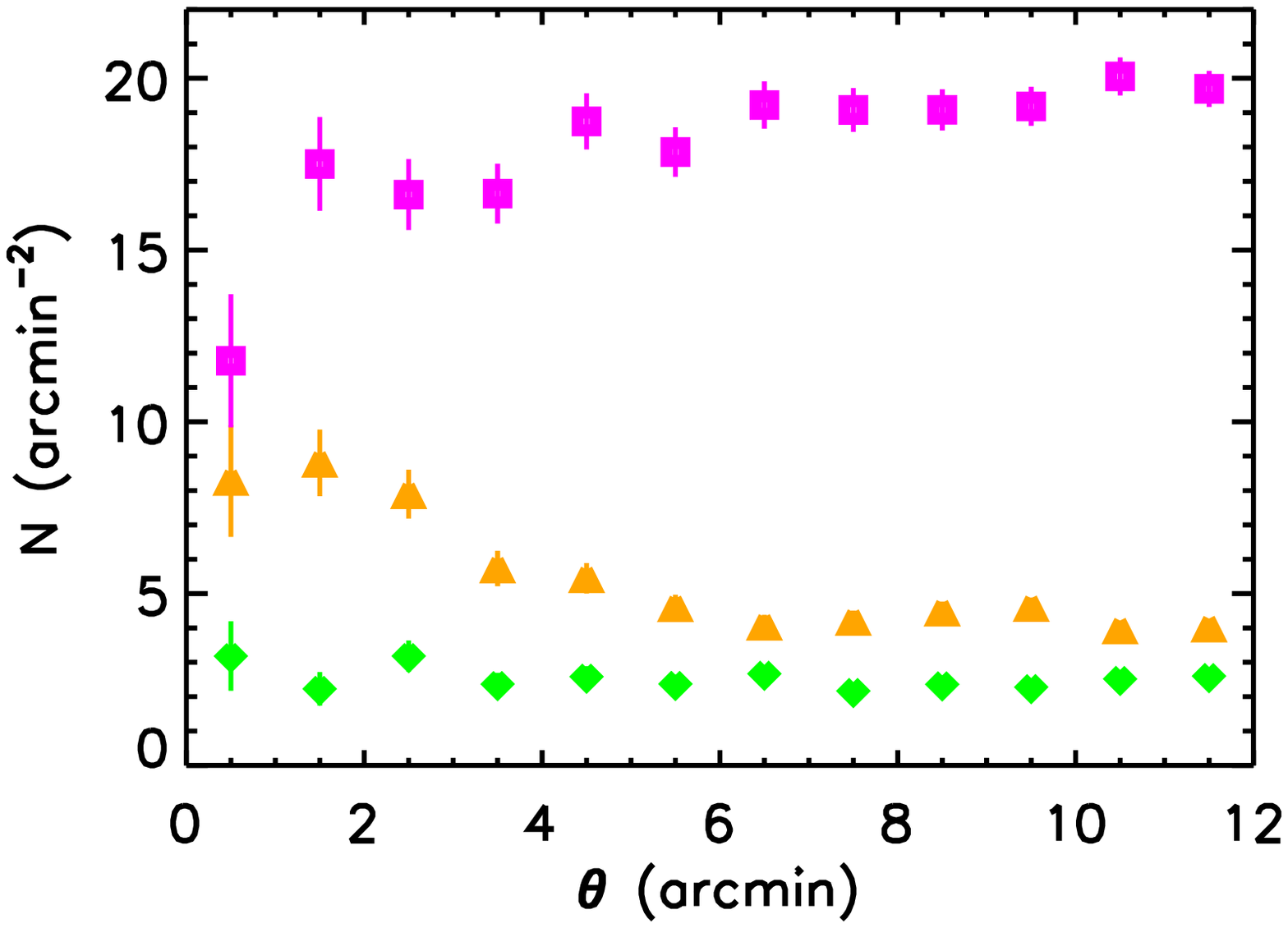}
    \includegraphics[scale=.3]{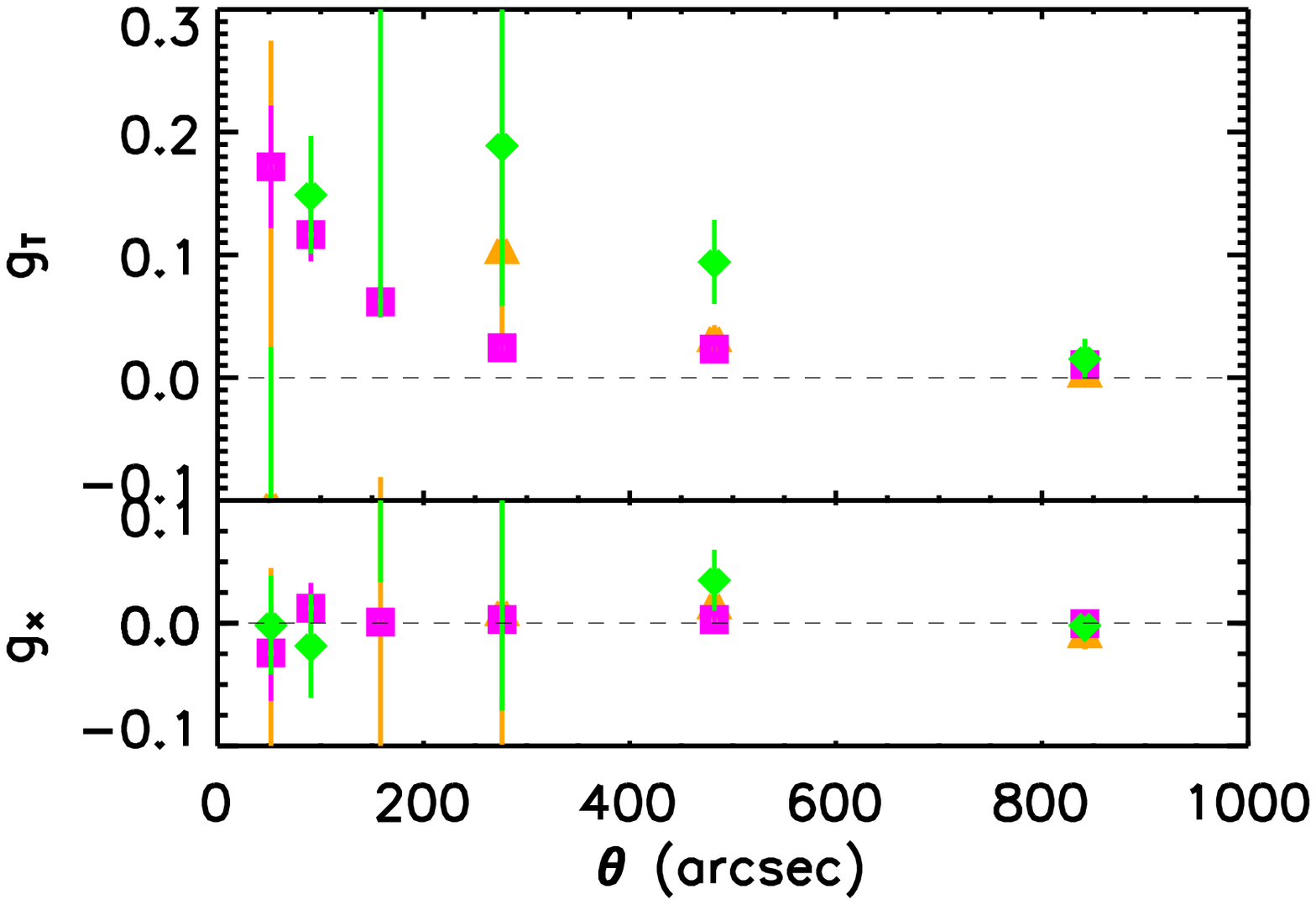}

\caption{Second step of the selection process applied to the simulations. The five rows of panels refer to the  test cases 1-5 for which we give a lengthy description in Sect.~\ref{sel_met_sim_uf}. The left panels show the identification of background (magenta), foreground (green),  and cluster galaxies (orange) in color space. Unclassified galaxies are indicated by black dots.
The dot-dashed lines in the left panels correspond to the initial separation between foreground and background galaxies based on COSMOS data. 
Dashed lines correspond to this separation  after the fine-tuning with the shear signal. 
 In the middle panels, we show the radial number density profiles of background, foreground, and cluster galaxies. The results for the three galaxy sub-samples are shown using the same color-coding as in the left panels. The right panels show the corresponding radial profiles of the tangential ($g_T$) and cross components ($g_\times$) of the reduced shear.}
\label{fig_col_col_sim}
\end{center}
 \end{figure*}

 As outlined above,  in all these scenarios, we start by deriving  the CC diagrams of the COSMOS galaxies for the corresponding filter combinations.
  As an example, Fig. \ref{fig_cos_abel} shows the \textit{r}-\textit{i} vs \textit{g}-\textit{r} diagram of the COSMOS  galaxies  and the foreground/background galaxy separation in the case of a cluster at redshift  $z_l=0.45$ (case 5). Galaxies with photo-zs   $ > 0.47$ are plotted in red, while those with photo-zs $\le 0.47$ are plotted in yellow.  A margin of 0.02 in redshift is used to account for photo-z accuracy when we select galaxies at a redshift higher than the cluster redshift. 
  Density contours of background and foreground galaxies are shown in blue and in green respectively. Also displayed is the ellipse fitting the 20\% foreground contour; the dot-dashed line is the tangent to this ellipse.
Fig.~\ref{fig_col_col_sim} summarizes the full procedure. The dot-dashed lines in the left panels correspond to the initial values of the $a,c$ coefficients, derived for all the test cases here considered. Obviously, the results depend on the filter combination and on the lens redshift. The coefficients $a$ and $c$ obtained for test cases 1-5 are listed in the Table \ref{tab_coeffic}.
Test cases 6-7 are omitted because the results are identical to test case 3 (the selection based on COSMOS data does not depend on the lens mass).
The second step of our procedure consists in the fine-tuning of the selection based on the amplitude of the shear profile derived from the galaxies which are classified as background sources. Therefore, we consider now the CC diagrams built using all  galaxies in the simulated cluster fields.  We start by identifying the background sources based on the lines derived in the first step. For each galaxy, we compute the distance $d_i$ from these  
 lines  and  looking at the amplitude of the shear profile of sources with $d_i < 0$ we derive new coefficients $(a,c)$ as explained in the Sect. \ref{sel_met}. The adjusted lines separating the background from the foreground and cluster galaxies  are given for the test cases 1-5 by the dashed lines in the left panels of Fig. \ref{fig_col_col_sim}. The lines derived from the COSMOS data are also shown for comparison (dot-dashed lines). The optimized coefficients after the fine tuning with the shear signal are listed in the Table \ref{tab_coeffic}.
Given that fine tuning is done on the CC diagrams, which also include the cluster galaxies, being the lens different, the lines optimized for the test cases 6 and 7 differ slightly from that derived for the test case 3.
\\ The final results of the selection process  are shown in the left panels of Fig.~\ref{fig_col_col_sim} for the test cases 1-5, where background, foreground, and cluster galaxies are shown in magenta, green, and orange, respectively. 
The detailed classification of the  sources in the CC diagrams corresponding to each of the test cases is given below. \\

\begin{table}
%\begin{center}
 \begin{tabular}{ccc}
\hline
\hline
Simulation  & initial coefficients  & optimized coefficients \\

\hline
\hline
case 1&$a=-2.6$, $c=0.07$&$a=-2.25$, $c=-0.12$ \\
case 2&$a=-2.2$, $c=-0.06$&$a=-1.60$, $c=-0.16$ \\
case 3&$a=-2.9$, $c=1.00$&$a=-2.65$, $c=0.90$ \\
case 4&$a=-3.0$, $c=1.10$&$a=-2.75$, $c=1.07$ \\
case 5&$a=-3.3$, $c=1.20$&$a=-2.75$, $c=1.35$ \\
case 6& - &$a=-2.69$, $c=0.96$\\
case 7& - &$a=-2.67$, $c=0.93$\\ 
\hline
 \end{tabular}
 %\end{center}
\caption{Coefficients of the line separating the background from the foreground and cluster member galaxies obtained following our procedure outlined above for each test case analyzed. Column 1: simulated case; Column 2: line coefficients found using Cosmos photometry; Column 3: line coefficients found after the optimization with the amplitude of the shear profile of background sources.}
\label{tab_coeffic}
\end{table}
 
%%%%%%%%%%%%%%%%%%%%%%%%%%%%%%%%%%%%%%%%%%%%%%%%%%%%%%%%%%%%%%%%%%%%%%%%%%%%%%%%%%%%% 
\noindent{\bf cases 1, 2:}  when using the colors $col1=$\textit{V}-\textit{R} and $col2=$\textit{B}-\textit{R}  to separate background, cluster and foreground galaxies, we find that the selection which maximizes the amplitude of the  tangential reduced shear profile  is obtained using the following criteria for the background galaxies:
\begin{enumerate}
\item $d_{i}  \geqslant  0$,  $24.2 \leqslant R \leqslant  26$ mag;
\item $d_{i}  < 0$,  $21< R \leqslant  26 $ mag;
\item $d_{i}  \geqslant  0$, \textit{V}-\textit{R}  $< 0.1 $ and  $21< R < 24.2 $ mag.
\end{enumerate}
The foreground and the cluster galaxies are separated  as follows:
\begin{itemize}
\item foreground galaxies:  $d_{i} \geqslant 0$,  $0.7 \leqslant $ \textit{B}-\textit{R} $< 1.3$, $0.12 \leqslant$ \textit{V}-\textit{R} $ < 0.35$ and  \textit{R}  $< 24.2$ mag;
\item cluster galaxies: $d_{i} \geqslant  0$,  $0.2 <$ \textit{V}-\textit{R} $<  0.6$, $1.3\leqslant$  \textit{B}-\textit{R} $< 2.2$   and  \textit{R}  $< 24.2$ mag.
\end{itemize} 
If the selection is performed  in the color space  \textit{R}-\textit{i} vs \textit{B}-\textit{V}, the criteria  (i) and (ii) above obviously remain valid. In this case,
the background sample is composed by galaxies satisfying the criteria (i) and (ii) and by those 
  satisfying the following conditions:  $d_{i}  \geqslant  0$, \textit{R}-\textit{i}$< 0.03 $ and  $21< R < 24.2 $ mag.
\\In addition, we identify  the foreground and the cluster galaxies as follows:
\begin{itemize}
\item foreground galaxies:  $d_{i}  \geqslant   0$,  $0.03 \leqslant$ \textit{R}-\textit{i} $<  0.16$  and   \textit{R}  $ <  24.2$ mag.
\item cluster galaxies: $d_{i} \geqslant  0$,  $0.16 \leqslant$ \textit{R}-\textit{i} $<  1.1$ and  \textit{R} $ <  24.2$.
\end{itemize} 

%%%%%%%%%%%%%%%%%%%%%%%%%%%%%%%%%%%%%%%%%%%%%%%%%%%%%%%%%%%%%
\noindent {\bf cases 3, 6, 7:}
in these cases, the cluster redshift is $z_l=0.23$ and the filter combination is $gri$. However, the cluster mass is different. The selection is performed in color space $col1=$\textit{r}-\textit{i} and $col2=$\textit{g}-\textit{r}.
In the test case 3, the background sources are selected following the criteria:
\begin{itemize}
\item  $d_{i}\ge 0$, $25 \leqslant r \leqslant 26 $ mag;
\item  $d_{i} < 0$,  $22 < r \leqslant 26$ mag;
\item  $d_{i} \geqslant 0$,    $22< r <25$ mag, \textit{g}-\textit{r} $> 1.7$.
\end{itemize}
The  foreground galaxies and the cluster members  are instead identified as follows:
\begin{itemize}
\item foreground galaxies:  $d_{i} \geqslant 0$, $0.1<$ \textit{g}-\textit{r} $< 0.5$,   \textit{r} $< 24.5$ mag;
\item cluster  galaxies:  $d_{i} \geqslant 0$, $ 0.5 \leqslant$ \textit{g}-\textit{r} $ \leqslant 1.5$,   \textit{r} $< 24.5$ mag.
\end{itemize} 
We find that these criteria change very little when considering the test cases 6 and 7.
%%%%%%%%%%%%%%%%%%%%%%%%%%%%%%%%%%%%%%%%%%%%%%%%%%%%%%%%%%%%%%%%%%%%%%%%%%%%%%

%%%%%%%%%%%%%%%%%%%%%%%%%%%%%%%%%%%%%%%%%%%%%%%%%%%%%%%%%%%%%
\noindent {\bf case 4:}
in the case of this cluster at $z_l=0.35$, the criteria to identify the background galaxies are very similar to those found for the test case 3. Indeed, the filter combination is identical and the selection must be modified only accounting for the shift of the cluster galaxies in the CC diagram. The optimal criteria for identifying the background galaxies are:
\begin{itemize}
\item  $d_{i}\ge 0, 25 \leqslant r \leqslant 26 $ mag;
\item   $d_{i} < 0$,  $22< r  \leqslant 26$ mag;
\item  $d_{i} \geqslant 0$,    $22< r <25$ mag, \textit{g}-\textit{r} $> 1.8$.
\end{itemize}

The foreground and  cluster galaxies are instead selected based on the conditions:
\begin{itemize}
\item foreground galaxies:  $d_{i} \geqslant 0$, $0.1<$ \textit{g}-\textit{r} $< 0.65$,  \textit{r} $< 24.5$ mag;
\item cluster  galaxies:  $d_{i} \geqslant 0$, $ 0.65 \leqslant$ \textit{g}-\textit{r} $ \leqslant 1.78$,  \textit{r} $< 24.5$ mag.
\end{itemize} 

%%%%%%%%%%%%%%%%%%%%%%%%%%%%%%%%%%%%%%%%%%%%%%%%%%%%%%%%%%%%%%%%%%%
\noindent {\bf case 5:} finally, in the case of the cluster at the highest redshift among those we have considered ($z_l=0.45$), operating in the same color space of cases 3-4, we optimize the identification of the galaxies in the cluster background by applying the following color and magnitude cuts:
\begin{itemize}
\item $d_{i}\ge 0,  25 \leqslant r \leqslant 26 $ mag;
\item  $d_{i} < 0$,  $22< r \leqslant 26$ mag;
\item $d_{i} \geqslant 0$,    $22< r <25$ mag, \textit{g}-\textit{r} $> 1.85$.
\end{itemize}
Again, we distinguish between cluster members and foreground galaxies:
\begin{itemize}
\item foreground galaxies:  $d_{i} \geqslant 0$, $0.1<$ \textit{g}-\textit{r} $< 0.7$,   \textit{r} $< 24.5$ mag;
\item cluster member galaxies:  $d_{i} \geqslant 0$, $ 0.7\leqslant$ \textit{g}-\textit{r} $\leqslant 1.82$,  \textit{r} $< 24.5$ mag.
\end{itemize}
In all cases, for a fraction of galaxies, the identification is uncertain. In Fig. ~\ref{fig_col_col_sim}, background galaxies that are located too close to foreground galaxies in the  CC diagrams, are marked as black dots.
 The radial profiles of the galaxy number densities and of the tangential and cross components of the reduced shear, measured in each of the three sub-samples of background, foreground, and cluster galaxies, are shown in the middle and right panels of Fig.~\ref{fig_col_col_sim}. Again, we use the magenta, green, and orange colors to distinguish between the three classes of sources.
\\As expected, in all cases we notice that the number density of background galaxies drops at small radii. The decrement of number counts is in large part due to magnification, which becomes stronger when approaching the cluster critical lines \citep{broad_2008}. In addition, a fraction of sources at the smallest distances from the cluster center remains hidden behind the brightest cluster members.
\\On the contrary, at least in the test cases 3, 4 and 5, the number density profiles of the cluster members have the opposite trend, rising towards the center of the cluster. This is less evident in the test cases 1 and 2, corresponding to the lowest cluster redshift among those we have investigated. In these cases, the profiles are quite flat and a small increment of galaxy counts is seen only in the innermost radial bin.
\\The tangential shear profiles of the background galaxies nicely rise towards small radii, as expected in the case of centrally concentrated lenses.  The profiles of the cross component of the shear are all consistent with zero, indicating that the measurements are not affected by systematics. 
\\The tangential shear profiles of foreground and cluster galaxies should be consistent with zero. We find this not to be the case in several of the test cases. In the next section, we will see that the origin of this  behaviour is due to the contaminations of these two samples by background galaxies. We want to stress that the goal of our selection method is to obtain a sample of background galaxies with a low contamination by unlensed sources, this implies to lose  a fraction of background galaxies lying in the same regions of the CC space populated by foreground and cluster members.

 %%%%%%%%%%%%%%%%%%%%%%%%%%%%%%%%%%%%%%%%%%%%%%%%%%%%%%%%%%%%%

\subsection{Contamination}
\label{cont_bia}

Working with simulated data, we can easily quantify how precisely we are able to distinguish between background, foreground, and cluster galaxies.  We quantify the performance in terms of contamination of the subsample of background galaxies by other kind of galaxies. Although identifying the cluster and the foreground galaxies is not the main goal of this analysis, we can verify also what is the level of contamination by background galaxies in the sub-samples of foreground and cluster galaxies.
\\We compute the contamination of the background sample by foreground galaxies as:
\begin {equation} 
f_{for}=\frac{n_{fb}}{n_b},
\end{equation}
where  $n_{fb}$ is the number of galaxies with redshift $z_t \le z_l$ ($z_t$ input redshift) identified as background galaxies and $n_b$ is the total number of  background selected galaxies.
\\Similarly, the contamination of the background sample by cluster galaxies  is given by
\begin {equation} 
f_{clus}=\frac{n_{cb}}{n_b},
\end{equation}  
where  $n_{cb}$ is the number of cluster galaxies mis-identified as background sources.
\\We also define the fraction of background galaxies which are incorrectly assigned to the foreground sample,
\begin {equation} 
f_{back_{f}}=\frac{n_{bf}}{n_f},
\end{equation}  
where  $n_{bf}$ is the number of galaxies with $z_t > z_l$ mis-identified as foreground galaxies and $n_f$ is the total number of  foreground galaxies, and the fraction  of background galaxies contaminating the sub-sample of cluster galaxies,
\begin {equation} 
f_{back_{c}}=\frac{n_{bc}}{n_c},
\end{equation}  
where  $n_{bc}$ is the number of galaxies with $z_t > z_l$ erroneously assigned to the sub-sample of cluster members, which contains  $n_c$ galaxies.
\\These quantities are listed in columns 3-7 of Table~\ref{tab_cont} for the seven test cases described above (column 5 shows the total contamination, resulting from the both foreground and cluster galaxies, $f_{tot}=f_{for}+f_{clus}$). Using the method that we have developed, we find that the contamination  by cluster galaxies is always very small, of the order of $\sim 1\%$ in all the test cases. Instead, the major contaminants of the sample of background galaxies are sources in the cluster foreground. In this case, the contamination ranges between  $\sim 6\%$ and $\sim 10\%$, being larger for lenses at larger redshift, as expected.  Given that the total level of contamination is always $\lesssim 11\%$, we can conclude that the method works well.
\\The level of contamination of the sub-samples of foreground and cluster galaxies allows us to interpret some of the results shown in Fig.~\ref{fig_col_col_sim}. As noted earlier,  the tangential shear profile measured from galaxies in these two sub-samples often varies as a function of radius, while, in absence of lensing the profiles should be flat and consistent with zero.   From Table~\ref{tab_cont}, we see that a significant number of background galaxies are indeed mis-identified as cluster members or foreground galaxies. The contamination of these two sub-samples varies from $\sim 32\%$ to $\sim 64\%$. The level of contamination depends not only on the lens redshift, but also on the filter combination used to define the color space. For example,  considering the test case 1, we see that the number of background galaxies included in the samples of foreground or cluster galaxies amounts to $\sim 63-64\%$ of the total. The addition of the $i$-band allows us to reduce the contamination to $\sim  50\%$.  Given that our goal is to optimize the selection of background sources for the subsequent lensing analysis, we consider the contamination of the sub-sample of foreground and cluster galaxies a fair price to pay. 
\\It is also  important to note that the method produces the same results when applied to lenses of different masses. This is evident comparing the results for the test cases 3, 6, and 7. While in Sect.~\ref{sel_met_sim_uf}, we showed that changing the mass of the lens only weakly impacted on the choice of the coefficients $a$ and $b$, we see here that the contamination of the sample of background sources is also independent on the lens. We have to bear in mind however that the same prescriptions were used to populate the lenses with galaxies and to assign them a SED.

\subsection{Comparison to other selection methods}
\label{method_al_sim}
\begin{figure}
       \begin{center}
\includegraphics[scale=.42]{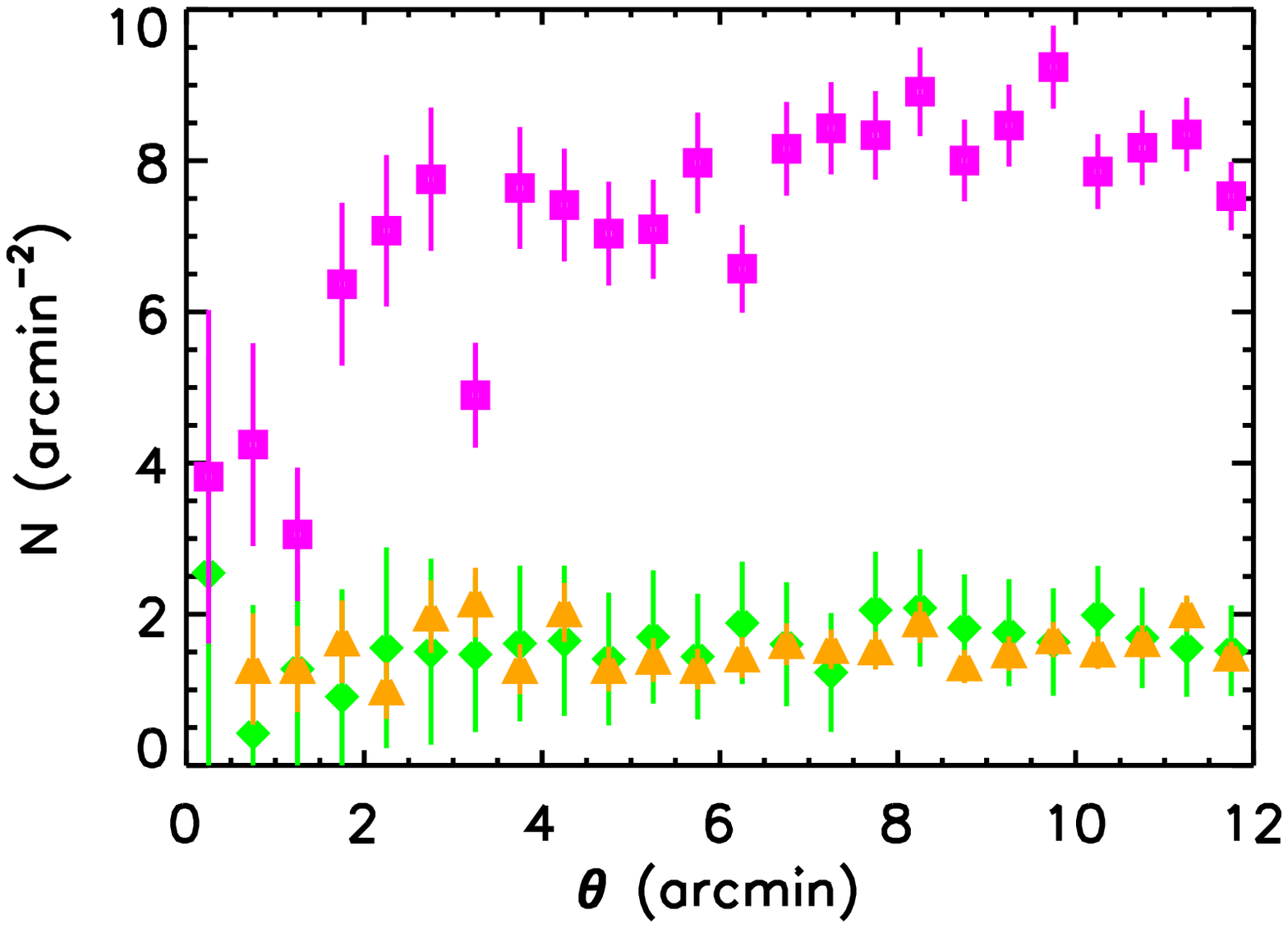}\\
      \includegraphics[scale=.42]{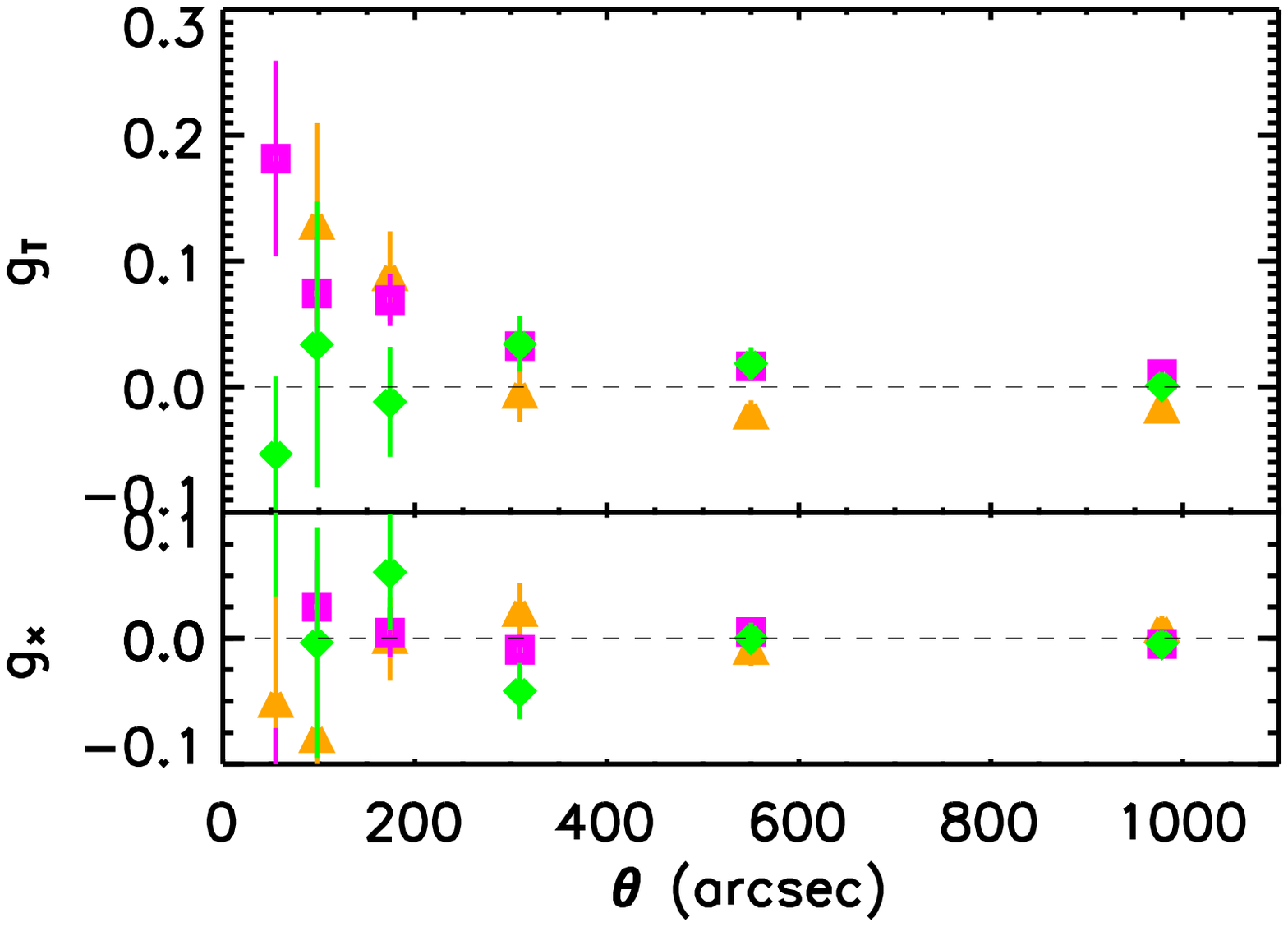}
\caption{Selection with photometric redshifts.  Background galaxies  having  $z_{phot} > 0.27$ are marked with magenta
  squares, cluster galaxies  having  $0.19 \le z_{phot} \le 0.27$ with orange triangles
  and foreground galaxies having  $z_{phot} < 0.19$ with green diamonds. 
The rise  towards the centre of the $g_T$ profile of the cluster member galaxies   
is due to the  limited accuracy of the photometric redshifts derived. 
 }
       \label{fig_zphot_sel_sim_mod}
      \end{center}
\end{figure}

An interesting question is how the performance of the method proposed here compares to that of other techniques to identify lensed galaxies behind clusters. To answer this question, we implement some of these other methods and apply them to our simulations. 
\\We consider here three further approaches, which we apply to the lens at $z_l=0.23$ used in the test cases 1 e 2 above. The first is based on the identification of the cluster members through the location of the red-sequence in the color-magnitude diagram \citep[see e.g.][]{okabe}. Even in this case,  some boundaries on both sides of the red sequence must be defined in order to separate the candidate background sources from the cluster members. 
This is done again by maximizing the amplitude of the tangential shear profile derived from the  sources ending up in the background sample. Therefore  we identify the cluster red sequence in the \textit{V}-\textit{R} vs \textit{R} color-magnitude diagram and we apply the Okabe's method.
\\We further consider a selection based on a simple magnitude cut, selecting the galaxies with  $22 \le R \le 26$. The idea behind this simple method is that the faintest galaxies are likely to be in the cluster background, while the brightest ones are in large fraction foreground and cluster galaxies.
\\Finally, we consider the identification of background sources based on photometric redshifts. Of course, the precision of the  measurements depends on the number of bands available and on the depth of the corresponding exposures. In order to make a fair comparison to our previous results, we use the simulated \textit{BVRi} photometric data produced for the test cases 1 and 2. We use  the ZEBRA code in Maximum-Likelihood mode to fit the source SEDs with a library of templates and derive the photometric redshifts \citep{feldman}. Given the limited number of bands used, we decide not to include galaxies with photometric redshifts higher than $z_{phot}=2$ and with measurement error $\sigma_z/(1+z) > 0.1$. We select  as background galaxies those having  $z_{phot} > 0.27$. 
\\Fig. \ref{fig_zphot_sel_sim_mod} shows the radial profiles of the galaxy number counts (upper panel) and of the tangential and cross components of the reduced shear (bottom panel) for the sub-samples of background, foreground, and cluster galaxies identified using the photometric redshift technique. Comparing to the test case 2  in Fig.~\ref{fig_col_col_sim}, we notice that our conservative cuts lead to a lower number density of background galaxies ($\sim 8$ gal/sq. arcmin) compared to the test case 2 ($\sim 10$ gal/sq. arcmin). 
\\The tangential shear profile derived from the sub-sample of background sources is steeper than measured in test cases 1 and 2. We also notice that the corresponding profile obtained using those galaxies classified as cluster members grows towards the cluster center, leading to the suspect that this sub-sample is contaminated by background galaxies.
 In the last three rows of Table~\ref{tab_cont}, we list the contamination estimates obtained by analyzing the simulations with the three alternative methods outlined above. It appears that the selection based on the photometric redshifts leads to a contamination of the sample of background sources similar to that found in test case 2 (the total contamination is $\sim 6\%$). The contaminations of the sub-samples of foreground and cluster galaxies are smaller using the photometric redshift selection than found using our selection in color space ($\sim 37-38\%$ vs $49-51\%$). 
\\The methods based on magnitude cuts and on the identification of the red-sequence lead to define sub-samples of background sources which have the highest contamination (at the level of $\sim 10-13\%$). In these two cases, we did not attempt to separate foreground and the cluster galaxies.

\subsection{Mass estimates} 
\label{mas_fit_sim_new}
As a last test based on simulations, we quantify now how the selection of background sources impacts on the estimate of the lens mass. 
\\Mass estimates are obtained by fitting the reduced  tangential shear profiles. Obviously, only the sources classified as background galaxies are used for this analysis. The fitting model is based on  the NFW density profile \citep{navarro}: 
\begin{equation}
\rho(r)=\frac{\rho_s}{r/r_s(1+r/r_s)^2} \;.
\end{equation}
The model depends on two parameters, namely the scale radius $r_s$ and the characteristic density $\rho_s$. Most commonly, the NFW profile is parametrized in terms of the concentration, $c=r_{vir}/r_s$, and of the virial mass:
\begin{equation}
M_{vir}=4\pi\rho_sr_s^3\left[\ln(1+c)-\frac{c}{1+c}\right]\;.
\end{equation}
The analytic formulas describing the radial profile of the shear for such a model can be found in \citep{bartelmann_1996}. The fit is performed minimizing the log-likelihood function (Eq. \ref{log_lik} \citep{schneider_2000}):
\begin{equation}
l_\gamma =\sum^{N_\gamma}_{i=1}\left[\frac{|\epsilon_i-g(\vec{\theta_i})|^2}{\sigma^2[g(\vec{\theta_i})]}+2 \ln  \sigma[g(\vec{\theta_i})]\right];
\label{log_lik}
\end{equation}
where $\epsilon_i$ are the observed ellipticities of the background sources, $g(\vec{\theta_i})$ is the reduced shear  predicted by the model at the position $\vec{\theta_i}$, and  $\sigma[g(\vec{\theta_i})]=(1-|g(\vec{\theta_i})| ^2)\sigma_e$.  The variance of the intrinsic source ellipticity is assumed to be $\sigma_e=0.3$.
The computation of the cluster mass requires to know the critical surface density $\Sigma_{cr}$ of the lens,
\begin{equation}
 \label{sigma_crit}
 \Sigma_{cr}=\frac{c^2}{4\pi G} \frac{D_S}{D_{L}D_{LS}}=\frac{c^2}{4\pi G} \frac{1}{D_{L}\beta};
\end{equation}
 where  $D_S$, $D_L$ and $D_{LS}$ are the angular diameter distances between the observer and the sources, between the observer and the lens, and between the lens  and the source, respectively. While $D_L$ is fixed for a given lens, the sources are not all at the same redshift. They have a redshift distribution which should be accounted for. The dependence of $\Sigma_{cr}$ on the source redshift  $z_s$ is contained in the function $\beta=D_{LS}/D_S=\beta(z_s)$. 
We adopt the approximation of a constant $ \langle \beta \rangle$ for all galaxies. Its value, $ \langle \beta \rangle=\frac {1}{N} \sum^{N}_{i=1}\left( \frac{D_{LS_{i}}}{D_{S_{i}}}\right )$, is computed using 
 the COSMOS photometric redshifts, applying to the COSMOS catalog the color and magnitude cuts which led to the identification of the  background galaxies. 
  The values of $ \langle \beta \rangle$ in all test cases are listed in Column 12 of Table~\ref{tab_cont}.
 We verify that the $ \langle \beta \rangle$ value computed in this way is equivalent to that obtained assuming that all galaxies lie
at the same mean redshift $ \langle  z_s \rangle$: $\beta=\frac{D_{LS}(\langle z_s \rangle)} {D_{S}(\langle z_s \rangle)}$.
 In Column 11 of Table \ref{tab_cont}, we summarize the measurements of the virial masses for all the test cases considered. In all cases, the fit is performed  keeping the concentration constant and equal to the input value used to generate the lens mass distributions with MOKA ($c=3.8$). Thus the fit of the tangential reduced shear profiles is done assuming only one free parameter, namely the virial mass. The fit is performed in the radial  range of  $30^{''}< \theta < 1000^{''}$.
The results show that the masses obtained from the galaxies selected in color space differ from the true masses by $\lesssim 10-15\%$. On average, the ratio between estimated and true masses is $\sim0.98\pm0.09$, thus consistent with no mass bias.  
The values of $\langle\beta\rangle$ based on the COSMOS photometric redshifts are similar in all the test cases where the lens has the same redshift ($\langle\beta\rangle\sim0.74$). This value is consistent with the true value obtained from the input  source redshift distribution, when truncated to mimic the depth of the simulated observations. As expected, $\langle\beta\rangle$ is smaller for lenses at higher redshift (test cases 4 and 5) when the depth of the observation is fixed. 
\\In Table \ref{tab_cont}, we report also the mass estimates obtained in the three cases where we have used alternative methods to identify the background galaxies (bottom rows). The selections based on the identification of the  cluster red sequence  and on the magnitude cuts give the highest contaminations by foreground and cluster galaxies, thus resulting in masses that under-estimate the true masses more significantly ($-17\%$ and $-22\%$, respectively). When we select sources based on photometric redshifts derived with the ZEBRA code, we find that the value of $\langle\beta\rangle$ estimated from the photometric redshifts themselves is smaller than the true value ($\langle\beta\rangle\sim0.62$). In part, this is due to the choice of using only sources with $z_{phot}<2$. However, we notice that, for a significant fraction of galaxies, the photometric redshifts under-estimate the true redshifts. Indeed, the median photometric redshift of the selected sample is $z_{phot,med}=0.86$, while the median true redshift of the same galaxies is $z_{true,med}=0.99$. This results in under-estimating $\beta$. Since this appears in the denominator of Eq.~\ref{sigma_crit}, the mass of the lens is over-estimated by $\sim 11\%$.

\begin{table*}
\begin{center}
\begin{tabular}{ccccccccc|ccc}
\hline
\hline
Simulation & Sel. Method & $f_{for}$  & $f_{clus}$  & $f_{tot}$ & $f_{back_f}$ & $f_{back_c}$ & $z_{l}$ & Mass & Density & Lensing mass & $\langle \beta \rangle$ \\
& & & & & & & & $[10^{15} M_\odot]$ & $[$arcmin$^{-2}]$ & $[10^{15}  M_\odot]$ & \\
\hline
\hline
case 1 & \textit{V}-\textit{R}  vs \textit{B}-\textit{R} & $8\%$ & $1\%$ & $9\%$ &  $63\%$ &$64\%$ & $0.23$ & 1& 7 & $0.96_{-0.12}^{+0.16}$ & 0.74\\
case 2 & \textit{R}-\textit{i}    vs \textit{B}-\textit{V} & $6\%$ & $1\%$ & $7\%$ & $51\%$ &$49\%$ & $0.23$ & 1 & 10 & $0.85_{-0.10}^{+0.11}$& 0.74 \\
case 3 & \textit{r}-\textit{i} vs \textit{g}-\textit{r}  & $6\%$ & $1\%$ & $7\%$ & $64\%$ & $38\%$ & 0.23 & $1.55$ & 22 & $1.40_{-0.08}^{+0.08}$& 0.74\\
case 4 & \textit{r}-\textit{i} vs \textit{g}-\textit{r} & $8\%$ & $1\%$ & $9\%$ & $58\%$ &$40\%$ & $0.35$ & $1.55$ & 20&$1.50_{-0.09}^{+0.08}$ & 0.60\\
case 5 & \textit{r}-\textit{i} vs \textit{g}-\textit{r} & $10\%$ & $1\%$ & $11\%$ & $35\%$ &$32\%$ & $0.45$ & $1.55$ & 19 & $1.50_{-0.10}^{+0.11}$& 0.46\\
case 6 & \textit{r}-\textit{i} vs \textit{g}-\textit{r} & $6\%$ & $1\%$ & $7\%$ & $65\%$ &$36\%$ & $0.23$ & 0.5 & 22 & $0.55_{-0.05}^{+0.06}$& 0.74 \\
case 7 & \textit{r}-\textit{i} vs \textit{g}-\textit{r} & $6\%$ & $1\%$ & $7\%$ & $64\%$ &$37\%$ & $0.23$ & 0.75 & 22 & $0.86_{-0.06}^{+0.07}$& 0.74\\
\hline
case 1-2 & $z_{phot}$ & $5\%$ & $1\%$ & $6\%$ & $37\%$ &$38\%$ & $0.23$ & 1 & 8 &$1.11_{-0.14}^{+0.14}$ & 0.62 \\
case 1-2 & $22 \le R \le 26$ & $12\%$ & $1\%$ & $13\%$ & $-$ & $-$& $0.23$ & 1 & 13& $0.78_{-0.08}^{+0.07}$& 0.72\\
case 1-2 & \textit{VR} & $9\%$ & $1\%$ & $10\%$ & $-$ & $-$& $0.23$& 1 & 8 &$0.83_{-0.10}^{+0.11}$ &0.73 \\
\hline
 \end{tabular}
\end{center}
\caption{ Summary of the simulation analysis. Column 1: simulated case; Column 2: selection method. Selection in color space is used in the first seven cases. The bottom three  rows refer to the analysis of the cluster field used in test cases 1-2 when the selection of the background galaxies is done with alternative methods: based on photometric redshifts, magnitude cut, and identification of the cluster red-sequence. Columns 3-4: fractions of foreground galaxies, $f_{for}$, and member galaxies, $f_{clus}$, contaminating the sample of background galaxies. Column 5: total fraction of non-background galaxies contaminating the sample of background galaxies; Columns 6-7: fractions of background galaxies contaminating the samples of foreground and cluster galaxies; Column 8: redshift of the lens; Column 9: mass of the lens; Column 10: number density of background galaxies used in the weak lensing analysis; Column 11: mass recovered from the weak lensing analysis; Column 12: mean value of $\beta$.
}
\label{tab_cont}
\end{table*}

\subsection{Effects of mis-centering}
\begin{figure}
       \begin{center}
\includegraphics[scale=.40]{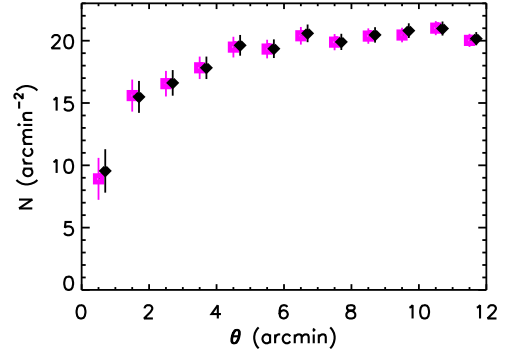}\\
\caption{ Radial number density profiles of background galaxies using the BCG as the shear centre (magenta squares) and adding an offset of $30''$ from the BCG position (black squares). For clarity of the plot we apply an offset of $0.2$ arcmin to the black squares. }
       \label{off_shear}
      \end{center}
\end{figure}

Finally we verify if the selection of background sources can be significantly affected by the choice of the cluster centre, which is used to measure the tangential shear profile of the background galaxies.  For this purpose, we repeat the procedure by assuming a center which is off with respect to the true center (BCG position) by $\sim30''$. The radial density profiles of background sources selected in both cases of correct and wrong identification of the cluster center are shown in Fig. \ref{off_shear}.  They are barely distinguishable, indicating that the mis-centering has little effect on the identification of the background galaxies.

%%%%%%%%%%%%%%%%%%%%%%%%%%%%%%%%%%%%%%%%%%%%%%%%%%%%%%%%%%%%%%%%%%%

\section{Abell 2219}
\label{sec_a2219_prop}
Having evaluated the performance of the method on simulated data, we apply our procedure to select the galaxies in the background of Abell 2219. The choice of this cluster is motivated by the fact that this is one of the best studied gravitational lenses in the literature. Some of the previous analyses will be discussed in the following sections.

\subsection{General properties}
Abell 2219   is a galaxy cluster at $z_{l}=0.228$ with X-ray luminosity  $L_x=2  \times 10^{45}$  ergs s$^{-1}$ in the 0.1-2.4 keV  energy range \citep{ebelin}. This makes it  one of the most luminous X-ray clusters in the northern sky. 
Its X-ray surface brightness distribution  has quite elongated structure. 
\\The core is dominated by a massive cD galaxy. \\ \cite{smail} reported the discovery of two systems of giant arcs which allowed them to constrain the (dark) matter distribution in the central region. They found that this is not well aligned with the gas, interpreting this result as the signature of an on-going merger event. 
On the basis of  other multiple images identified by \cite{smith}, \cite{richard} estimated a projected mass within a radius $R < 250$ kpc  
 $M_{SL}=2.33\pm0.23   \times 10^{14} M_{\odot}$.  The Einstein radius  was found to be $(15.6\pm 0.6)$ arcsec.
 \\The study of the internal dynamics of the cluster on the basis of TNG spectroscopic data  \citep{boschin} gives a  value for the line of sight velocity dispersion of  $\sigma_v = 1438^{+109}_{-86}$ km s$^{-1}$, while the mass within the virial region is $\sim 2.8\times 10^{15} h^{-1} M_\odot$.
On the basis of the multi-wavelength analysis obtained using  both optical and X-ray data, \cite{boschin}  also suggested that the cluster is not   dynamically relaxed and likely to be a merging system.
\\At X-ray  wavelengths, Chandra observations reveal  the presence of a large-scale shock front  \citep{milion} with $kT  \lesssim 16$ keV,
high pressure and entropy which can be associated with a merger-driven shock front. 
The X-ray mass of Abell 2219 was estimated by \cite{mahdavi_2008} on the basis of data available in the  Chandra X-ray Observatory public archive, obtaining a value of $M_{500}=17.32\pm6.50 \times 10^{14} M_{\odot}$, and by \cite{mantz_2010} who  obtained a value of  $M_{500}=18.9\pm2.5\times 10^{14} M_{\odot}$ from the analysis   of Chandra and ROSAT  images taken with the Position Sensitive Proportional Counters (PSPC). 

\subsection{Previous weak lensing analyses}

We briefly summarize some results of previous weak lensing analyses of Abell 2219. 
\cite{bardeau_07} investigated this cluster as part of  a sample of eleven X-ray luminous  clusters  selected from the 
XBACs catalog \citep{ebelin_6} in the narrow redshift range at $z = 0.21\pm 0.04$. The weak lensing analysis was based on wide-field images obtained with the CFHT12k camera at the  Canada-France-Hawaii-Telescope (CFHT) in the \textit{B} (5400 s), \textit{R} (6300 s), and \textit{I} (3000 s) bands.
The background galaxies were selected  by applying  the magnitude cut  $22.2< R <25.3$, followed by the color selection \textit{R}-\textit{I} $\gtrsim -0.7$  to isolate galaxies above the cluster red sequence. The resulting source number density is  $8-10$ gal arcmin$^{-2}$.  The mass and the concentration were estimated to be $M_{200}$ = $2094 \pm   435 \times h^{-1}_{70}$ 10$^{12} M_\odot$ and  $c_{200}=3.84 \pm 0.99$.
\\\cite{hoekstra_07} performed  an independent weak lensing analysis  using the same data of \cite{bardeau_07}. After selecting  galaxies in the magnitude range   $21< R <25$,  he used the photometry in $B$ and $R$ bands to remove galaxies that lie on the cluster red-sequence. The shear profile was corrected  for  the residual contamination by increasing the observed shear  by the factor $1+f_{cg}(\theta)$, where $f_{cg}(r)$ is the fraction of cluster galaxies at  radius $r$.
The value of $f_{cg}$ as  function of radius was determined by stacking several other clusters. Using a value of $\langle \beta \rangle=0.54$ inferred  from the  redshift distribution of  the Hubble Deep Field (HDF)  \citep{Fern}, and fitting the tangential shear profile with an NFW model, the resulting virial mass was 
$11.3^{+3.2}_{-2.7} \times h^{-1}$ 10$^{14} M_\odot$. The fit was done assuming one free parameter (the mass) and using the relation between concentration and mass proposed by \cite{bullock_2001}. 
\\\cite{okabe} also performed a weak lensing analysis of Abell 2219 as part of their study of  30 clusters from the 
Local Cluster Substructure Survey (LoCuSS) sample. 
They  used only SUBARU data in \textit{R}  and \textit{V} bands. The exposure times were 24 and 18 minutes, respectively.
The selection of background sources was done combining the magnitude cut $22 \le R \le 26$ with color cuts 
to identify galaxies redder and bluer than the cluster red sequence. 
The optimal selection was found by maximizing the mean amplitude of the tangential shear profile with respect to the color offset from the cluster red sequence.
The resulting number density of background sources was  found to be $\sim10$ gal arcmin$^{-2}$.   They estimate  the virial mass and concentration to be $M_{vir}=9.11^{+2.54}_{-2.06} \times h^{-1}_{72}$ 10$^{14} M_\odot$; $c_{vir}=6.88^{+3.42}_{-2.16}$.  However, given the quality of their fit to the tangential shear profile (reduced $\chi^{2}=2.26$ and the value of the significance probability $Q$ used to quantify the goodness of the
fit of the model), \cite{okabe} concluded that the NFW  model does not describe well the data.

%%%%%%%%%%%%%%%%%%%%%%%%%%%%%%%%%%%%%%%%%%%%%%%%%%%%%%%%%%%%%%%%%%%%%%%%%%%%%%

\subsection{Image reduction}
\label{sect_image_reduction}
\begin{figure}
 %\begin{center}
\centering
\includegraphics[scale=.45]{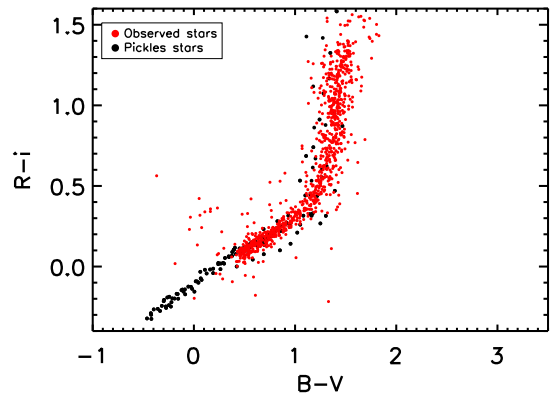}
\caption{ Star observed colors  (red dots) and colors  obtained by the convolution of the stellar observed spectra from 
the Pickles' library with the filter transmission curves  (black dots), after the offsets given in the 
text were applied.}
\label{color_off}
%\end{center}
\end{figure}

The observations of the cluster Abell 2219 were performed   with the  Suprime-Cam  mounted at the prime focus of the SUBARU telescope, an 8.2 meter telescope located  at the summit of the Mauna Kea. The camera   is a mosaic of ten 2048 x 4096 CCDs which covers a $34^{'} \times 27^{'}$ field of view with a pixel scale of 0.20$^{''}$.
We analyze data  publicly available in  the \textit{BVR} bands,    retrieved from the SMOKA Science Archive.
\\Images are reduced using the VST-Tube imaging pipeline \citep{lino}, 
 a  software  designed to work on optical astronomical images taken with different instruments and instrumental setups.
The data reduction consisted in the following steps: 
  overscan correction,  flat fielding,  correction of distortions due to optics and sky background subtraction.
The astrometric solution describing the distortions produced by the optics and telescope was computed for each exposure using the  ASTROMC 
code \citep{mario_2004}. This was then used in the  SWARP tool, which was used to resample and  coadd all exposures.
The  absolute photometric calibration of the images in each filter was  achieved  using standard  Stetson fields.   
The SDSS  \textit{i}-band
image of Abell 2219  was taken with the Large Binocular Camera  mounted on the prime focus of the 8.4m Large Binocular Telescope. 
 LBC has a  field of $23^{'} \times 23^{'}$  and provides images with a sampling of 0.225$^{''}$/pixel. The data were taken in 2010; 
 the data reduction  was done by the LBC Data Centre using a pipeline specifically designed for LBC data.
Magnitudes  in the \textit{BVRi} filters are in the AB system.
The  observation nights of the images in each filter,  
the total exposure time,  the photometric zero point of the final coadded images and the   average FWHM for point-like sources are given in Table \ref{tab1_a2219}. 
Finally, galaxy magnitudes are corrected   for galactic extinction using the maps produced by \cite{Schlegel}.
\\We use the deep \textit{R} band image  to perform  the weak lensing analysis,
 while  photometric catalogs in the  \textit{B} and  \textit{V}  bands are derived using SExtractor in dual-mode, with the \textit{R} band image  used as the detection image. The masking of reflection haloes and diffraction spikes near bright stars is performed with the  ExAM code described in  \cite{joy}.
 We  apply to the \textit{B} and \textit{V} magnitudes the offsets derived within the COSMOS survey by \cite{capak}:
+0.19 (\textit{B}), +0.04 (\textit{V}); according to \cite{joy}, no offset is required for the \textit{R} band.
We convolve the stellar spectra from 
the Pickles' library \citep{pickels} by the transmission curves. 
Comparing  the  colors obtained   with those derived for the stars in our catalog, 
we derive an offset in \textit{i} magnitudes of $+0.065$. Fig. \ref{color_off} shows this comparison 
  after adding the  offsets. We verified that the addition of these offsets improves the match of stellar colors in our data with  colors obtained from the convolution of the stellar  spectra from 
the Pickles' library by the filter transmission curves.

\begin{table}
\begin{center}

%{
%\footnotesize
\resizebox{\hsize}{!}{ 
\begin{tabular}{@{}lccccc}
\hline
 Date & Band& Exp.Time  & zp  & FWHM\\
 &&(s)&&(arcsec)\\
 \hline 25/06/06 & \textit{B}  &$720$ & 27.54&$0.90$ \\
18/07/04 & \textit{V} &  $1080$ & 27.56 &$0.68$ \\
  19/07/04, 14/08/07 & \textit{R}  &$3330$  & $27.59$&$1.09$  \\
   07/05/10 & SDSS \textit{i} &$3005.8$&27.57&$0.89$\\
\hline
\end{tabular}
}
 \caption{Abell 2219  observations summary.} 
  \label{tab1_a2219}
\end{center}
\end{table}

%%%%%%%%%%%%%%%%%%%%%%%%%%%%%%%%%%%%%%%%%%%%%%%%%%%%%%%%%%%%%%%%%%%%%%%%%%
\subsection{Weak lensing analysis}
\label{sect_ksb_abell}
\subsubsection{Shape measurement}

\begin{figure}
\begin{center}
\includegraphics[scale=.54]{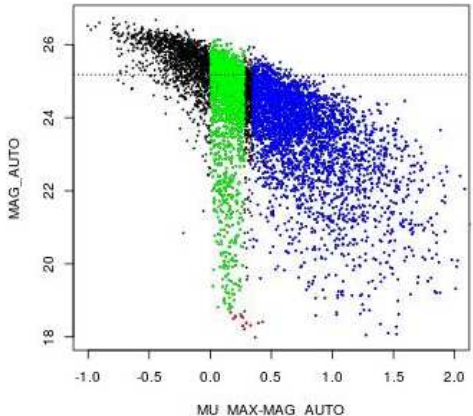}
\caption{ MAG AUTO vs. $\delta$ = MU MAX -MAG AUTO plot.  Stars are plotted with green dots, saturated stars with red dots and  galaxies with blue dots. Sources with size smaller than  stars and sources with ambiguous classification are plotted with  black dots.}
\label{a2219_sg}
\end{center}
\end{figure}

\begin{figure}
 \begin{center}
\includegraphics[scale=.40]{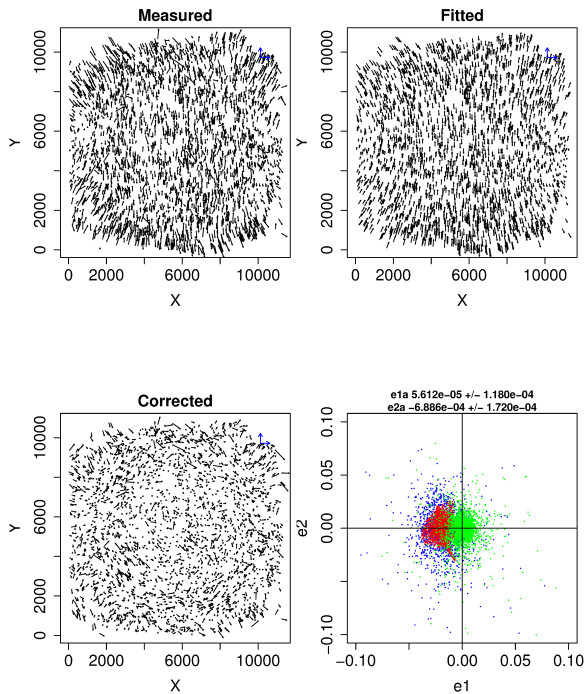}
\caption{PSF anisotropy correction: the first three panels  show the measured, fitted, and residuals
ellipticities as function of the  position (X and Y are in pixels). In the last panel, blue dots are the measured values, the red dots the modelled values and the green dots the values  after the correction.}
\label{a2219_emaps}
\end{center}
\end{figure}

The observed shear signal produced by the gravitational field of the cluster Abell 2219
is reconstructed using  our new KSB implementation  described in Sect. \ref{sect_ksb}. 
The selection of stars used to correct the distortions introduced by the PSF components is made 
in  the magnitude MAG AUTO  vs. $\delta$ = MU MAX-MAG AUTO space. 
\\In Fig. \ref{a2219_sg}, the star locus is defined by  objects populating the vertical branch (green dots). Objects with $\delta$ lower than stars are 
classified as spurious detections (black dots) while galaxies are marked with blue dots. Saturated stars are marked with red dots.
We use stars with \textit{R} magnitude in the range $[19, 22]$ to correct for the effects introduced by PSF anisotropy and seeing.
\\In Fig. \ref{a2219_emaps} the first three panels show stars measured, fitted and residuals ellipticities, 
while the last one shows stars measured ellipticities (marked with blue dots),  stars modelled ellipticities (marked with red dots) and their 
values after the correction (green dots).
 We obtain $ \langle e_{aniso,1}  \rangle = (0.6 \pm 1.2) \times 10^{-4}$,  $ \langle e_{aniso,2} \rangle = (-6.9 \pm 1.7) \times 10^{-4}$.
 \\The tangential and the cross components of the reduced shear, $g_{T}$ and $g_{\times}$,  are 
 obtained from the quantities $e_{1iso}$ and $e_{2iso}$ used, respectively, as  $g_{1}$ and  $g_{2}$ 
  in Eq. \ref{eq_1.0} and Eq. \ref{eqq_1.0}.
The radial profiles are obtained averaging both  shear components in annular regions centered on the BCG.
\\We  exclude  from the catalog  galaxies with  SNe $< 5$  (our choice of SNe $<5$ corresponds to about S/N $<10$), for which the ellipticity measurement is not meaningful, 
and  galaxies with corrected ellipticities  $e_{1iso}$ or $e_{2iso}$  $>1$.
 The final  catalog which will be used for the 
  analysis has a number density of  $\sim 20 $ gal arcmin$^{-2}$ .


%
%%%%%%%%%%%%%%%%%%%%%%%%%%%%%%%%%%%%%%%%%%%%%%%%%%%%%%%%%%%%%%%%%%%%%%%%%%%%%%%%%%%%%%%%%

\subsubsection{Selection of the background galaxies}
\label{sel_met_abell}
In this section, we describe  the selection of the background galaxies obtained with  our method based on colors (Sect. \ref{sel_met} and Sect. \ref{sel_met_sim_uf}). For comparison, we also use the alternative methods introduced earlier and based on  photometric redshifts, magnitude cuts, and identification of the cluster red-sequence.

       \begin{figure*}
  \begin{center}
\includegraphics[scale=.3]{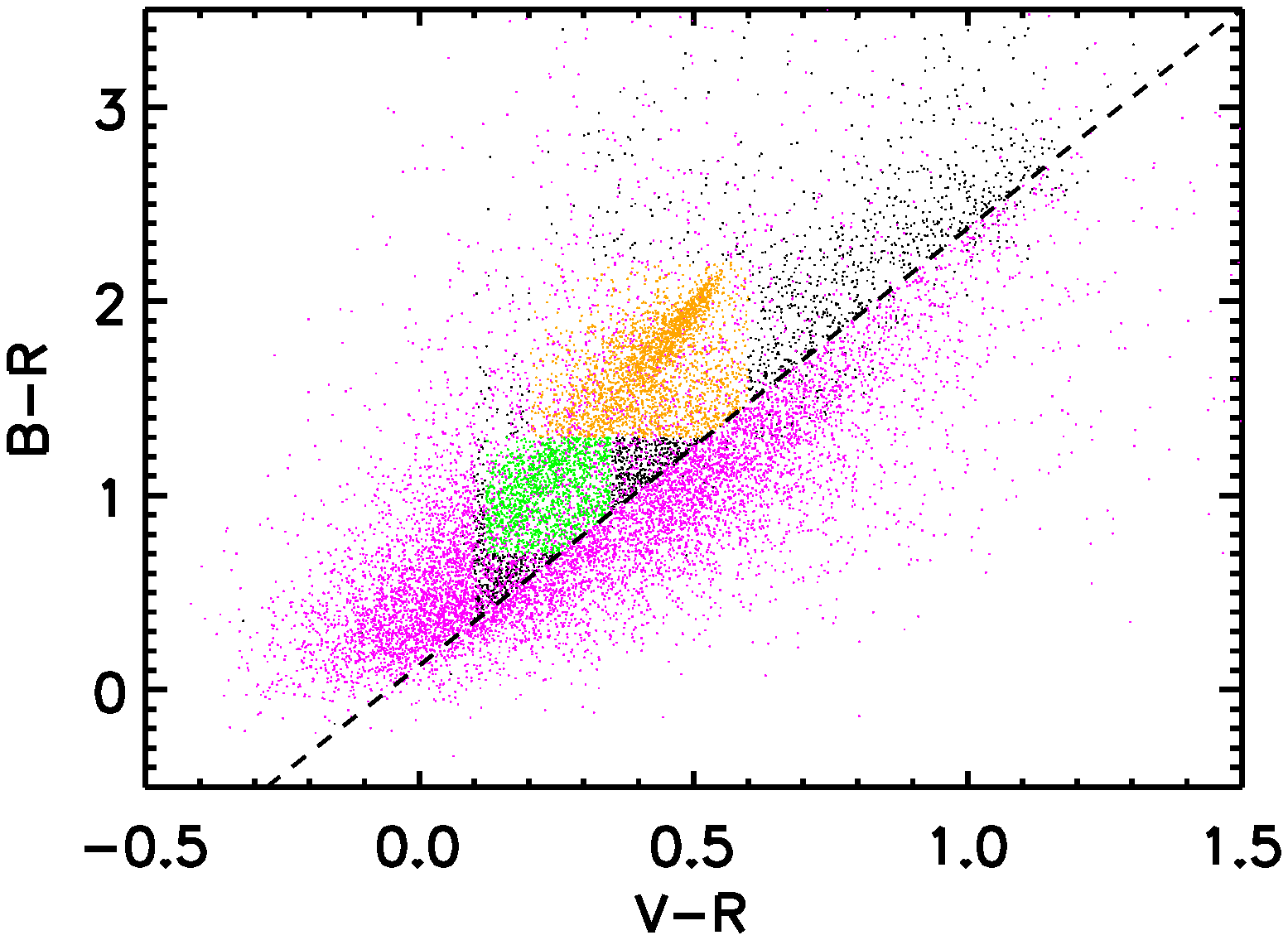}
\includegraphics[scale=.3]{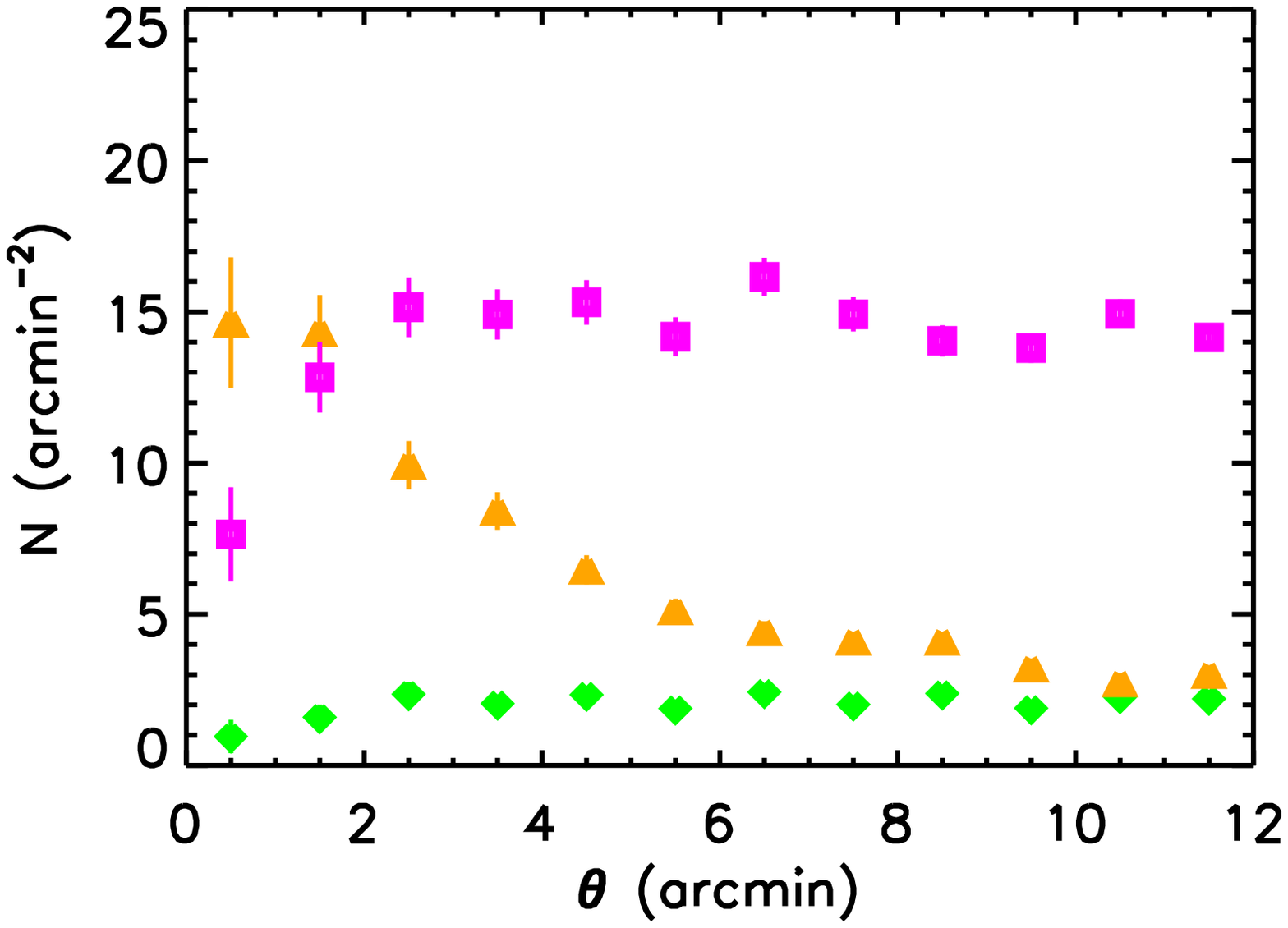}
   \includegraphics[scale=.3]{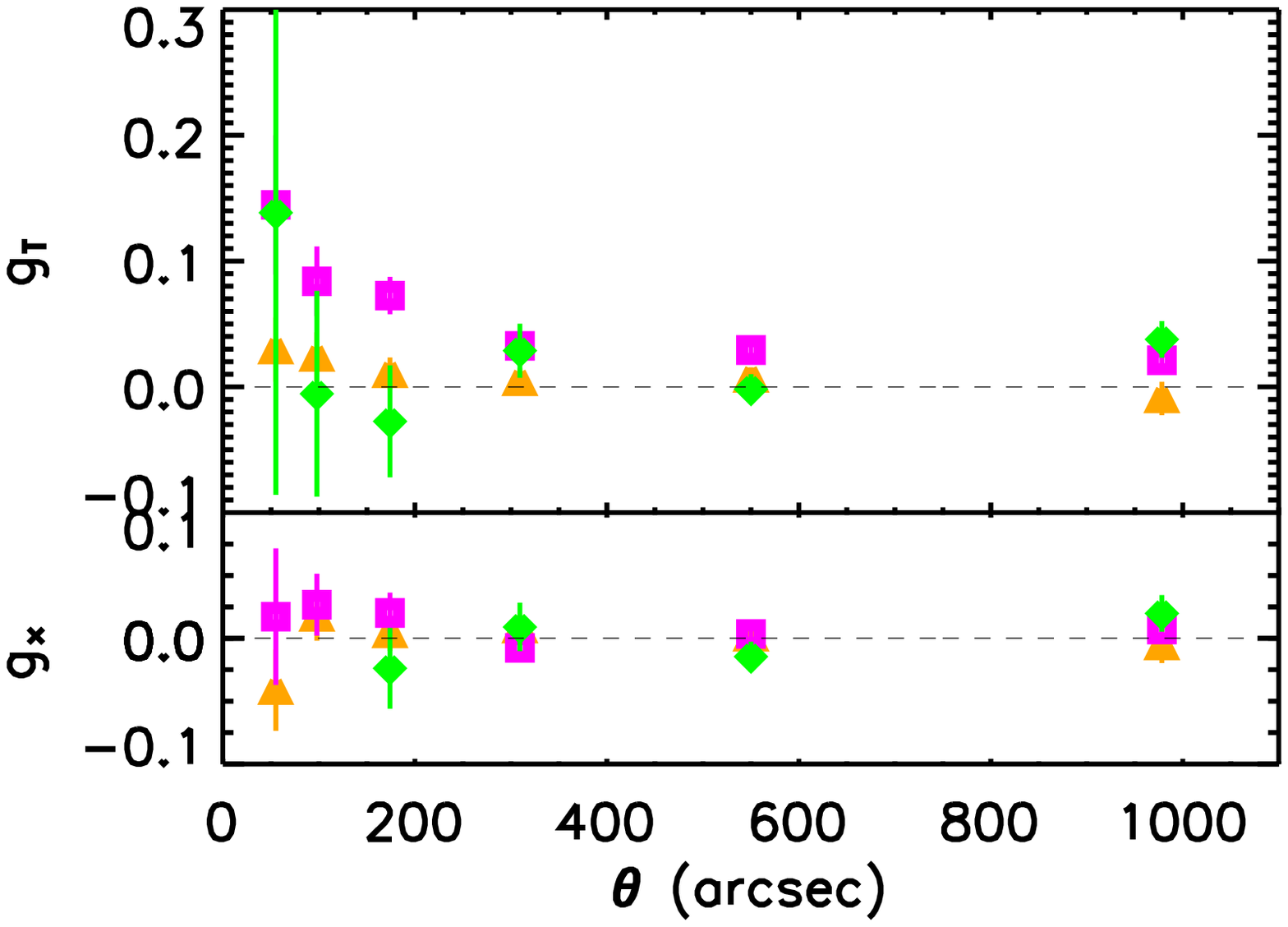}
 \includegraphics[scale=.3]{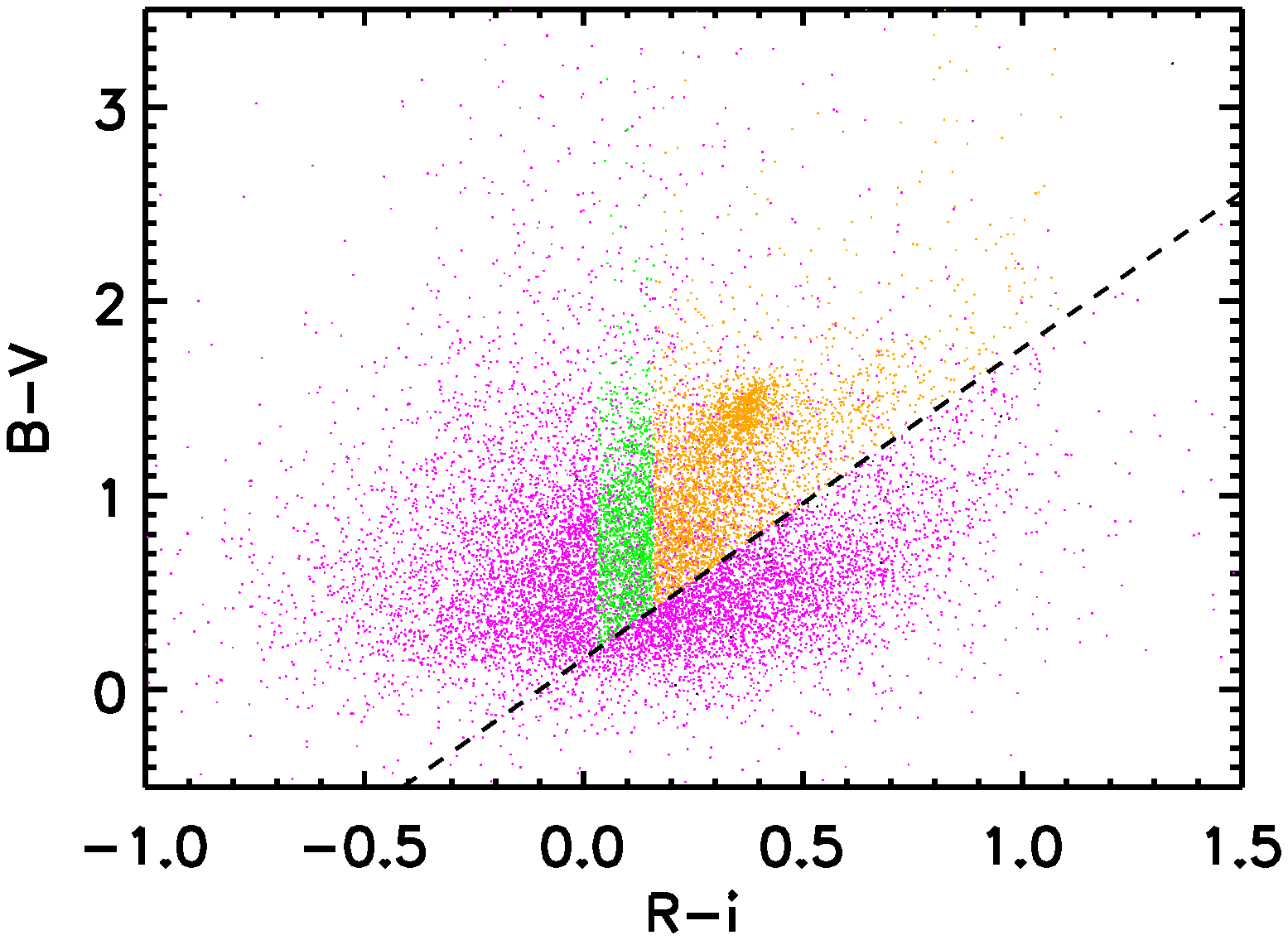}
\includegraphics[scale=.3]{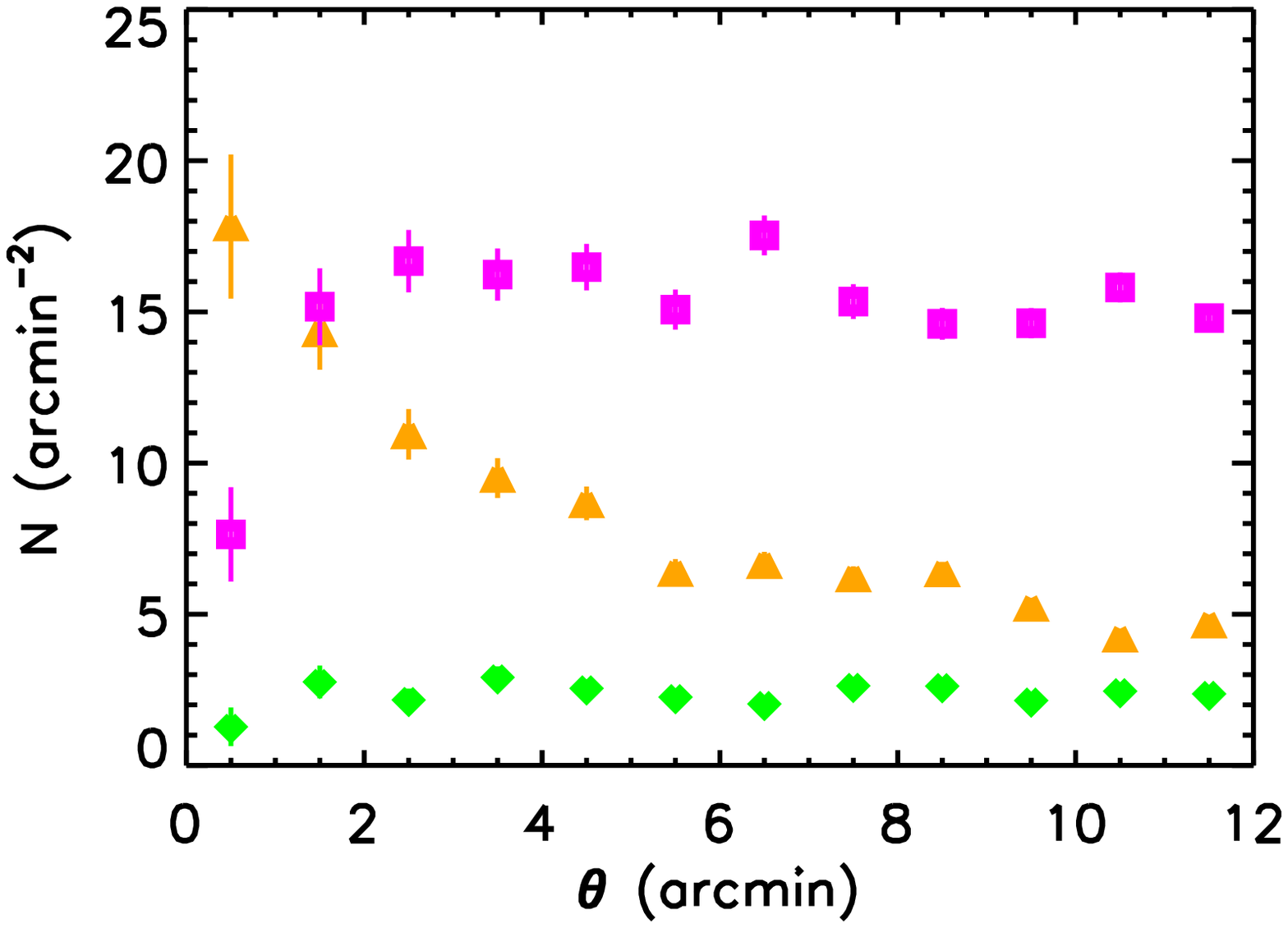}
    \includegraphics[scale=.3]{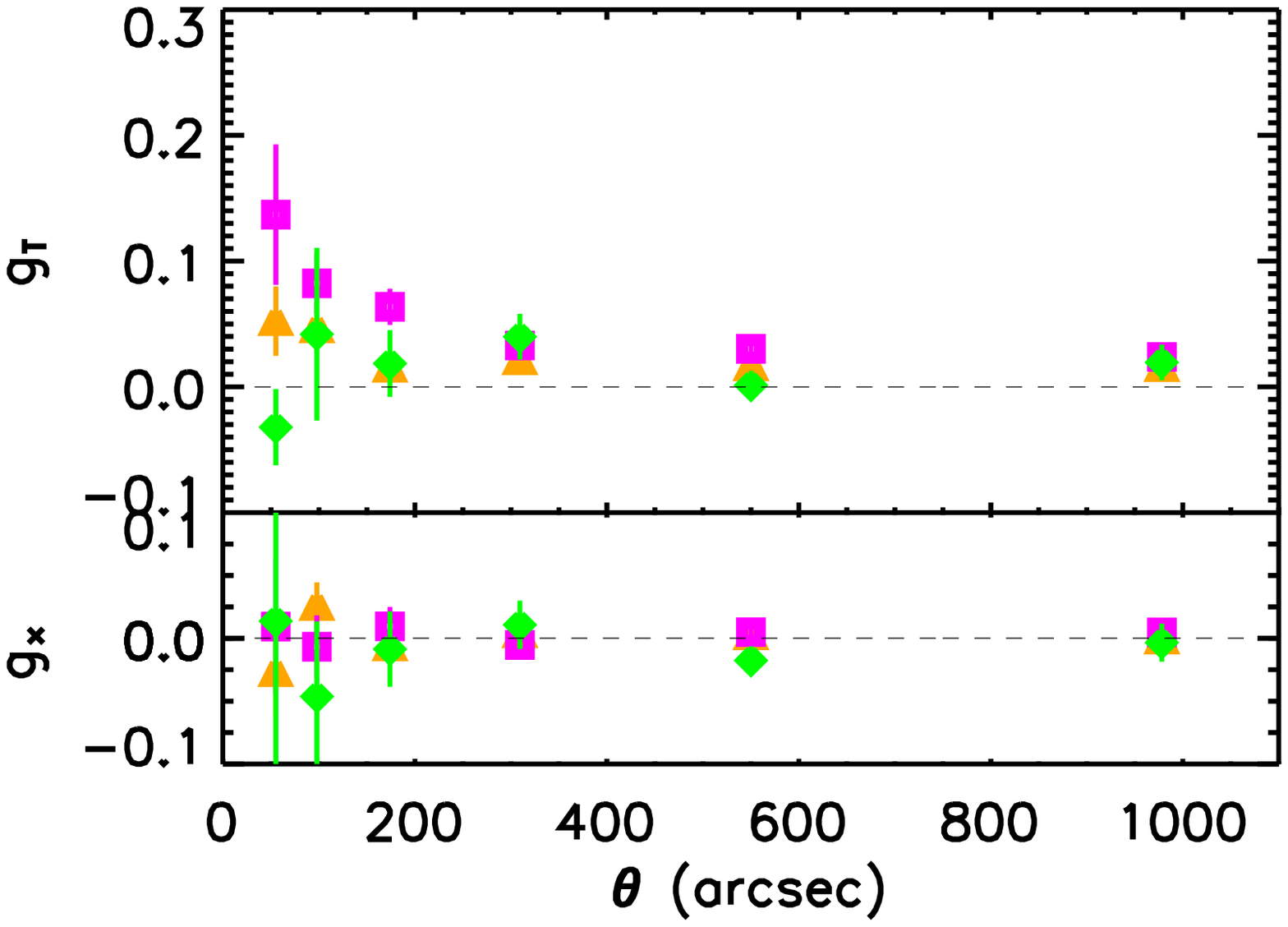}
\caption{Sample selections in CC space. Left-hand panels: CC diagrams for Abell 2219 derived choosing two combinations of photometric bands.
The samples corresponding to cluster member galaxies  are displayed with orange dots and the foreground and background   samples with green and magenta dots respectively. Unclassified sources are plotted with black dots.
Centre: radial number density profiles of galaxies. Background  density profiles (magenta squares) show 
a decrease in the central region (magnification bias and area loss by bright sources). The cluster density profiles (orange triangles)  rise  towards the centre  indicating a population of cluster members. The foreground density profiles are almost flat and are  represented with green diamonds. 
Right-hand panels: $g_T$ and $g_\times$  radial profiles of the corresponding samples.}
\label{fig_col_col_abel}
\end{center}
\end{figure*}

\begin{figure}
     \begin{center}
\includegraphics[scale=.47]{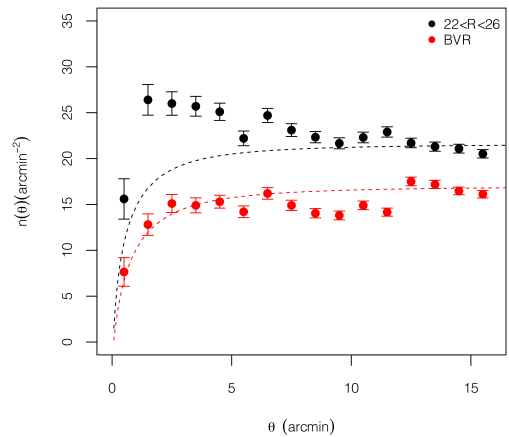}
\caption{Number  counts profile of the background objects selected  in \textit{V}-\textit{R}  vs \textit{B}-\textit{R} diagram  of Abell 2219 (red points) and with a magnitude cut (back points). Model curves are overplotted for comparison.}
\label{magnification_ila}
 \end{center}
\end{figure}

\begin{figure}
      \begin{center}
\includegraphics[scale=.35]{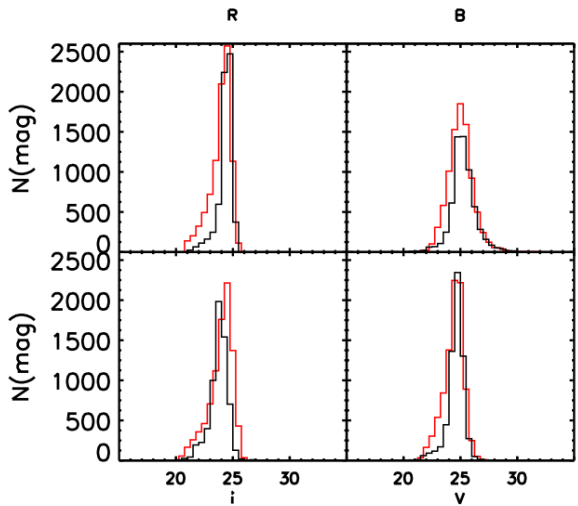}
\caption{Magnitude distributions of objects selected  in \textit{V}-\textit{R}  vs \textit{B}-\textit{R} diagram of the simulated images (black lines) and  of the images of Abell 2219 (red lines).}
\label{limit_mag}
  \end{center}
\end{figure}

\begin{figure}
      \begin{center}
\includegraphics[scale=.43]{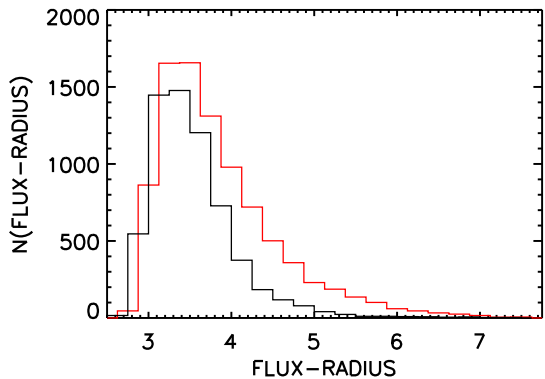}
\caption{Size distribution  for galaxies selected  in \textit{V}-\textit{R}  vs \textit{B}-\textit{R} diagram in the simulations (black line)  compared to that derived for galaxies  in the images of Abell 2219 (red line).}
\label{limit_siz}
  \end{center}
\end{figure}

\paragraph{Selection in color space}
We derive the  \textit{V}-\textit{R}  vs \textit{B}-\textit{R}  and the \textit{R}-\textit{i} vs \textit{B}-\textit{V}  CC diagrams of the galaxies in the field of Abell 2219.
Applying the selection described in the Sect. \ref{sel_met_sim_uf} for the \textit{V}-\textit{R}  and \textit{B}-\textit{R} colors  we obtain a 
background population with a number density
 of $\sim 13 $ gal arcmin$^{-2}$.
The selection in \textit{V}-\textit{R}  vs \textit{B}-\textit{R} space 
 is represented in the upper left panel of the Fig. \ref{fig_col_col_abel}.
In all panels of Fig. \ref{fig_col_col_abel}, galaxy sub-samples are shown using the same
color-coding of Fig. \ref{fig_col_col_sim}.
\\The selection  based on the colors \textit{R}-\textit{i} and \textit{B}-\textit{V} according to the criteria  in  Sect. \ref{sel_met_sim_uf}   is represented in Fig. \ref{fig_col_col_abel} (bottom left panel). The number density of background galaxies is 
$\sim 14$ gal arcmin$^{-2}$. 
\\The   radial number density profiles of cluster members clearly show  overdensities at small radii as expected for cluster member galaxies.  The   radial number density profiles of background  and foreground galaxies do not show clustering at small radii which rises from cluster member galaxies, while they  appear almost flat  since they  are equally distributed on the sky except for  a decrease  at small radii due to area loss by  bright sources and the magnification bias \citep{gray, broad_2008} clearly evident in the background samples and with a lower amplitude in  the foreground samples indicating the presence of background sources in foreground selected samples. 
  We check that the  decrease  at small radii is effectively due to the magnification introduced by lensing  measuring its signal  in the case of  background galaxies selected in the \textit{V}-\textit{R}  vs \textit{B}-\textit{R} space:
   \begin{equation}
        \label{mag_bias}
        n_\mu(\theta) = n_0\mu(\theta) ^{2.5s-1};
  \end{equation}
  with $n_0 = dN_0 (< m_{cut} )/d \Omega$  the unlensed mean number density of background sources for a given magnitude cutoff, described with a power-law  with slope,
  $s = d \log_{10} N_0(< m)/dm > 0.$
  The normalization and slope are estimated  in the outer region i.e. for  $\theta > 10'$.
 We obtain the following values for $s=0.02\pm 0.001$ and $n_0=16.8\pm 0.3$ galaxies arcmin$^{-2}$  that we use to derive the expected magnification for a lens described by a NFW density profile as shown in Fig.\ref{magnification_ila} (red dot line). 
  For comparison, we derived the  expected magnification applied to the sample selected with the magnitude cut $22\le R \le 26$ (back points in Fig.\ref{magnification_ila}). In this case, the rise in the number counts at small radii is a clear evidence of the presence of cluster members in the sample. The model curve (black dot line) is overplotted to data.
 \\The  $g_T$ profiles  of  the cluster and foreground galaxies show  a positive value in some bins.  This is similar to what found in simulations, where we noticed that the subsamples of cluster and foreground galaxies are contaminated by background sources.
  The  $g_T$ profiles of background sources nicely rise at small radii.
The $g_\times$ profiles for each identified population  are consistent with zero, indicating the lack of systematic errors arising from an imperfect PSF correction.
\\The consistency of the real and simulated data is further verified  showing in the Fig. \ref{limit_mag}
 the magnitude  distributions of the background sources selected  in \textit{V}-\textit{R}  vs \textit{B}-\textit{R} diagram in the two cases;
  and in the Fig. \ref{limit_siz},  their size distribution plotting the FLUX RADIUS derived from the \textit{R} images.


\begin{figure}
       \begin{center}
\includegraphics[scale=.42]{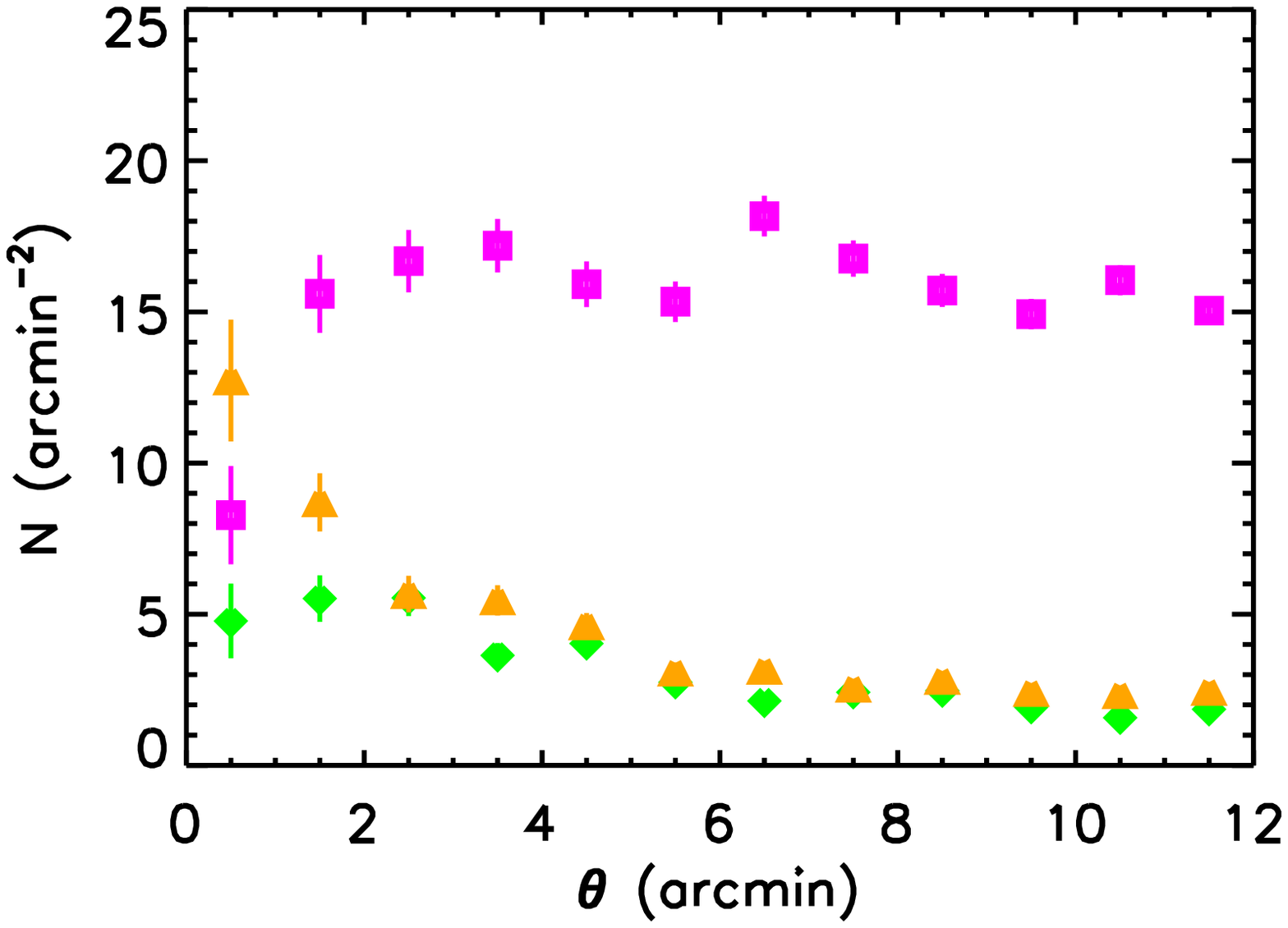}\\
\includegraphics[scale=.42]{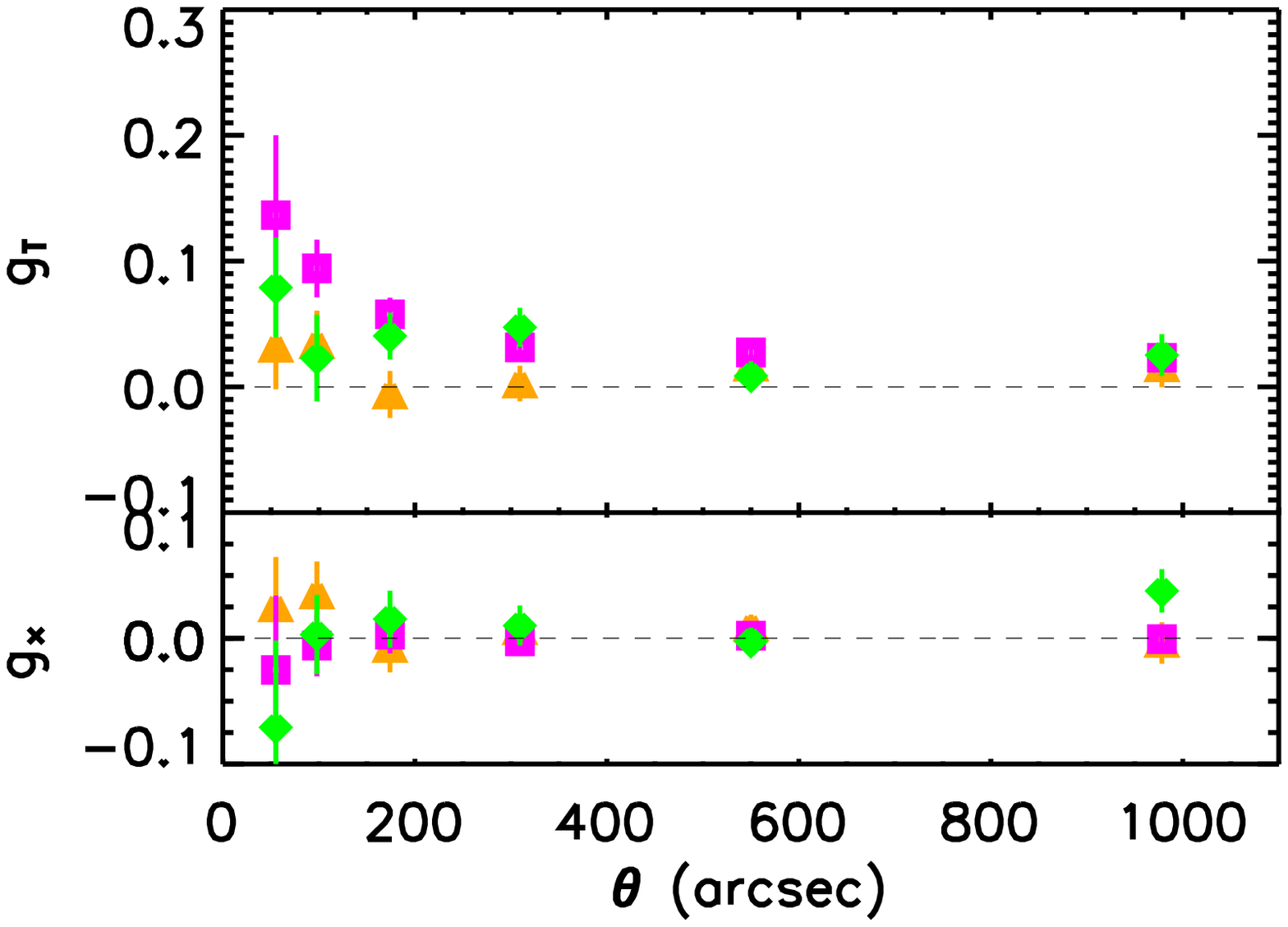}
\caption{Selection with photometric redshifts.  Background galaxies  having  $z_{phot} > 0.27$ are marked with magenta
  squares, cluster galaxies  having  $0.19 \le z_{phot} \le 0.27$ with orange triangles
  and foreground galaxies having  $z_{phot} < 0.19$ with green diamonds. The rise in the  radial number density profile of foreground galaxies is due to the low accuracy
  of photometric redshifts derived caused by a limited spectral coverage.}
       \label{fig_zphot_sel_abel}
      \end{center}
\end{figure}

\paragraph{Alternative selection methods} 
We compare results obtained with our selection method to those obtained with other techniques
to identify lensed galaxies: the identification based on a magnitude cut, that based on the  identification of the cluster red-sequence and finally the use of photometric redshifts. \\The selection based on magnitude cuts consists in considering as background galaxies those having magnitude in the range  $22\le R \le 26$.
\\The selection of background galaxies  on the base of the colors respect to the cluster red-sequence  is done as described in  Sect. \ref{method_al_sim} (Okabe's method).
\\In the following we reported our analysis with photometric redshifts. 
We use the photometry in four bands to derive   photometric redshifts  using   the ZEBRA code in  Maximum-Likelihood mode as described in  Sect.    \ref{method_al_sim}. 
  As done analyzing the simulations, we do not use galaxies with photo-zs $>2$  and $\sigma_z/(1+z) > 0.1$.
  For the galaxies with spectroscopic redshift given in \cite{boschin} we
    find $\langle z_{phot}-z_{spec} \rangle = -0.017 \pm 0.003$.
    The sample given in \cite{boschin} is characterized by galaxies with $z \le 0.4$ so it is not possible to verify the accuracy of sources which have higher  redshift value.
  The accuracy of the photometric redshifts derived for $z >0.4$ is checked  from the comparison with the Sloan Digital Sky Survey  photometric redshifts of galaxies in the Data Release 9, although the latter are obtained from shallower images. 
  We 
    find $\langle z_{phot}-z_{sloan} \rangle = -0.066 \pm 0.007$.
 We select  as background galaxies those having  $z_{phot} > 0.27$ (Fig. \ref{fig_zphot_sel_abel}, both panels, magenta squares), as foreground galaxies those having  $z_{phot} < 0.19$ (Fig. \ref{fig_zphot_sel_abel}, both panels,  green diamonds) and as cluster galaxies those having  $0.19 \le z_{phot} \le 0.27$ (Fig. \ref{fig_zphot_sel_abel}, both panels, orange triangles). The number density of background galaxies selected in this way  is $\sim 14$ gal arcmin$^{-2}$.
The  radial number density profile of foreground galaxies shows an overdensity at small radii indicating the presence of cluster galaxies in the sample (Fig. \ref{fig_zphot_sel_abel}, upper panel). 
This result reflects the low accuracy of the photometric redshifts obtained, due  to the limited number of bands available and in particular to the lack of photometry in the \textit{u} passband.
\\For comparison, we compute the fraction of background  sources common  to both the \textit{BVR} and  the photo-z selections. We find that the two samples overlap at the level of $68\%$. In the case of the  \textit{BVRi} and to the photo-z selections, we find that the $72\%$ of the sources are in common.

%%%%%%%%%%%%%%%%%%%%%%%%%%%%%%%%%%%%%%%%%%%%%%%%%%%%%%%%%%%%%%%%%%%%%%%%%%%%%%%%%%%%

\subsubsection{Mass estimate} 
\label{mas_fit_A2219}
\begin{figure*}
\includegraphics[scale=.3]{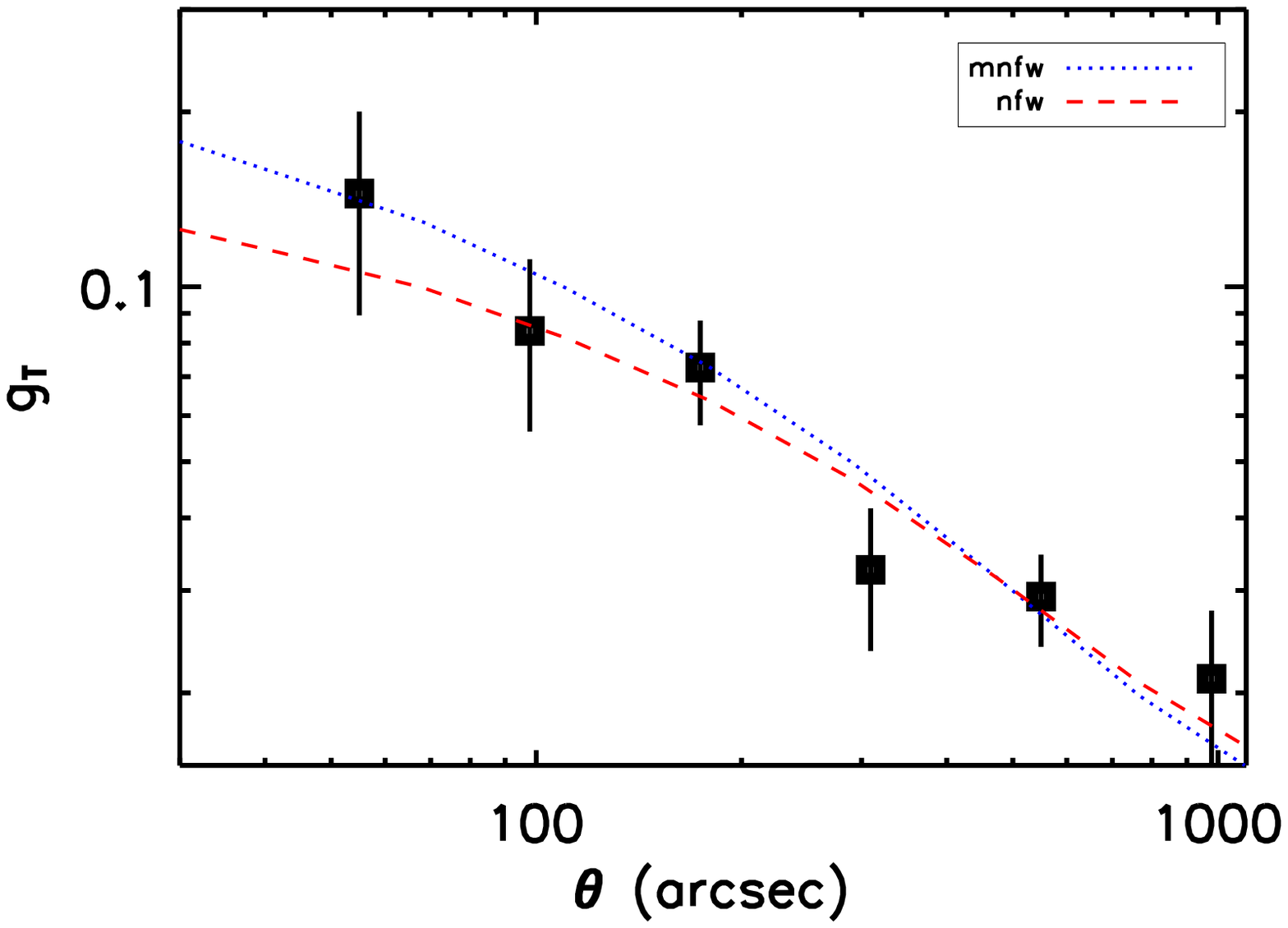}
\includegraphics[scale=.3]{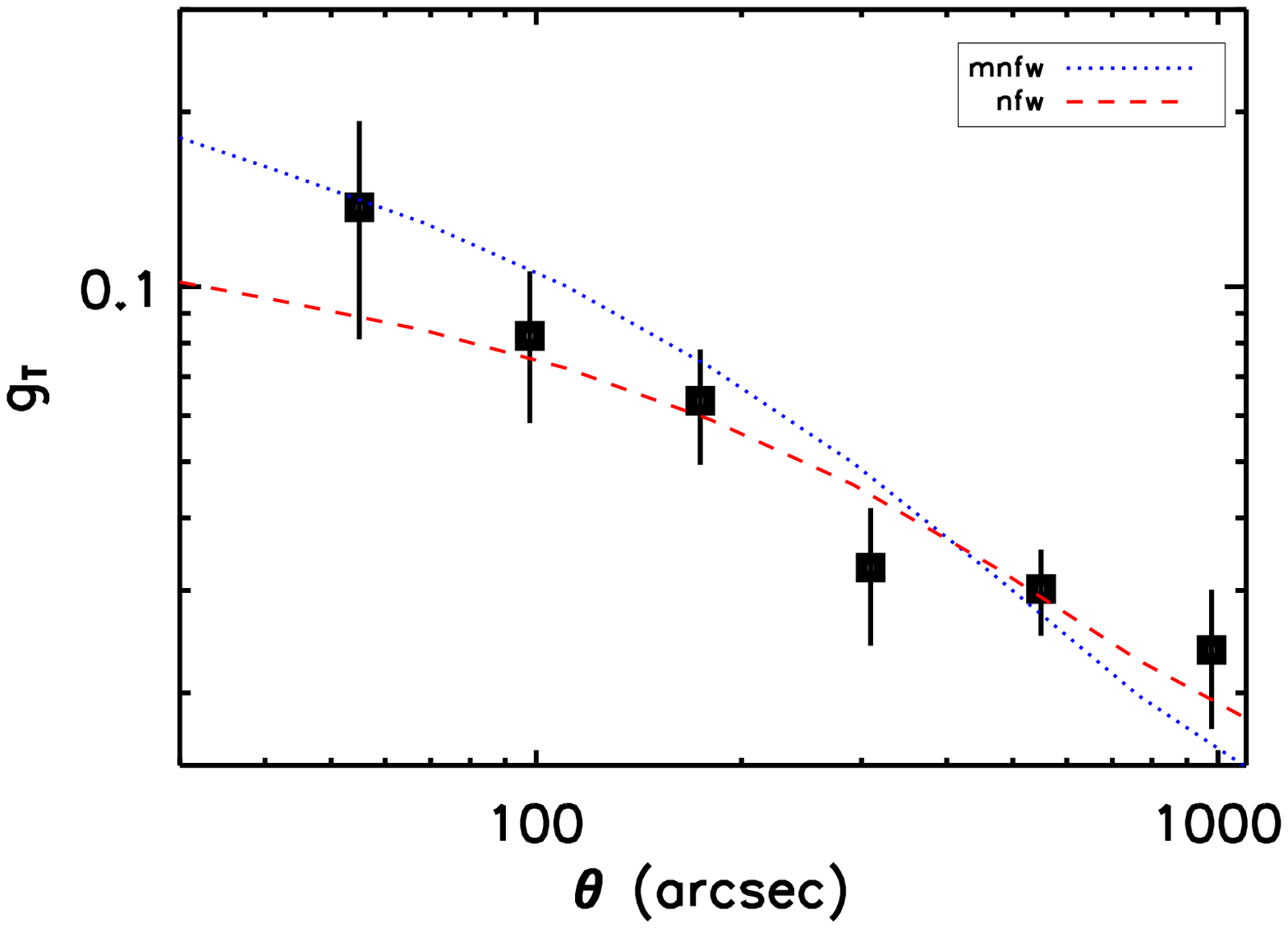}
    \includegraphics[scale=.3]{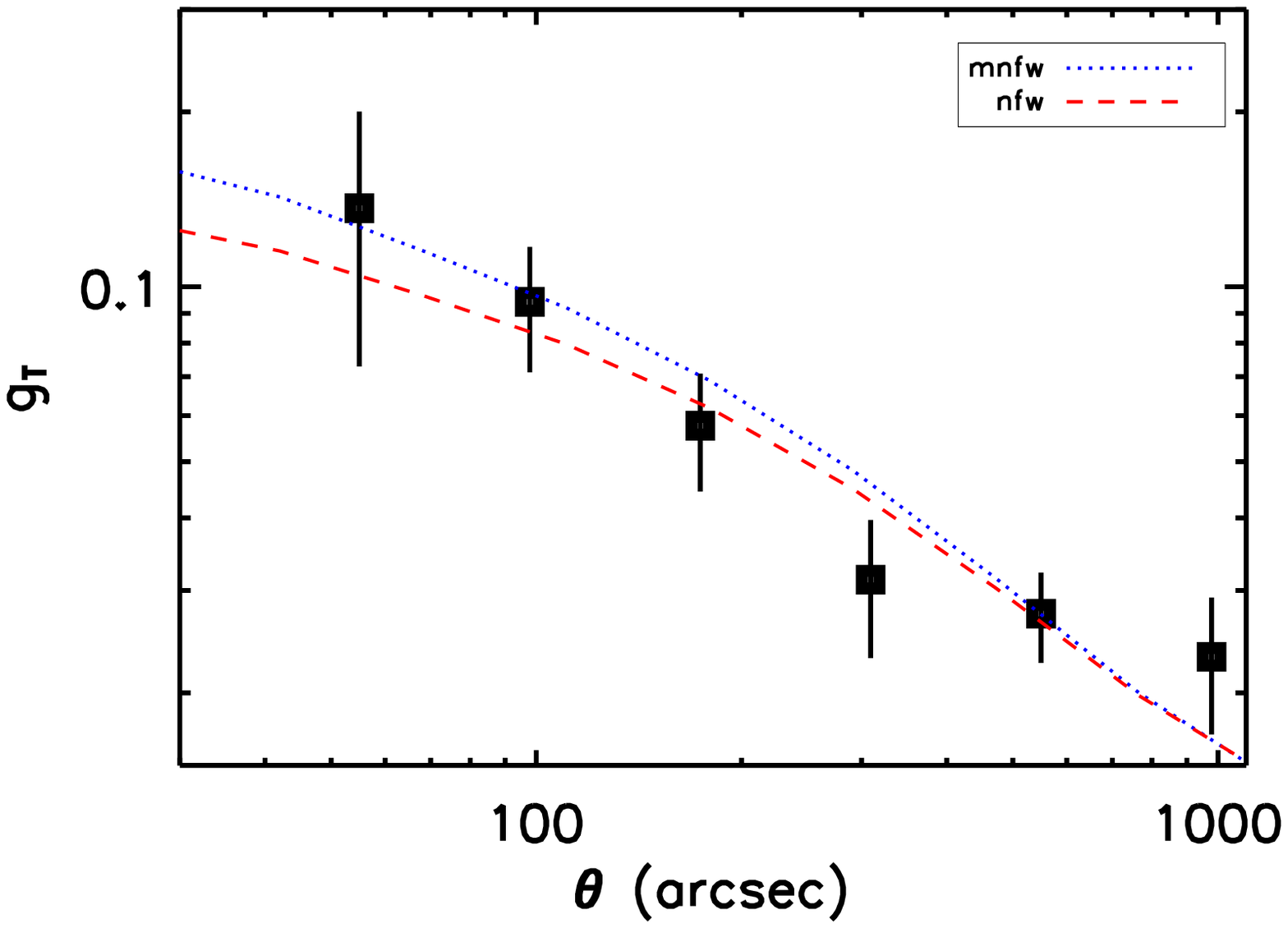}

\caption{ Shear profiles obtained with the different selection methods. Left panel: selection in  \textit{V}-\textit{R}  vs \textit{B}-\textit{R}  space. Central panel: selection in \textit{R}-\textit{i} vs \textit{B}-\textit{V} space. Right panel: selection with photo-zs. The  two models nfw and mnfw are overlapped.}
\label{sshear_fit}
\end{figure*}

Mass estimates  of Abell 2219 are derived 
by fitting the  reduced tangential shear profile with an NFW lens model, as described in the Sect. \ref{mas_fit_sim_new}. In this case, 
 we perform the fit in two different ways in order to compare our results with those presented in literature. In the first case, we use $c_{vir}$ and $M_{vir}$ as free parameters. In the second, we
adopt the mass-concentration relation proposed by \cite{bullock_2001}  to get rid of one of the two parameters. We denote these two methods to estimate the mass with the acronyms {\em nfw} and {\em mnfw}, respectively.
\\In Table \ref{tab_fit_abel}, we summarize the results of our analysis. The first column describes the selection method applied. The second column reports the number densities of background sources selected in each case. The recovered masses and concentrations are given in columns 4-8. For comparison we add the results in the case of a magnitude cut selection and in the case of color-magnitude selection as described in \cite{okabe}. We perform all  the fits in a  range of  $30^{''}< \theta < 1000^{''}$ since 
 the lower limit for the radius ($ \theta=30^{''}$) is chosen sufficiently large  so that the weak lensing limit $g_T \approx \gamma_T $ holds. 
 The shear profiles obtained from the selection in CC spaces and with photo-zs and the corresponding fitted models are displayed in Fig. \ref{sshear_fit}.
In the following, we compare our estimates of the mass of Abell 2219 with previous analyses found in literature.
  The weak lensing analysis made by \cite{bardeau_07} gives a value of
 $M_{200}$ = $2094 \pm   435 \times h^{-1}_{70}$ 10$^{12}$ $M_\odot$ and  $c_{200}=3.84 \pm 0.99.$
\\The comparison of our results with those found in the paper of \cite{bardeau_07} is done performing the NFW fit with the  fixed value of $c_{vir}=4.83$, corresponding to  $c_{200}=3.84$. 
The values of $M_{200}$ found in our work are smaller by a factor of $2.4-3$ with respect to that found in \cite{bardeau_07}.
\\The weak lensing analysis of  \cite{hoekstra_07}  gives a virial mass of  
$11.3^{+3.2}_{-2.7} \times h^{-1}$ 10$^{14}$ $M_\odot$ 
derived using the relation of \cite{bullock_2001}, in agreement with our results.
\\\cite{okabe} derived the weak lensing mass on the base of SUBARU data in only two photometric bands (\textit{R} and \textit{V}).
They found a value of $M_{vir}=9.11^{+2.54}_{-2.06} \times h^{-1}_{72}$ 10$^{14}$ $M_\odot$ and a  concentration of $c_{vir}=6.88^{+3.42}_{-2.16}$. 
\\In order to get a comparison with the result obtained by \cite{okabe}
 we performed  all the fits using the same values of the cosmological parameters of their paper and fixing  $c_{vir}$ to the value found 
  by the authors. The value of the virial mass found by  \cite{okabe} is in agreement with our results except for that derived selecting in the  \textit{R}-\textit{i} vs \textit{B}-\textit{V} space.
However,   the authors concluded that the fit of the  NFW  model to data does not give an acceptable 
result 
on the base of  the
significance probability $Q$ used to quantify the goodness of the fit of the model, 
unlike
 the results of this work. This can be due to the different selections of background sources in both works.
 This is further supported by results on simulations in which the Okabe's method is applied.
 In this case, the fitted mass is lower 
 than
  those derived by the selection based on colors (Tab. \ref{tab_cont}).  \cite{okabe_2013} presented a new analysis with a significantly revision of their color-cut technique
used to select background sources.
The paper shows the results of the
stacked analysis of 50 galaxy clusters but the individual cluster masses are not reported. They claim that the new masses are higher by $14\%-20\%$ 
compared to
 the previous analysis, suggesting that such difference could arise from 
 systematics in the shape measurement methods and contamination by unlensed sources.

  \begin{table*}
\begin{center}
\footnotesize
\begin{tabular}[width=50mm]{ccccccccccc}
\hline
 Selection & density &method  & $M_{vir}$ &  $c_{vir}$ & $M_{200}$ &$c_{200}$ &$M_{500}$  & $\langle  \beta \rangle $ \\
 &  (arcmin$^{-2}$) & &$(10^{15}M_\odot)$&&$(10^{15}M_\odot)$&&$(10^{15}M_\odot)$  &\\

     \hline \textit{V}-\textit{R} vs \textit{B}-\textit{R}   & $13$& mnfw &$1.48_{-0.12}^{+0.14}$ & $4.27$  &  $1.23_{-0.09}^{+0.11}$ & $3.39$ & $0.82_{-0.06}^{+0.07}$& $0.737$ \\      
                             & & nfw& $1.65_{-0.19}^{+0.21}$ & $3.16_{-0.40}^{+0.47}$ & $1.32_{-0.13}^{+0.16}$& $2.47_{-0.32}^{+0.39} $ & $0.81_{-0.07}^{+0.07}$ & \\

              \hline \textit{R}-\textit{i} vs \textit{B}-\textit{V}   & $14$& mnfw & $1.46_{-0.11}^{+0.12}$ & $4.30$  &  $1.21_{-0.09}^{+0.10}$ & $3.40$ & $0.81_{-0.06}^{+0.06}$ &  $0.742$\\
                          
                  & & nfw& $1.93_{-0.24}^{+0.32}$ & $2.48_{-0.40}^{+0.41}$ & $1.50_{-0.17}^{+0.21}$& $1.92_{-0.32}^{+0.33} $ & $0.85_{-0.07}^{+0.07}$ & \\

   \hline $z_{\rm phot}$   & $14$& mnfw & $1.91_{-0.15}^{+0.16}$ & $4.15$  &  $1.58_{-0.12}^{+0.14}$ & $3.29$ & $1.04_{-0.08}^{+0.09}$ &   $0.622$\\
                           & & nfw& $1.93_{-0.26}^{+0.28}$ & $3.50_{-0.54}^{+0.60}$ & $1.56_{-0.19}^{+0.19}$& $2.75_{-0.44}^{+0.49} $ & $0.98_{-0.09}^{+0.09}$  &  \\
                           
                            \hline     $22 \le R \le 26$ & $18$& mnfw & $1.10_{-0.08}^{+0.08}$ & $4.44$  &  $0.91_{-0.07}^{+0.07}$ & $3.52$ & $0.61_{-0.05}^{+0.05}$ &   $0.720$\\
       & & nfw& $1.71_{-0.20}^{+0.24}$ & $2.14_{-0.32}^{+0.36}$ & $1.29_{-0.14}^{+0.17}$& $1.64_{-0.29}^{+0.31} $ & $0.71_{-0.07}^{+0.08}$  &  \\

 \hline    \textit{VR} & $10$& mnfw & $1.33_{-0.11}^{+0.13}$ & $4.33$  &  $1.11_{-0.10}^{+0.10}$ & $3.43$ & $0.74_{-0.07}^{+0.06}$ &   $0.735$\\
       & & nfw& $1.48_{-0.22}^{+0.27}$ & $3.30_{-0.63}^{+0.68}$ & $1.20_{-0.17}^{+0.17}$& $2.58_{-0.51}^{+0.56} $ & $0.73_{-0.08}^{+0.08}$  &  \\

 \hline

\end{tabular}
%}
\end{center}
\caption{Summary of the fit parameters derived for the cluster A2219. Column 1: selection criterium;  Column 2: number density of the sample; 
Column 3: fitting method; Column 4:  $M_{vir}$; Column 5:  $c_{vir}$; Column 6: $M_{200}$; Column 7: $c_{200}$; Column 8: $M_{500}$;  Column 9: $\langle  \beta \rangle $.
For each selection, in the first row we give the results obtained using the relation found by \protect  \cite{bullock_2001} (labelled mnfw) while in the second row (labelled nfw) we keep both $M_{vir}$ and $c_{vir}$ as free parameters. }
\label{tab_fit_abel}
\end{table*}

\section{Conclusion}
\label{conclus}
 Dilution of the lensing signal due to unlensed sources can crucially affect the estimation of the cluster mass. 
  We developed  a  new selection method for the background lensed sources based on the simultaneous
analysis of the shear signal and of the colors of the galaxies,  with the photometry  from the COSMOS used as training set. 
Our selection method can be applied to the weak lensing analysis of clusters,  under different cluster redshifts, masses and filter combinations.
Its performance in the selection  of the background sources  and its effect on
 the dilution bias  are tested using simulations.
To this end, we
  produced realistic SUBARU and LBT  images containing lensing signal with
 the SkyLens code under different  cluster redshifts ($z=0.23 - 0.45$), masses ($M_{vir}$ = $0.5- 1.55 \times10^{15}$ $M_\odot$) and filter combinations.
 We showed with simulations that three photometric bands are enough to obtain a safe sample of
background sources for the weak lensing analysis.
The values of $M_{vir}$ obtained from the fit of the
expected   NFW  profile to the shear  for the simulated clusters support a low contamination by unlensed sources in the samples selected by cuts in the color-color space.
\\We find that the fitted  values of $M_{vir}$ obtained with our selection  agree with the true masses within $\lesssim 10-15\%$. On average, the ratio between estimated and true masses is $\sim0.98\pm0.09$. 
\\The simulations also allowed us to  estimate the residual fraction of unlensed galaxies in the samples of background galaxies, and  the fraction of background galaxies misidentified as foreground and cluster members.
Color selections and photometric redshifts give less than the $ 11\%$ of unlensed sources contaminating the background selected samples, even if the photo-zs  have a low accuracy due to the limited spectral coverage.
 A  fraction of background lensed sources is lost since they are misidentified as foreground and cluster galaxies in a percentage varying for each kind of selection done.
 However, the low accuracy of the derived photo-zs could
 lead to an additional bias in the cluster mass reconstruction due to the wrong identification
 of the value of $\langle\beta\rangle$ when  estimated from the photometric redshifts themselves. Its value is smaller than the true value resulting in a  over-estimate of the  mass of the lens   by $\sim 11\%$  with respect to the input value.
  \\Selections based on a simple magnitude cut or on the analysis of the color-magnitude diagram following the technique described in \cite{okabe} lead to a
  lower mass due to a residual contamination by unlesed sources.
  \\ We further apply the developed selection method to Abell 2219. 
  In our work, we used a new KSB implementation based on the  PSFex code \citep{mario_2015} to 
  measure 
the observed shear signal produced by the gravitational field of the cluster. 
As shown by simulations the high purity of the selected samples obtained  allows us  to derive 
fiducial samples of background sources on which
 an accurate cluster mass reconstruction can be performed. 
 \\The virial mass of Abell 2219  was obtained for each kind of selection by fitting the NFW model to data. 
Our results are in agreement with the previous analysis  of  \cite{hoekstra_07} and  \cite{okabe} (for this last one  makes exception the selection in the  \textit{R}-\textit{i} vs \textit{B}-\textit{V} space).
However differently from our results, \cite{okabe} found that the NFW shear model does not fit the observed shear.
This can arise from  
 different selection criteria and source redshifts identification used in the analyses, which can origin the other different results found in literature.
\\To conclude, in this paper we showed how using colors from three photometric bands can improve the selection of background galaxies for the weak lensing analysis of galaxy clusters: the selection is driven both by an external photometric catalogue with known redshifts (the COSMOS catalogue), and the maximization of the amplitude of the shear in the observed cluster. The procedure proposed in this paper can be easily automated and applied to large datasets.

\section{Acknowledgements}
We acknowledge the support from the LBT-Italian Coordination Facility for the
execution of observations, data distribution and reduction. CG's research is part of the project GLENCO, funded under the
European Seventh Framework Programme, Ideas, Grant Agreement n. 259349. CG thanks CNES for financial support.
We acknowledge support from PRIN-INAF 2014 1.05.01.94.02
Con il contributo del Ministero degli Affari Esteri e della Cooperazione
Internazionale, Direzione Generale per la Promozione del Sistema Paese.

\newpage

\begin{appendix}
 \appendix 
\section{Bias in the estimated ellipticities}
\label{ksb_bias_ell}
\begin{figure}
 \begin{center}
  \includegraphics[scale=.40]{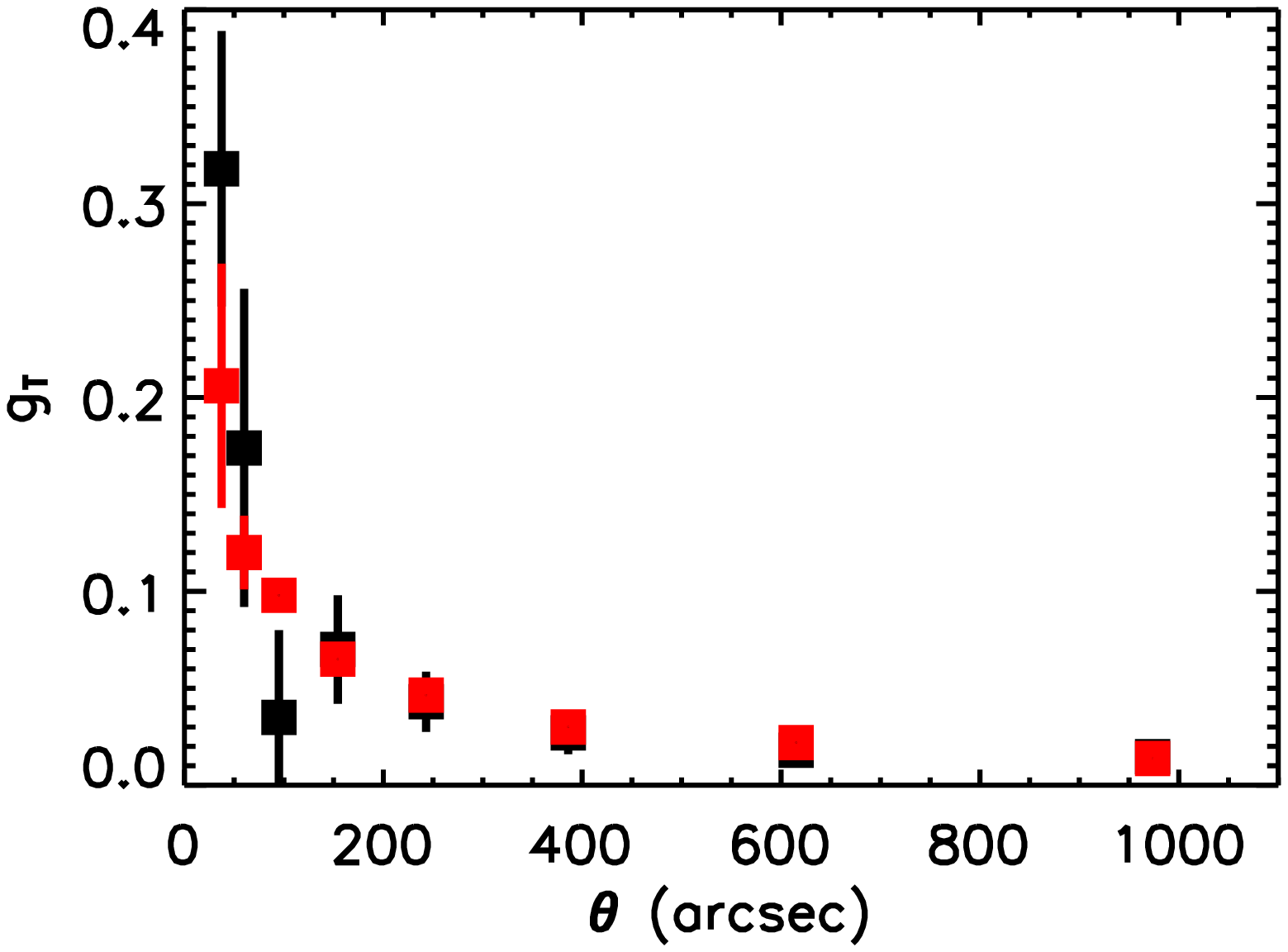}\\
\includegraphics[scale=.40]{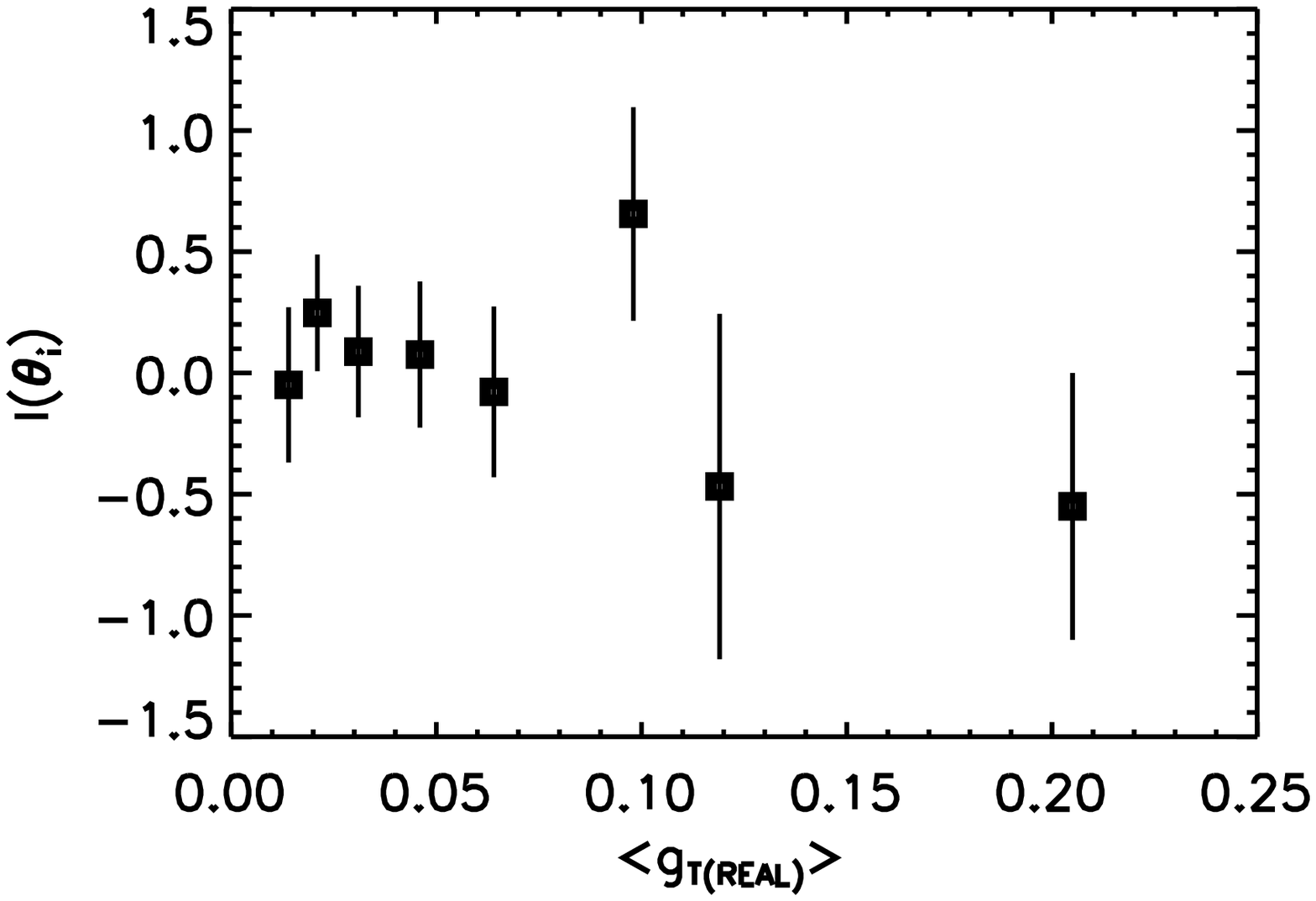}
\caption{ Top panel: $g_T$ radial  profiles  for the simulated lensed galaxies obtained using the measured KSB ellipticities (black squares) and  their input  values (red squares).
Background  galaxies  are selected on the base of their  input value of the redshift in the simulation. Bottom panel:  $l(\theta_i)$ as function of the amplitude of the mean true reduced shear  value  $ \langle g_{T_{(REAL)}} \rangle$.}
\label{fig_bias}
\end{center}
\end{figure}

\begin{figure}
      \begin{center}
\includegraphics[scale=.40]{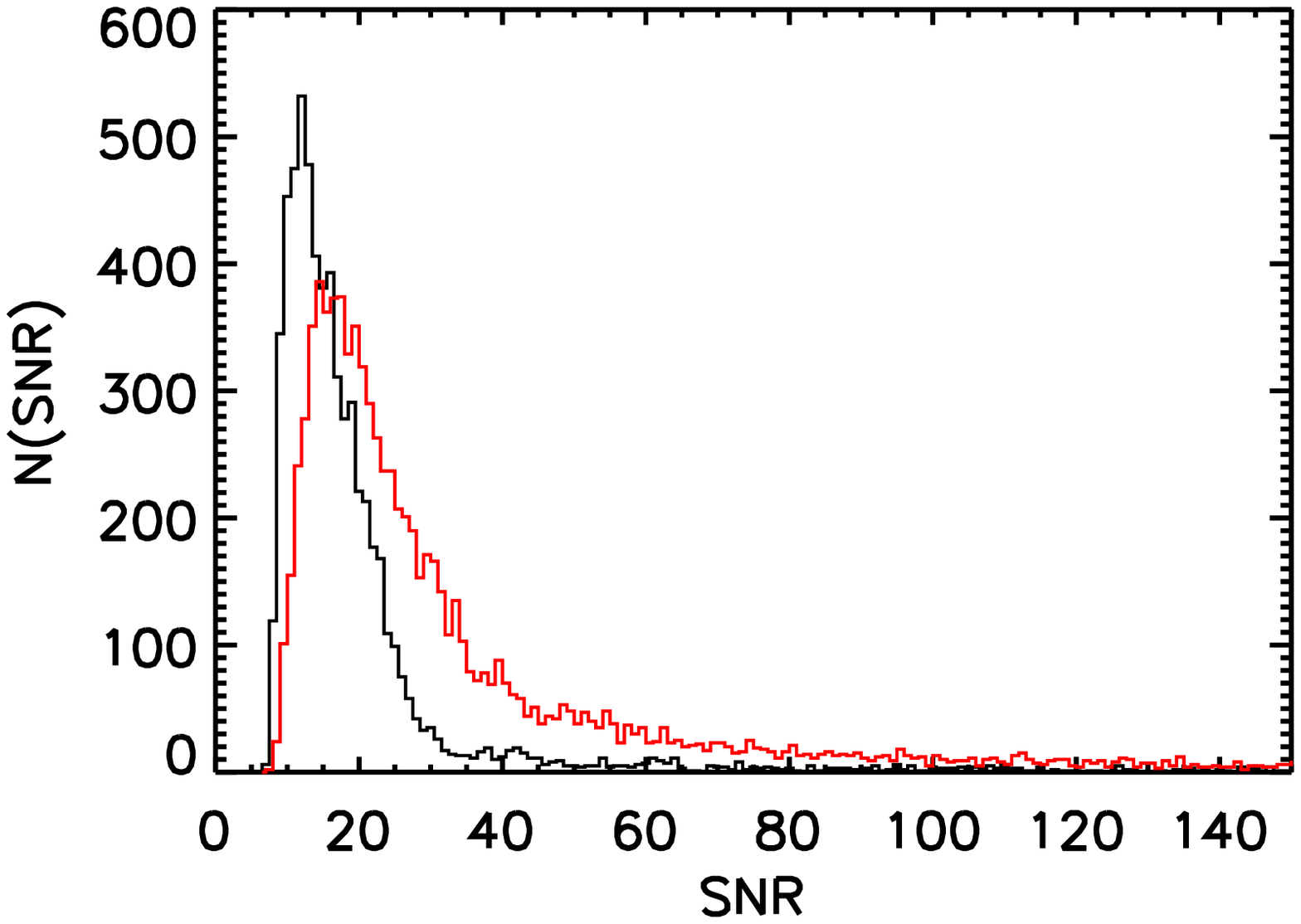}
\caption{Distribution of the signal-to-noise ratio of the background galaxies selected in the \textit{V}-\textit{R} vs \textit{B}-\textit{R} CC diagram  in the simulations (black line) and  in the images of Abell 2219 (red line).}
\label{limit_snr}
  \end{center}
\end{figure}

The accuracy of the weak lensing analysis crucially  depends  on the performance of the method used to correct for the distortions introduced by the PSF components. 
Tests on mock images were  launched in the past to exploit this  issue \citep{erben_2001, heymans, massey, bridle, kitch}.
Ellipticities derived by different  methods for the shape measurement are affected by a multiplicative (m) and an additive (c) bias defined as: $e_{obs}-e_{true} = m e_{true}+c$ \citep{heymans}.
 We  try to quantify this multiplicative bias measuring the shear from the simulated images and comparing it to the true input value.
To this end we use the set of simulations described in the Sect. \ref{sec_sim} tailored on Abell 2219  observations in the \textit{BVRi} bands with exposure times as listed in Table~\ref{tab1_a2219}. Simulated images are characterized by a spatially constant gaussian PSF with FWHM $0.6''$.
We derive radial  profiles of the tangential  component of the reduced shear,  obtained using the measured ellipticities (Fig. \ref{fig_bias} top panel, black squares) and  their input  values (Fig. \ref{fig_bias} top panel, red squares) for background  galaxies  selected on the base of the  input value of the redshift in the simulations. 
In each radial bin $\theta_i$ we compute the difference between the mean  tangential reduced shear obtained using the measured KSB ellipticities
 $\langle g_{T_{(KSB)}} \rangle$ and that derived using their  true input  values $\langle g_{T_{(REAL)}} \rangle$:
\begin {equation} 
l(\theta_i)=\frac{\langle g_{T_{(REAL)}}\rangle - \langle g_{T_{(KSB)}}\rangle}{\langle g_{T_{(REAL)}}\rangle}.
\end{equation}
We derive the mean value of $l$ avoiding the three inner bins characterized by a low statistic. 
In Fig. \ref{fig_bias} (bottom panel), we show how $l(\theta_i)$ changes with the amplitude of the mean true reduced shear  value $\langle g_{T_{(REAL)}} \rangle$.
We find that,  for these simulations, our KSB pipeline on average underestimates the measured shear of about $5\%$ with respect to the input values.
Since the background sources derived  in the analyses  of A2219 and of the simulated data are both dominated by a SNR $\le 40$ (Fig. \ref{limit_snr}), we apply the correction factor of 1.05 to the measured ellipticities and use the so corrected ellipticities to derive the cluster mass.
A dependence of the amplitude of the multiplicative shear calibration factor  on the amplitude of the shear signal 
could arise in  intermediate lensing regimes. We will further  investigate this aspect in a future work.

\end{appendix}

\end{document}